\documentclass[twocolumn,showpacs,superscriptaddress,prb,aps,floatfix]{revtex4-1}
\usepackage{amsmath}
\usepackage{hyperref}
\usepackage{graphicx}
\usepackage{dcolumn}
\usepackage{threeparttable}
\usepackage{multirow}
\usepackage{booktabs}
\usepackage{txfonts}
\usepackage{xcolor}
\usepackage{bm}
\usepackage{amssymb}
\usepackage{amsmath}
\usepackage{latexsym}
\usepackage{epsfig}
\usepackage{amsbsy}
\usepackage{array}
\usepackage{tabularx}
\usepackage{adjustbox}
\usepackage{mathtools}
\usepackage{oubraces}

\begin{document}

\title{Converged GW quasiparticle energies for  transition metal oxide perovskites}

\author{Zeynep Erg\"onenc, Bongjae Kim, Peitao Liu, Georg Kresse, Cesare Franchini}

\email[Corresponding author: ]{cesare.franchini@univie.ac.at}
\affiliation{Faculty of Physics, Computational Materials Physics, University of Vienna, Vienna A-1090, Austria}
\email[Corresponding author: ]{cesare.franchini@univie.ac.at}
\begin{abstract}
The \emph{ab initio} calculation of quasiparticle (QP) energies is a technically and computationally challenging problem.
In condensed matter physics the most widely used approach to determine QP energies is the GW approximation. Although the GW method has been widely applied to many typical semiconductors and insulators, its application to more complex compounds such as transition metal oxide perovskites has been comparatively rare, and its proper use is not well established from a technical point of view.
In this work, we have applied the single-shot G$_0$W$_0$ method to a representative set of transition metal oxide perovskites including 3$d$
(SrTiO$_3$, LaScO$_3$, SrMnO$_3$, LaTiO$_3$, LaVO$_3$, LaCrO$_3$, LaMnO$_3$, and LaFeO$_3$),
4$d$ (SrZrO$_3$, SrTcO$_3$, and Ca$_2$RuO$_4$) and 5$d$ (SrHfO$_3$, KTaO$_3$ and NaOsO$_3$) compounds with different electronic configurations,
magnetic orderings, structural characteristics and bandgaps ranging from 0.1 to 6.1 eV.
We discuss the proper procedure to obtain well converged QP energies and accurate bandgaps within single-shot G$_0$W$_0$ 
by comparing the conventional approach based on an incremental variation of a specific set of parameters (number of bands, energy cutoff for the plane-wave expansion and number of \textbf{k}-points) and the basis-set extrapolation scheme [Phys. Rev. B \textbf{90}, 075125 (2014)]. Although the conventional scheme is not supported by a formal proof of convergence, for most cases it delivers QP energies in reasonably good agremeent with those obtained by the basis-set correction procedure and it is by construction more useful for calculating band structures. In addition, we have inspected the
difference between the adoption of norm-conserving and ultrasoft potentials in GW calculations 
and found that the norm-violation for the $d$ shell can lead to less accurate results in particular for charge-transfer 
systems and late transition metals.
A minimal statistical analysis indicates that the correlation of the GW data with the DFT gap is more robust than the correlation
with the experimental gaps; moreover we identify the static dielectric constant as alternative useful parameter for the approximation of GW gap in high-throughput automatic procedures. Finally, we compute the QP band structure and spectra within the random phase approximation and compare the results with available experimental data.
\end{abstract}

\maketitle

\section{INTRODUCTION}

\begin{figure}[ht!]
\includegraphics[width=0.48\textwidth]{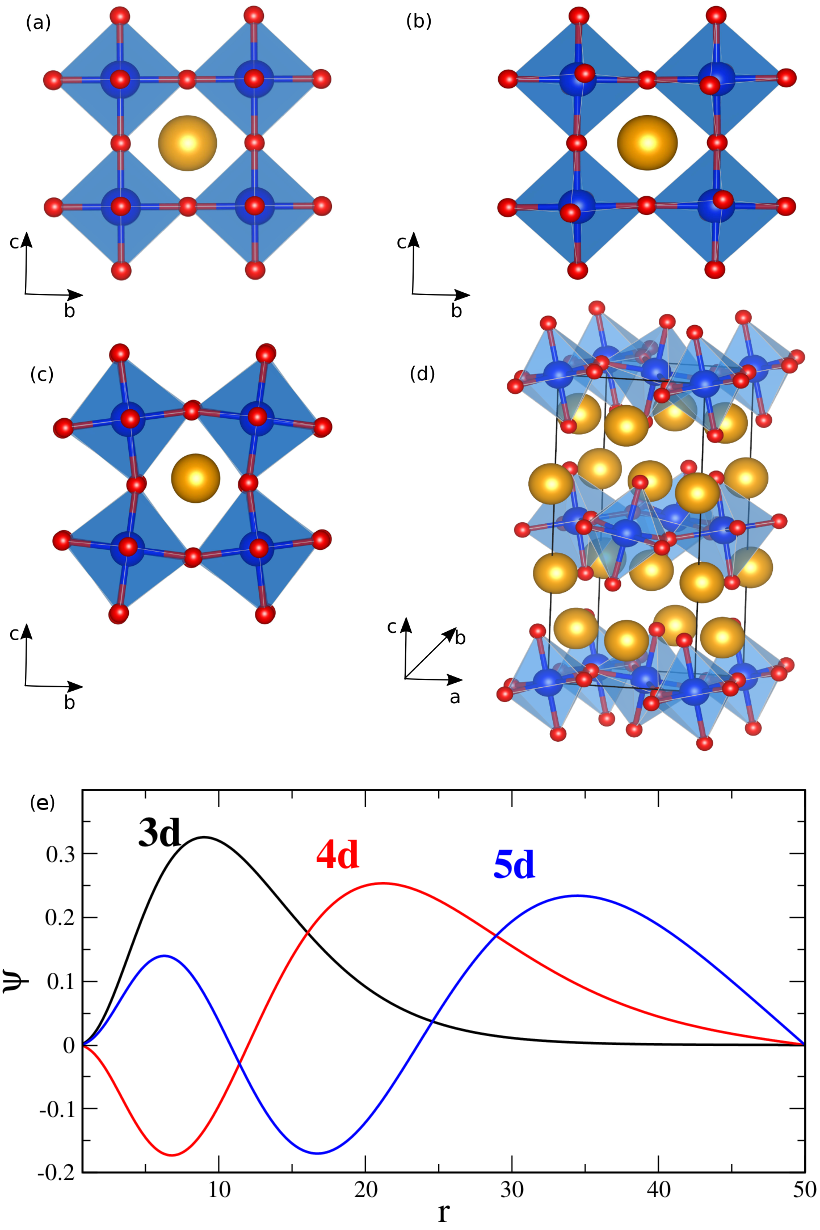}
\caption{(Color online) Different types of lattice distortions (a-d) and TM $d$ orbitals (e) in the oxide perovskites studied in this paper. (a) P$_{m\bar{3}m}$ for SrMO$_3$ (M= Sr, Hf, Zr), SrMnO$_3$ and KTaO$_3$; (b) P$_{nma}$ for LaScO$_3$, LaTiO$_3$, LaCrO$_3$, LaMnO$_3$, LaFeO$_3$, SrTcO$_3$, NaOsO$_3$; (c) P$_{21/b}$ for LaVO$_3$; (d) P$_{bca}$ for Ca$_2$RuO$_4$. The blue and red balls represent the TM and O ions, respectively. (e) Different degree of spatial extension in 3$d$, 4$d$, and 5$d$ orbitals (derived from atomic calculations).}
\label{fig:01}
\end{figure}

Transition metal oxide (TMO) perovskites are a widely studied class of materials owing to the wide spectrum of interesting physical and chemical properties including colossal magnetoresistance~\cite{PhysRevLett.71.2331,RevModPhys.73.583}, metal-insulator
transitions~\cite{Imada1998}, superconductivity~\cite{Bednorz1986,Tokura2003}, two-dimensional electron
gas\cite{Ohtomo2004}, multiferroicity~\cite{Wang09}, spin and charge ordering~\cite{Rao}, bandgaps ranging from the visible to the ultraviolet wavelength~\cite{arima}, as well as chemical and catalytic activity\cite{Zhu}.
Many of these fundamental properties have found technological applications in fields as diverse as fuel cells, spintronic,
oxide electronics and thermoelectricity~\cite{applications}.
More recently, oxides perovskites incorporating 4$d$ and 5$d$ transition metals have attracted increasing attention due to many novel
electronic and magnetic quantum states of matter observed in these compounds, originating from spin-orbit coupling effects. Notable examples are relativistic-Mott iridates~\cite{diracsoc}, Lifshitz magnetic insulators~\cite{Lifshitz}, and different types of anisotropic magnetic interactions~\cite{Jackeli, DM, Kim2017ER}.
This impressive range of properties and functionalities is the result of two main factors: (i) chemical and structural flexibility
and (ii) the occupation and spatial extension of the transition metal $d$ orbitals (see Fig.~\ref{fig:01}).
Oxide perovskites can be formed with cations of different sizes and many different types of lattice and structural distortions
can occur depending on the value of the tolerance factor. The specific type of $d$ orbitals, instead, modulates the degree
of electronic correlation (stronger for localized 3$d$ states), electron and spin itinerancy (larger for 5$d$) and
spin-orbit coupling strength (larger for 5$d$ orbitals). The strong interplay between lattice, spin and orbital degrees of freedom
leads to a rich structural, electronic and magnetic phase diagram, characterized by highly tunable phase transitions.

One of the most important quantities of materials in general, and specifically for TMO perovskites, is the bandgap, which is essential for the characterization and understanding of the electronic structure and is crucial for virtually all possible practical functionalizations. Experimentally, the optical bandgap is measured using spectroscopy techniques such as photoemission, inverse photoemission, X-ray absorption, Electron Energy Loss Spectroscopy, to name a few. Spectroscopy experiments can be interpreted and simulated using the Green's function formalism which allows the treatment of excited states beyond the single particle picture.

Density functional theory (DFT)~\cite{kohn1964} has been the method of choice for decades to estimate the ground state properties of many materials. Despite its great success in interpreting existing results and predicting experimentally difficult to access properties, DFT
is not capable to accurately account for the band gap due to the approximation in treating many-body exchange-correlation effects which
hinder the accurate description and calculation of excitation processes~\cite{VanSchilfgaarde2006b}. An elegant and increasingly popular
method to overcome the limitations of DFT is the GW approximation, originally proposed by Hedin~\cite{Hedin}. This method uses
single-particle Green's functions and many-body perturbation theory to obtain the excitation spectrum by explicitly computing
the self-energy $\Sigma$ of a many-body system of electrons. This is done by expressing $\Sigma$  in terms of the single particle Green's function $G$ and the screened Coulomb interaction $W$, i.e. $\Sigma = iGW$~\cite{Hedin}. The resulting GW bandgaps are much improved compared to the DFT ones and often very close to the measured values~\cite{VanSchilfgaarde2006b, PhysRevLett.45.290, PhysRevB.25.2867, Hybertsen1985, Hybertsen1986, aulbur, GWKresse}.

In GW calculations, it is common to start from DFT orbitals, with which the initial G and W are constructed. There exist different GW schemes depending on the way W and G are updated. The most common choice is the so-called single-shot G$_0$W$_0$ starting from DFT orbitals.
This usually delivers band gaps in good agreement with experimental measurements~\cite{GWGK1, GWGK2}.
The practical disadvantage of the GW method is the large computational cost and memory requirements due to the high number of unoccupied bands (and therefore number of plane wave (pw) basis functions, $N_{pw}$) required for the accurate calculation of the self-energy and the response function.
The convergence of the QP energies with respect to the number of basis functions $N_{pw}$ is, therefore, a particularly crucial issue:
even for small systems, such as ZnO, over thousand bands are necessary to achieve well-converged results~\cite{Shih2010, Klimes2014}.
To address this issue, Klime{\v{s}} \emph{et al.} have recently  proposed a finite-basis-set correction scheme~\cite{Klimes2014}
based on the formal proof that QP energies converge like 1/$N_{pw}$~\cite{Harl2008, Bjorkman2012,PhysRevB.87.075104}. Within this scheme, well-converged QP energies extrapolated to the infinite-basis-set limit were obtained for a representative material dataset including 24 elemental and binary semiconductors and insulators~\cite{Klimes2014}. Moreover, the authors pointed out the advantage of using norm-conserving (NC) projector augmented wave (PAW) potentials,
instead of the commonly employed ultrasoft (US) ones, since US-PAWs were found to underestimate the scattering probability from occupied into high energy unoccupied orbitals~\cite{Klimes2014}.

These computational limitations have inhibited the application of GW for larger systems like perovskites, despite some efforts 
devoted to speeding up GW calculations~\cite{PhysRevB.94.165109,arXiv:1707.06752}. While there are relatively many
GW studies on (non-TMOs) hybrid halide perovskites~\cite{Umari2014, PhysRevB.90.245145, IvanoAPL2014, Bokdam2016},
the assessment of GW for TMOs perovskites is scarce~\cite{Friedrich2010, Nohara2009, BBO, CF1, Kang2015a, CF2, Lany},
in particular for 4$d$ and 5$d$ perovskites~\cite{Sousa2007, PhysRevB.93.075125, He2014}.
The scope of the present paper is the calculation of accurate QP energies at the G$_0$W$_0$ level for a representative dataset of 3$d$, 4$d$, and 5$d$ TMOs perovskites with different types and fillings of the TM $d$-orbitals, different crystal structure and lattice distortions and different magnetic orderings (see Tab.~\ref{tab1:structures}).
Specifically, we will consider:
1) Non-magnetic (NM) d$^0$ cubic perovskites: Sr$M$O$_3$ ($M$= Ti, Hf, Zr) and KTaO$_3$;
2) non-magnetic and structurally distorted 3d$^0$ LaScO$_3$;
3) magnetic d$^3$ cubic perovskites SrMnO$_3$. Note that to model the G-AFM ordering it is necessary to adapt a supercell containing 4 formula units;
4) magnetic and structurally distorted systems:
(a) 3$d$ La$M$O$_3$ ($M$= Ti, V, Cr, Mn, Fe),
(b) 4$d$  SrTcO$_3$ and Ca$_2$RuO$_4$, and
(c) 5$d$ NaOsO$_3$ (in this case we have included spin-orbit coupling, SOC).

We will inspect and compare two different procedures to compute QP energies and bandgaps: 
(i) In the first scheme the QP energies are not explicitely extrapolated to the infinite basis set limit, instead the convergence is inspected with respect to
the number of basis functions $N_{pw}$, the number of \textbf{k}-points and the total number of bands $N$~\cite{Setten}, and (ii) the basis set corrected method with QP energies extrapolated to $N_{pw}  \rightarrow \infty$.

Also, we will determined a minimal technical set-up to achieve sufficiently well-converged values in standard GW calculations without basis-set extrapolation,
which we will adopt to compute band structures, obtained by employing Wannier-function fitting of the QP energies (not feasible within the basis-set correction scheme), and optical spectra calculated from the frequency dependent dielectric tensor. In addition, we will test and discuss the choice of the PAW by comparing US- and NC-based results.  As we will see, the convergence rate is generally highly system dependent, 
as already pointed out in recent studies~\cite{Klimes2014, Gulans, Setten}, and is largely influenced 
by the orbital character and by the type of gap.

The paper is organized as follows: the first part is focused on the description of the two convergence procedures and on the computational setup. The main core of this article is the result section that is divided into three parts dedicated to the analysis of the convergence criteria, to the correlation analysis and to the discussion of the electronic structure and optical spectra.

\begin{table*}[ht!]
\caption{Fundamental characteristic of the TMO perovskites dataset used in this study. Crystal structures: C = cubic,  T = tetragonal, O = orthorombic, M = monoclinic; electronic configuration of the transition metal $d$-shell decomposed over $t_{2g}$ and $e_g$ states;
ground state magnetic ordering: NM = nonmagnetic, and different type of antiferromagnetic (AFM) spin configurations~\cite{PhysRevB.86.235117, PhysRevB.90.039907}.
The crystal structures and atomic positions are taken from the following experimental studies:
SrTiO$_3$ Ref.~\onlinecite{Swanson1954},
SrZrO$_3$ Ref.~\onlinecite{Smith1960},
SrHfO$_3$ Ref.~\onlinecite{Kennedy1999},
KTaO$_3$  Ref.~\onlinecite{Sigman2002},
SrMnO$_3$ Ref.~\onlinecite{PhysRevB.74.144102},
LaScO$_3$ Ref.~\onlinecite{Geller1957},
LaTiO$_3$ Ref.~\onlinecite{Cwik2003},
LaVO$_3$  Ref.~\onlinecite{Bordet1993},
LaCrO$_3$ Ref.~\onlinecite{States1977},
LaMnO$_3$ Ref.~\onlinecite{Elemans1971},
LaFeO$_3$ Ref.~\onlinecite{Dann1994},
SrTcO$_3$ Ref.~\onlinecite{Rodriguez2011},
Ca$_2$RuO$_4$ Ref.~\onlinecite{Braden1998},
NaOsO$_3$ Ref.~\onlinecite{Shi2009}.
For SrMnO$_3$ we have adopted the calculated lattice constant for the G-type AFM cubic phase, 3.824~$\AA$~\cite{PhysRevB.74.144102}, slightly larger than the corresponding experimental value, 3.80~$\AA$~\cite{doi:10.1143/JPSJ.37.275}.
}
\begin{adjustbox}{width=\textwidth,center}
\begin{tabular}{cccccccccccccccc}
\hline\hline

SrTMO$_3$ (TM= Ti, Zr, Hf) & KTaO$_3$          & LaScO$_3$   & SrMnO$_3$         & LaTiO$_3$   &  LaVO$_3$     & LaCrO$_3$   & LaMnO$_3$            & LaFeO$_3$           & SrTcO$_3$   & Ca$_2$RuO$_4$        & NaOsO$_3$   \\
\hline
C-P$_{m\bar{3}m}$          & C-P$_{m\bar{3}m}$ & O-P$_{nma}$ & C-P$_{m\bar{3}m}$ & O-P$_{nma}$ &  M-P$_{21/b}$ & O-P$_{nma}$ & O-P$_{nma}$          & O-P$_{nma}$         & O-P$_{nma}$ & O-P$_{nma}$          & O-P$_{bca}$ \\
$d^0$                      &       $d^0$       &     $d^0$   & t$^3_{2g}$        & t$^1_{2g}$  & t$^2_{2g}$    &  t$^3_{2g}$ &  t$^3_{2g}$e$^1_{g}$ & t$^3_{2g}$e$^2_{g}$ & t$^3_{2g}$  & t$^3_{2g}$e$^1_{g}$  &   t$^3_{2g}$\\
NM                         & NM                &  NM         &  G-AFM            & G-AFM       &  G-AFM        &     G-AFM   &  A-AFM               &    G-AFM            &      G-AFM  &      AFM             &    G-AFM  \\
\hline\hline
\label{tab1:structures}
\end{tabular}
\end{adjustbox}
\end{table*}


\section{TECHNICAL AND COMPUTATIONAL DETAILS}

The calculations presented in this paper were conducted using the Vienna Ab Initio Simulation Package ({\tt{VASP}})~\cite{Kresse1993,Kresse1996}
in the framework of the PAW method~\cite{Blochl1994}. The many-body Schr\"odinger equation was solved within the single-shot G$_0$W$_0$ approximation starting from DFT orbitals obtained using the generalized gradient parametrization introduced by
Perdew, Burke, Ernzerhof (PBE)~\cite{PhysRevLett.78.1396}. When the GGA was not able to open the gap a small on-site Hubbard $U$ was added following the scheme of Dudarev~\cite{Dudarev1998}
(LaTiO$_3$ and LaVO$_3$, $U-J=2$~eV).
The one particle Green's functions constructed from PBE eigenfunction and the dynamically screened Coulomb interaction $W$ was computed
from $G_0$ within the random phase approximation (RPA). The details of the implementation can be found in Ref.~\onlinecite{PhysRevB.74.035101}.
For the calculation of the polarizability, we have used a discretized frequency grid
with about 70 frequency points. This choice should guarantee a reasonably good convergence
of the gap with error of the order of $\approx$ 50 meV.
We have used crystal lattices and atomic positions derived from the experiment, all references are listed in Tab.~\ref{tab1:structures}.

The convergence criteria followed to calculate the response function and the correlation part of the self-energy, which requires
a summation over many empty states, as well as the dependence of the results with respect to the \textbf{k}-point sampling
are discussed in the next subsections \ref{ss:conentional} and \ref{ss:basis}. We have followed and compared two alternative strategies to reach converged results:
(i) so-called non-extrapolated method because it does not involve any extrapolation to large $N_{pw}$;
we refer to this method as \emph{conventional}, since this is the scheme typically used in GW calculations.
(ii) The basis set corrected method which does include an extrapolation of the QP energies to $N_{pw} \rightarrow \infty$.

\subsection{The conventional non-extrapolated method} \label{ss:conentional}
\begin{figure}[hb!]
\includegraphics[width=0.48\textwidth]{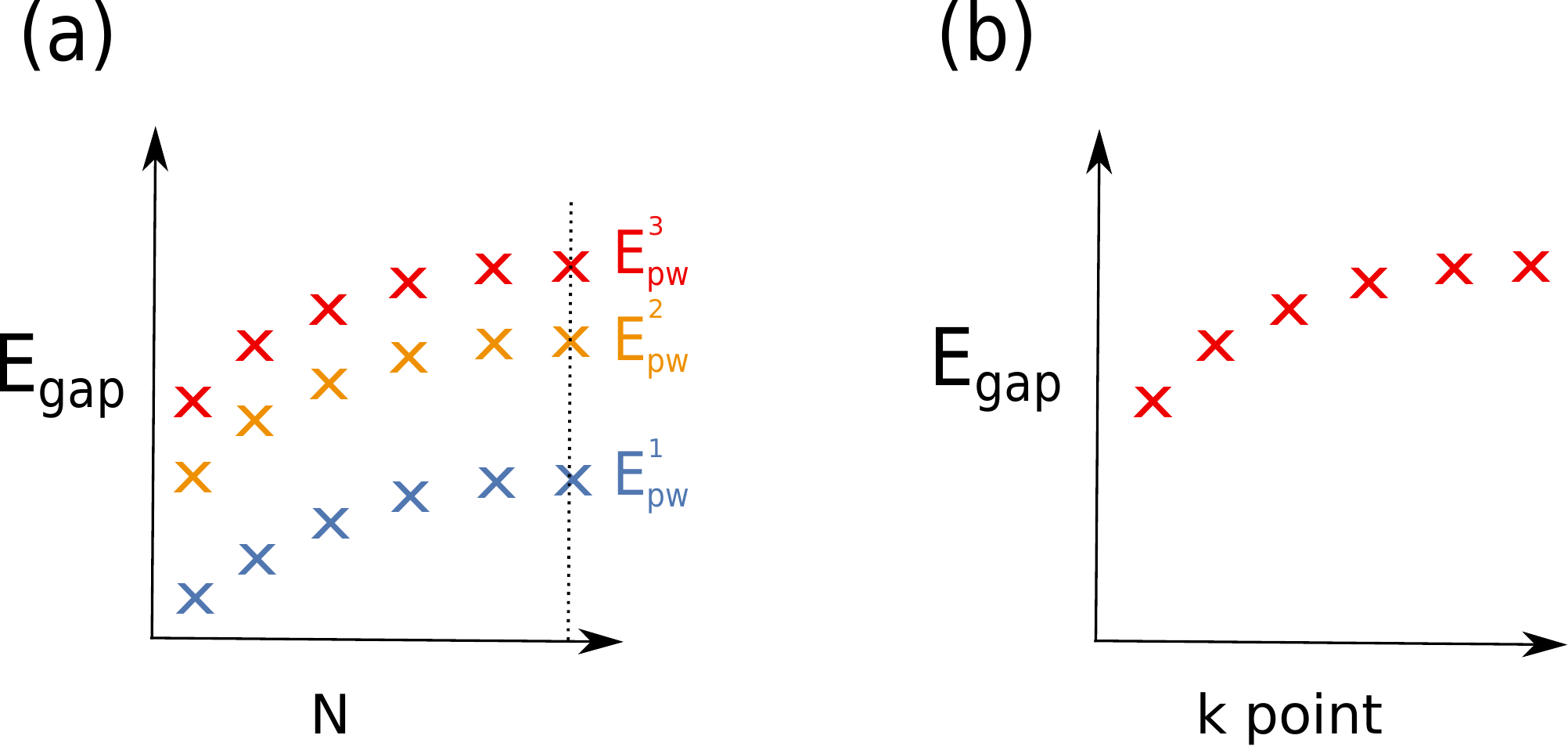}
\caption{(Color online) Schematic representation of the conventional non-extrapolated method. Convergence of the QP energy gap as a function of $N$ and $E_{pw}$ 
The convergence inspected as a function  (a) $N$ (at fixed \textbf{k}-points and $E_{pw}$) and $E_{pw}$ (for fixed \textbf{k}-points and $N$) 
where $E^3_{pw}$ $>$ $E^2_{pw}$ $>$ $E^1_{pw}$, and (b) shows the convergence of the QP gap with respect to the number of \textbf{k}-points (for fixed $N$ and $E_{pw}$).}

\label{fig:02}
\end{figure}

The \emph{conventional} method attempts to converge the QP energies (and therefore the QP energy gap $E_{g}$) with respect to a set of three parameters: number of bands ($N$), energy cutoff for the plane wave expansion for the orbitals $E_{pw}$ (which determine the total number of plane waves $N_{pw}$) 
and the number of \textbf{k}-points.
This procedure is schematically shown in Fig.\ref{fig:02}. First, $E_{g}$  is computed as a function of the number of orbitals $N$ for fixed energy cutoff for a given plane wave expansion (fixed $N_{pw}$) and \textbf{k}-points [see Fig.\ref{fig:02}(a)]. Then, by fixing $N$ and $N_{pw}$ to the optimum values that seemingly guarantee converged results within the required accuracy,  $E_{g}$ is converged with respect to the number of \textbf{k}-points [see Fig.\ref{fig:02}(b)].
This scheme can lead to reasonably well-converged results (as we will see later on); however, it neglects the exceedingly slow convergence of the QP energies with respect to the number of virtual orbitals. Since the conduction band minimum (CBM) and valence band maximum (VBM) converge at about the same rate, errors below 100 meV are often obtained even without explicit extrapolation to the infinite basis set limit (this conclusion does not apply to absolute QP energies, i.e. electron affinities and ionicities).
Within the conventional method some fitting procedures for extrapolating the QP energies for
$N \rightarrow \infty$ have been used in literature ~\cite{Kang2000, Friedrich2010, Kang2015a}, the exactness of this, however, is not supported
by a mathematical proof.

Recently, inspired by similar convergence problems occurring in quantum chemistry calculations~\cite{Kato}, Klime{\v{s}} \emph{et al.}
have provided an explicit derivation, that demonstrates that QP energies show a convergence proportional to the inverse of the number of basis functions
and introduced finite-basis-set extrapolation method~\cite{Klimes2014}. This is briefly described in the next section.
An important difference between these two approaches is that in order to perform a precise extrapolation, it is necessary to work with
the complete set of unoccupied orbitals compatible with the given energy cutoff, implying that varying $N$ for a fix energy cutoff as done
in the conventional scheme is not a formally correct practice.

\subsection{The basis-set extrapolation}\label{ss:basis}

The core aspect of the finite-basis-set correction method derived in Ref.~\onlinecite{Klimes2014} is that
the (orbital-dependent) leading-order error of the QP energy decays asymptotically with the inverse of the number of plane waves:

\begin{equation}
 \Delta{E}_m = -\frac{2}{9\pi}\frac{\Omega^2}{N_{pw}^\chi} \sum_{\textbf{g}} \rho_m(\textbf{g})\rho(-\textbf{g}).
 \label{eq:N}
\end{equation}
Here, $m$ is the orbital index, $\textbf{g} = \textbf{G} - \textbf{G'}$, where the vectors $\textbf{G}$
are three-dimensional reciprocal lattice vectors of a cell with volume $\Omega$,
$\rho$ and $\rho_m$ are the total and orbital density in reciprocal space, respectively, and
$N_{{pw}}^\chi$ is the number of auxiliary basis-set functions used to represent density related quantities, that
is controlled by a plane-wave cutoff $E_{pw}^\chi$~\cite{encutgw}.
This brings to another important result: both the total number of bands $N$,  the corresponding orbital basis set $N_{pw}$, and the size of the auxiliary basis set $N_{{pw}}^\chi$  need to be increased simultaneously at the same rate, meaning that fixing $E_{{pw}}^\chi$ and converge only with respect to $E_{{pw}}$ is not a good protocol~\cite{Klimes2014}. 

In our work we adopt the choice $E_{{pw}}^\chi$ = 2/3$E_{{pw}}$ and we have used
the \emph{complete} basis set for the given $E_{{pw}}$,
meaning that the number of orbitals equals the number of plane waves.~\cite{encutgw}
$E_{{pw}}$ was initially set to the maximum  plane-wave energy cutoff used to build the element-specific PAWs in the considered material
(the values, for US and NC PAWs, are listed in Tab.~\ref{tab:pp}).
In practice, we have systematically increased {\tt{ENCUT}} until the corresponding total number of plane-waves became  twice as larger as the initial value
(corresponding to the default $E_{{pw}}$).

\begin{figure}[h!]
\includegraphics[width=0.48\textwidth]{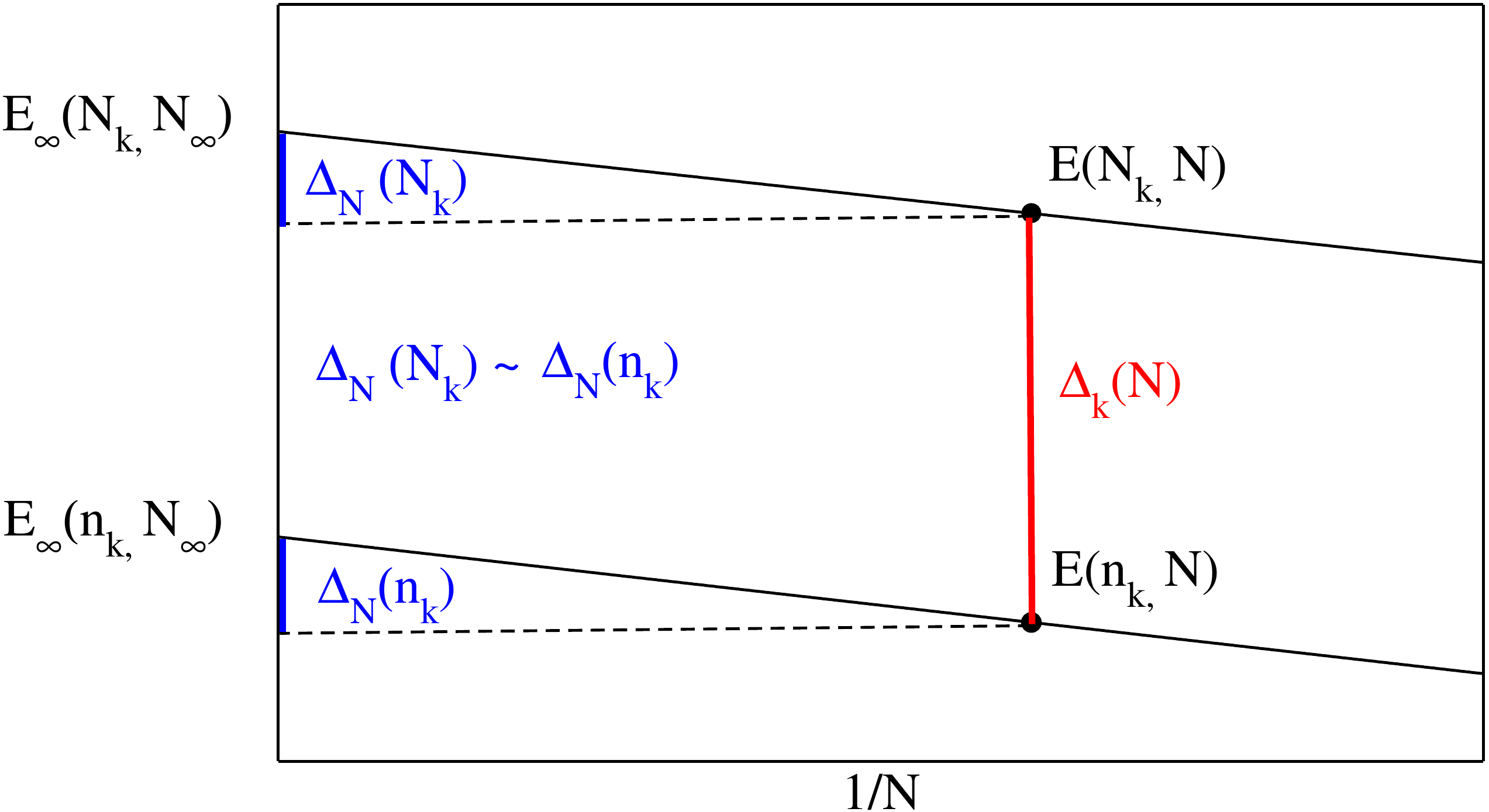}
\caption{(Color online) The schematic representation of the basis-set correction and the \textbf{k}-points correction for the QP gap
$E_{g}$. The labels indicate the contributions in Eq.~\ref{BSC_KC} and are defined in the text.
}
\label{fig:03}
\end{figure}

With Eq.~\ref{eq:N} at hand, it is formally possible to extrapolate the results for the QP energies obtained using a
finite-basis-set to the infinite-basis-set limit. To reduce the workload one can take advantage of the fact that
the convergence of the QP energies with respect to $N$ depends only weakly on the number of \textbf{k}-points~\cite{Klimes2014}.
Finally, the resulting basis-set correction formula reads:

\begin{equation}
E_\infty(N_k, N_\infty) \approx  \overunderbraces{&&\br{4}{\Delta_k(N)}}%
{&E_\infty(n_k, N_\infty) &- &E(n_k, N) &+ &E(N_k, N)}%
{&\br{3}{\Delta_N(n_k)}},
\label{BSC_KC}
\end{equation}
where $E(N_k, N)$ refer to the calculated QP energies with $N_k$ \textbf{k}-points and $N$ bands, whereas $E(N_k, N_\infty)$
refers to the corresponding extrapolated ($N\to\infty$) QP energies;
the variables $n_k$ and $N_k$ indicate the number of \textbf{k}-points in the small and large \textbf{k}-point mesh, respectively.
$\Delta_k(N)= E(N_k, N) - E(n_k, N)$ is the \textbf{k}-point correction
and $\Delta_N (n_k)= E_\infty(n_k, N_\infty) - E(n_k, N)$  the basis-set correction.
The graphical interpretation of the basis-set and \textbf{k}-point corrections  is given in Fig.~\ref{fig:03}.
Owing to the weak \textbf{k}-point dependence on the basis-set correction [$\Delta_N(n_k) \approx \Delta_N(N_k)$], in practice it is computationally more convenient to extrapolate  $\Delta_N$ using few \textbf{k}-points. Similarly, as $\Delta_k(N)$ is almost independent on $N$, the \textbf{k}-point correction computed for a small $N$ makes the computations less expensive.
For the calculations presented in this paper we have used $n_k$ = 2$\times$2$\times$2 and  $N_k$ = 6$\times$6$\times$6, with some exceptions, specified
in the text later on.

\subsection{PAW Potentials}

By extending the expression of the basis-set correction for the PAW method, Klime{\v{s}} \emph{et al.} recognized that
using US-PAW potentials the correction converges to the wrong value,  due to the incompleteness of the partial waves inside the atomic
spheres~\cite{Klimes2014}. The authors found that this error becomes smaller if the difference between the norm of the
all-electron partial waves and the pseudized partial waves is small, reaching the accuracy of full-potential linearized augmented 
plane-wave methods~\cite{PhysRevB.94.035118}.
This implies that the choice of the PAW potentials is critical and that
the best results are obtained by using NC-PAWs, for which the norm is almost fully conserved.
As shown in Tab.~\ref{tab:pp} the deviation between the all-electron and the pseudized norm,
quantified by the difference $\delta_d = |\psi_d|_{AE}^2 - |\psi_d|_{US}^2$ between the norm  of the all-electron (AE) and US-partial waves
of the $d$ orbitals, is larger for the more spatially localized 3$d$ orbitals and is substantially reduced for more extended
and smoother 4$d$ and 5$d$ orbitals (see also Fig.\ref{fig:01}(e)).
Therefore we expect that the basis-set correction error should be larger for 3$d$-based perovskites compared to 4$d$ and 5$d$ perovskites.

To inspect the influence of the choice of the PAWs on the basis-set correction results we have tested both types of PAWs, US, and NC.
For the TM ions, we have used GW-PAWs with the outermost $s$, $p$ and $d$ orbitals treated as valence states.
The NC PAW potentials were constructed following the prescription described elsewhere~\cite{Klimes2014}.

\begin{table}[ht!]
\caption{Collection of technical values related to the construction of the US and NC PAWs.
Difference $\delta_d = |\psi_d|_{AE}^2 - |\psi_d|_{US}^2$ between the all-electron (AE) and pseudized norm of
the $d$ partial waves for the 3$d$ (Sc, Ti, V, Cr, Mn, and Fe), 4$d$ (Zr, Tc, and Ru), and 5$d$ (Hf, Ta, and Os) TM ions considered in this study.
This value represents  the norm-violation in the construction of the PAW potentials. The data are extracted from the file FOUROUT, generated by the {\tt{VASP}}
PAW-generation package. Default energy cutoff for US ($E_{pw}^{US}$) and NC ($E_{pw}^{NC}$) PAWs (in eV), as given in the {\tt{VASP}} POTCAR files.
Additional details on the employed PAWs are given in the Appendix.}
\begin{tabular}{lcccccccccccc}
\hline\hline
              &  Sc  & Ti   & V    & Cr   & Mn  &  Fe  & Zr   & Tc  & Ru   & Hf   & Ta   & Os   \\\hline
$\delta_d$    & 0.13 & 0.20 & 0.27 & 0.33 & 0.4 & 0.45 & 0.02 & 0.1 & 0.17 & 0.03 & 0.04 & 0.1  \\
$E_{pw}^{US}$ & 379  & 384  & 384  & 219  & 385 & 388  & 346  & 351 & 348  & 283  & 286  & 319   \\
$E_{pw}^{NC}$ & 778  & 785  & 800  & 819  & 781 & 786  & 637  & 639 & 660  & 576  & 584  & 647    \\
\hline\hline
\end{tabular}
\label{tab:pp}
\end{table}


\section{RESULTS AND DISCUSSION}

This section presents and discusses the results obtained for the TMOs perovskites dataset (see Tab.~\ref{tab1:structures}). It is structured in three parts:
the first one focuses on the application of the convergence schemes described above to a subset of representative compounds. 
In the second one, we provide a minimal statistical interpretation of the data obtained and finally the third section
is dedicated to the calculation of the band structure and optical spectra for all compounds.

\subsection{Convergence tests and extrapolations}

In the following we show the results on the applications of the two convergence schemes,
conventional non-extrapolated and basis-set extrapolation, for selected 3$d$, 4$d$, and 5$d$ cases:
(i) cubic NM SrTiO$_3$ (3$d$), SrZrO$_3$ (4$d$) and SrHfO$_3$ (5$d$) and (ii) structurally distorted and magnetically ordered
SrMnO$_3$ (3$d$), SrTcO$_3$ (4$d$) and NaOsO$_3$ (5$d$). The complete set of results is given in the Supplemental Materials (SM)~\cite{SM}.

\subsubsection{Cubic non-magnetic systems}

We start showing and examining the results for 3$d$ SrTiO$_3$, and then we will extend the discussion by including the data for SrZrO$_3$ and SrHfO$_3$.
In all these materials the gap is opened between the filled O-$p$ states at the VBM and empty TM-$d$ states at the CBM.
\begin{figure}[h!]
\includegraphics[width=0.48\textwidth]{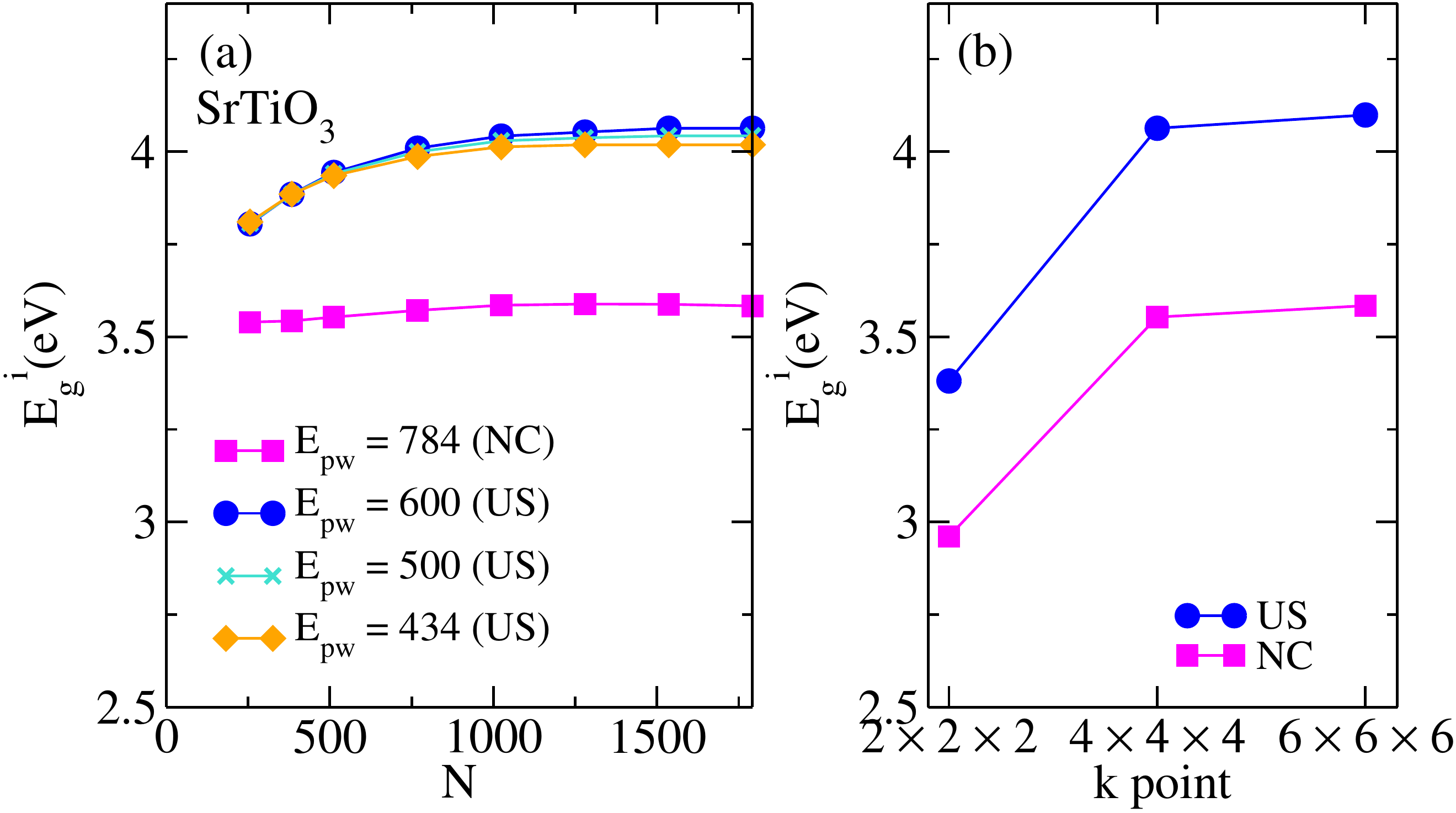}
\caption{(Color online) Conventional non-extrapolated method applied to SrTiO$_3$. Convergence of the indirect QP bandgap $E_g^i$ of SrTiO$_3$
with respect to (a) number of bands $N$ and plane-wave cutoff energy $E_{pw}$
(\textbf{k}-point mesh fixed to 4$\times$4$\times$4), and  (b) size of the \textbf{k}-point mesh for $N$ and $E_{pw}$  fixed to the optimum
values for US and NC potentials.}
\label{fig:04}
\end{figure}

First, we show the convergence behavior for SrTiO$_3$  using the conventional non-extrapolated scheme
by inspecting the variation of the QP indirect gap $E_g^i$ ($R-\Gamma$, highest occupied state at $R$ and the lower unoccupied state at $\Gamma$)
as a function of $N$, $E_{pw}$ and the number of \textbf{k}-points,
using both US and NC PAWs. Similar results and  conclusions are obtained for QP energies, but since we are    primarily
interested in the behavior of the bandgap, the discussion and analysis will be focused on the direct and indirect bandgap (i.e. differences of QP energies).

The results are displayed in Fig.~\ref{fig:04}. From panel Fig.~\ref{fig:04}(a) we note that using
US-PAWs the results are largely sensitive on $N$, and to a lesser extent on $E_{pw}$, and well-converged values are achieved for
$N \approx 1500$ and $E_{pw}$ $\approx$ 600 eV. By employing NC-PAW the convergence with respect to $N$ is much faster, $N \approx 1000$
is sufficient to obtain the same level of accuracy obtained at US level (as we will see below the faster convergence using NC-PAWs is in this
case related to similar convergence rates for CBM and VBM for NC-PAWs). 
However, owing to the generally larger default energy cutoffs for NC-PAWs (see Tab.~\ref{tab:pp})
it is computationally prohibitive to scan higher values of the cutoff energy. The dependence of $E_g^i$ on the number of \textbf{k}-points, displayed in
Fig.~\ref{fig:04}(b) shows that a 4$\times$4$\times$4 grid is sufficient to achieve an accuracy of about 0.03 eV.

The final values of the US and NC indirect bandgaps $E_g^i$, 4.06 eV (almost identical to the one reported in Ref.~\onlinecite{Kang2015a} using the same scheme)
and 3.55 eV, respectively, differ by about 0.5 eV and are both larger than the measured value, 3.3 eV~\cite{Takizawa2009} (see Tab.~\ref{tab:convergence}).
The difference between the US-PAWs and NC-PAWs is due to the relatively large norm-violation for the Ti US-PAW, 0.2, which causes a quite different QP shift of the
empty $d$ states at the bottom of the conduction band in US- and NC-based calculations (0.68 eV, see Tab.\ref{tab:convergence}).
On the other side, the difference in the QP shift between NC and US calculations is substantially smaller for the top of the valence band, mostly populated
by O-$p$ state, 0.11 eV, see Tab.\ref{tab:convergence}. This issue will be further discussed in the context of the data obtained using the basis-set extrapolation,
at the end of this subsection.

\begin{table*}[ht!]
\caption{Collection of data related to the convergence tests for selected 3$d$, 4$d$, and 5$d$ perovskites (cubic-NM and distorted-AFM, see text).
Energy differences between the US and NC QP energies at the CBM and VBM at $\Gamma$
$\Delta{E}_{QP}^{V} = |E_{QP-VBM}^{NC} - E_{QP-VBM}^{US}|$ and $\Delta{E}_{QP}^{C} = |E_{QP-CBM}^{NC} - E_{QP-CBM}^{US}|$, the
norm-violation  $\delta_d$ (same as in Tab.~\ref{tab:pp}), the non-extrapolated (nE) and extrapolated (E) value of the indirect bandgap $E_g^i$,
the basis-set correction $\Delta_N$ and the \textbf{k}-point correction $\Delta_k$
(evaluated with a reduced number of  \textbf{k}-points $n_k={2}\times{2}\times{2}$ and $N{\approx}400-500$, respectively, and $N_k={6}\times{6}\times{6}$, see Eq.~\ref{BSC_KC}).
$E_g^i$, $\Delta_N$ and $\Delta_k$ are provided for both type of PAWs (US and NC).
Within the conventional method, we have used NC PAW only for the representative case of SrTiO$_3$.
For non-$d^0$ compounds SrMnO$_3$, SrTcO$_3$, NaOsO$_3$ (with SOC), the amount of $d$ character in the valence ($d_V$) and conduction ($d_C$) band is also given.
Available experimental data for the gap are also listed.
All energies are given in eV.
}
\begin{ruledtabular}
\begin{tabular}{lcccccccccccccc}
Compound    & $\Delta{E}_{QP}^{V}$ & $\Delta{E}_{QP}^{C}$ & $\delta_d$ & $d_V$ & $d_C$ &\multicolumn{2}{c}{$\Delta_N$} &\multicolumn{2}{c}{$\Delta_k$} &\multicolumn{4}{c}{$E^i_{g}$} & $E^{Expt}_{g}$\\
            &                      &                      &            &       &       &  US       &   NC              &  US   & NC                    &  US  & NC     &   US  & NC       &               \\
            &                      &                      &            &       &       &          &                   &       &                       & (nE) & (nE)   &   (E) & (E)      &               \\
\hline
SrTiO$_3$   & 0.11                 &      0.68            &    0.20    &   --  &  --   &  0.20    &    0.01           &  0.56 & 0.59                  &  4.08& 3.55   &  4.06 & 3.55     &  3.3~\cite{Takizawa2009,Tezuka1994,VanBenthem2001a}             \\
SrZrO$_3$   & 0.11                 &      0.04            &    0.02    &   --  &  --   &  0.14    &    0.18           &  0.36 & 0.36                  &  5.29& --     &  5.36 & 5.43     &  5.6~\cite{Lee2002}             \\
SrHfO$_3$   & 0.10                 &      0.09            &    0.03    &   --  &  --   &  0.19    &    0.12           &  0.34 & 0.36                  &  5.69& --     &  5.76 & 5.81     &  6.1~\cite{Sousa2007}             \\
SrMnO$_3$   & 0.27                 &      0.43            &    0.4     &  0.54 & 0.93   &  0.03    &   -0.03           &  0.23 & 0.31                  &  1.75& --     &  1.66 &  1.46    &               \\ 
SrTcO$_3$   & 0.30                 &      0.33            &    0.1     &  0.86 & 0.87  &  0.01    &    0.02           & -0.13 & -0.14                 &  1.14& --     &  1.18 &  1.20    &               \\
NaOsO$_3$   & 0.26                 &      0.27            &    0.1     &  0.73 & 0.78  &  0.03    &    0.01           & -0.23 & -0.26                 &  0.27& --     & 0.28  &  0.27    & 0.1~\cite{Vecchio2013}              \\
\end{tabular}
\end{ruledtabular}
\label{tab:convergence}
\end{table*}

\begin{figure}[ht!]
\includegraphics[width=0.48\textwidth]{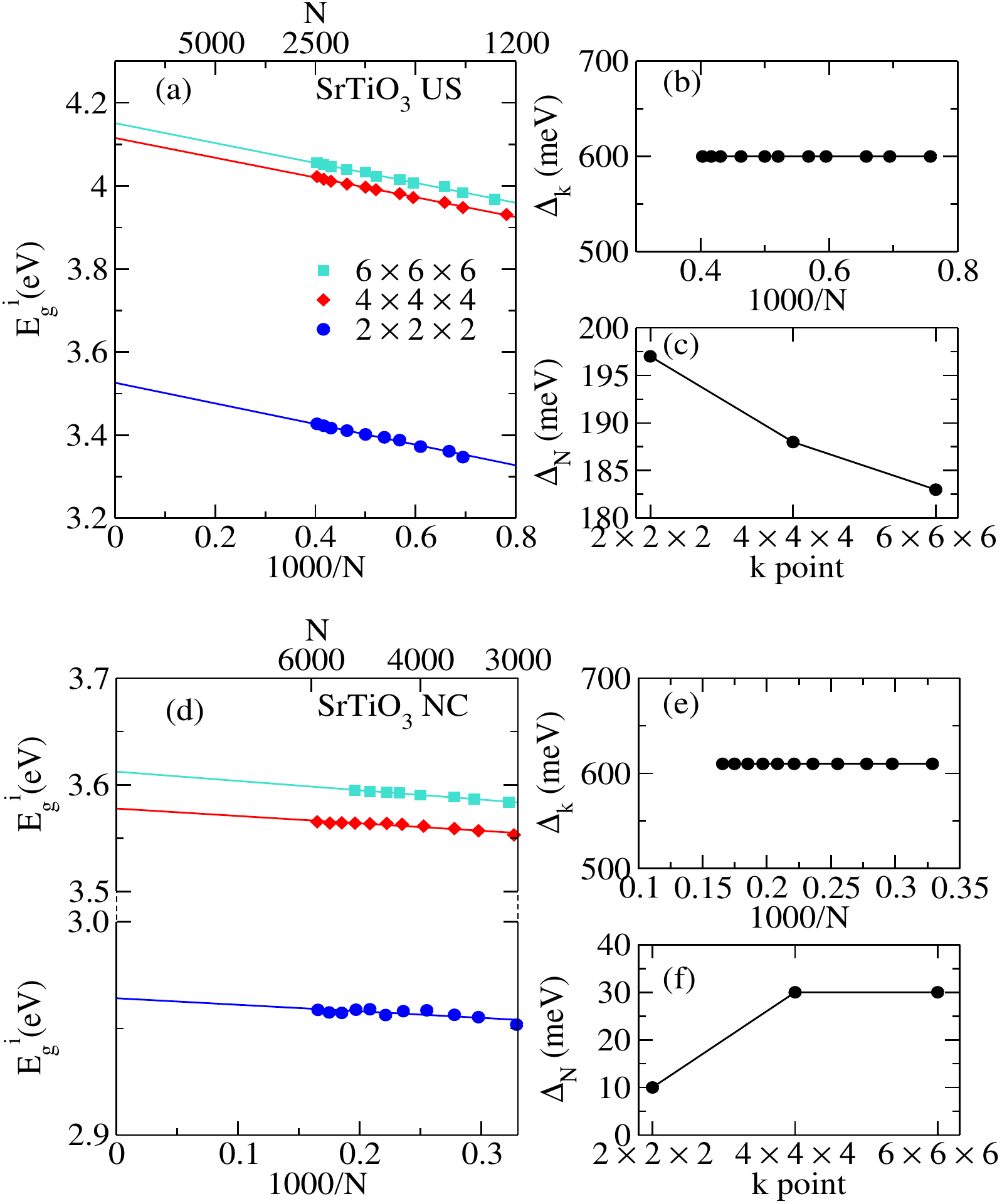}
\caption{(Color online) Basis-set correction data for SrTiO$_3$ using US (panels a, b, and c) and NC (panels d, e, and f) PAWs.
For each type of PAWs three different graphs are shown: (a,d) convergence of the QP band gap $E_g^i$ with respect to the inverse of the
number of bands (1000/$N$); (b,e) \textbf{k}-point correction $\Delta_k$ as a function of 1000/$N$;
(c,f) basis-set correction $\Delta_N$ as a function of  \textbf{k}-points.}
\label{fig:05}
\end{figure}

\begin{figure}[ht!]
\includegraphics[width=0.49\textwidth]{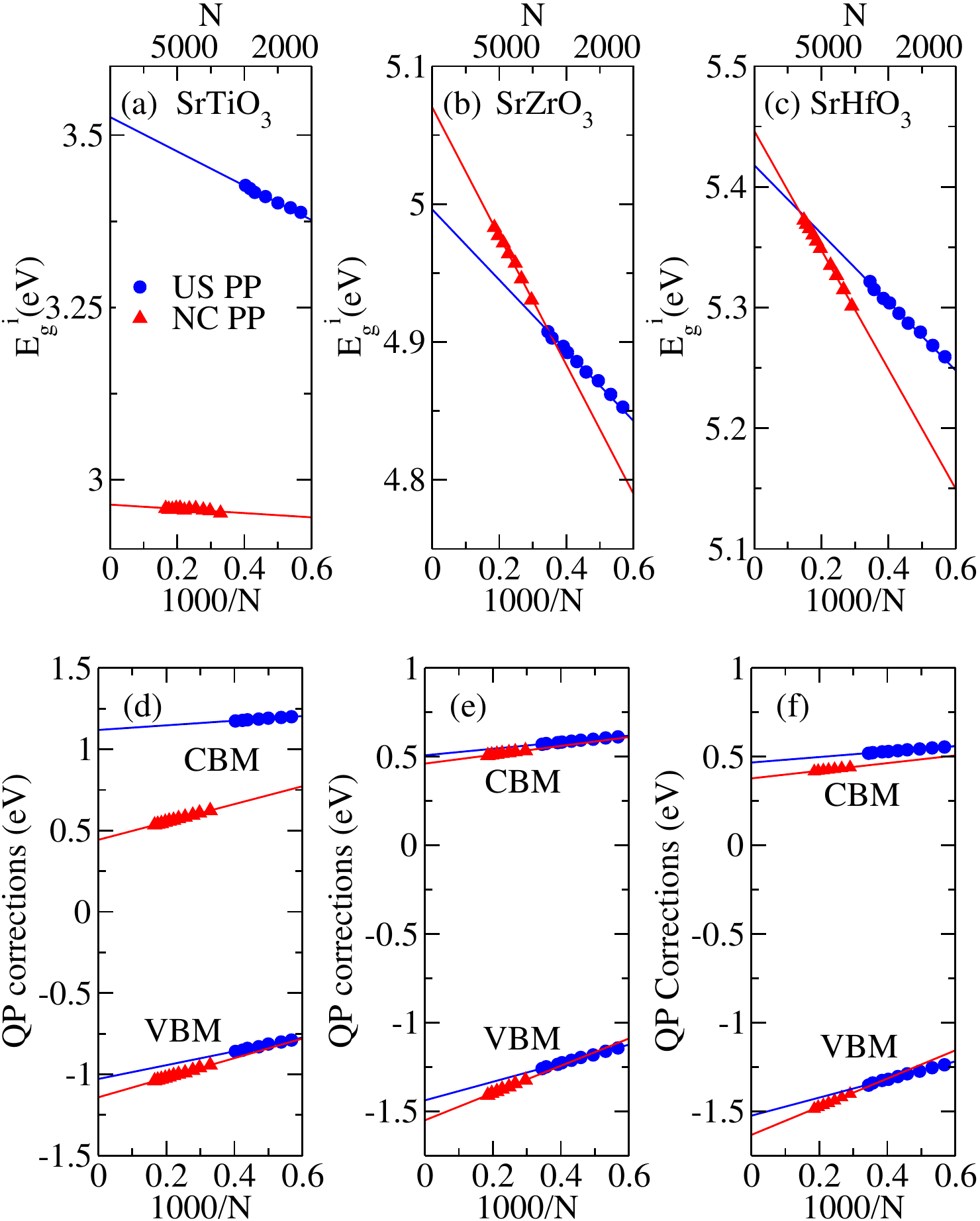}
\caption{(Color online) $E_g^i$ and QP corrections to Kohn-Sham eigenvalues ($E^{QP} - E^{KS}$) for the CBM and VBM at $\Gamma$
as a function of $N=N_{pw}$ for  SrTiO$_3$ (a,d), SrZrO$_3$ (b,e) and SrHfO$_3$ (c,f), computed within the basis set correction scheme
using a 2$\times$2$\times$2 \textbf{k}-point mesh. }
\label{fig:06}
\end{figure}

The basis-set extrapolation data for SrTiO$_3$ are collected in Fig.~\ref{fig:05}, where we show the evolution of $E_g^i$ and
the \textbf{k}-points corrections upon $N$, as well as
the basis-set correction $\Delta_N$ as a function of the size of the \textbf{k}-points mesh.
We highlight once more that here $N_{pw}$ refers to the maximum number
of plane-waves compatible to a given plane-wave cutoff energy $E_{pw}$. In this case (SrTiO$_3$), we have gradually increased $N$
from 1200 to about 2500 using US-PAWs ($E_{pw}$=434 eV), and from 3000 to 6000 using NC-PAWs (the minimally required $E_{pw}$ is significantly
larger for the NC-PAWs, 785 eV, which leads to a much larger number of basis functions). We have also inspected the convergence for three different \textbf{k}-point meshes: 2$\times$2$\times$2, 4$\times$4$\times$4, and 6$\times$6$\times$6.
The curves plotted in Fig.~\ref{fig:05}(a, d) clearly indicate that $E_g^i$ converges linearly with respect to 1/$N$ for both types of PAWs.
The values of  $E_g^i$, in particular its $N \rightarrow \infty$ extrapolation, varies with the number of \textbf{k}-points but
the \textbf{k}-points correction $\Delta_k$ ($\approx$ 600 meV) depends only marginally
on $N$ [see Fig.\ref{fig:05}(b,e) and Tab~\ref{tab:convergence}]. This
represents one of the great advantages of the extrapolation scheme: $\Delta_k$ can be determined using a small $N$ (the default value),
thereby reducing the computational cost of the calculation.
Moreover, $\Delta_k$ does not depend on the type of PAW potential used but it is sensitive to the specific \textbf{k}-point 
at which the QP energy correction is calculated: for $E_g^i$  $\Delta_k$ is 600 meV,
but the corresponding correction for the direct gap at $\Gamma$, $E_g^\Gamma$ ($\Gamma-\Gamma$ gap) is reduced by about 100 meV
(similar observations were made in Ref.~\onlinecite{Klimes2014} for most materials, specifically AlAs and GaAs).
A further positive aspect of this  scheme is that the basis-set correction
$\Delta_N$ does not  vary much with respect to the size of the \textbf{k}-point mesh [Fig.\ref{fig:05}(c,f)]:
$\Delta_N$ can be evaluated using a small \textbf{k}-point mesh, typically 2$\times$2$\times$2, which also helps in decreasing the CPU time.
However, unlike $\Delta_k$, which is essentially insensitive to the choice of the potential, $\Delta_N$ is one order of magnitude smaller for NC-PP
(0.01 eV against 0.2 eV, see Tab.~\ref{tab:convergence}).
Also, the NC value of the fundamental gap, 3.55 eV (the same for both the conventional and the extrapolated method), is in better agreement with the measured value, 3.3 eV, as compared to the US gaps,
which are substantially larger (3.94 eV and 4.08 eV, see Tab.~\ref{tab:convergence}).
The reason for the improved description of SrTiO$_3$ is the improved treatement of the 3$d$ CBM states, which show a larger slope
with the NC potentials (Fig.~\ref{fig:06}).

\begin{figure}[ht!]
\includegraphics[width=0.49\textwidth]{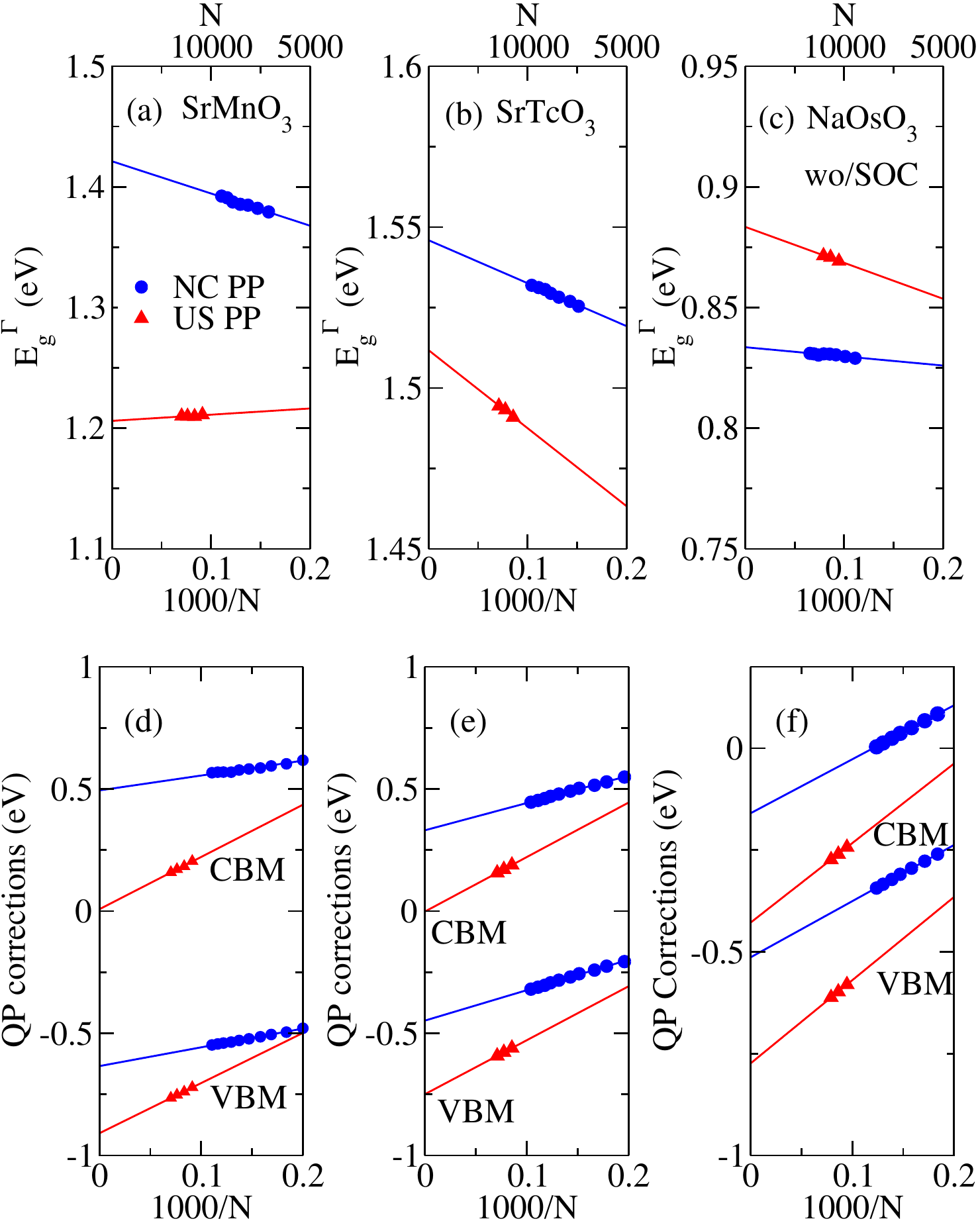}
\caption{(Color online) Band gap $E_g^\Gamma$ and QP correction $E^{QP} - E^{KS}$; similar to  Fig.~\ref{fig:06} but for
SrMnO$_3$, SrTcO$_3$, and NaOsO$_3$ (without, wo, SOC).
}
\label{fig:07}
\end{figure}

The convergence tests for the other two members of the cubic-NM 3$d$-4$d$-5$d$ series, SrZrO$_3$ and SrHfO$_3$, are displayed in Fig.~\ref{fig:06}
where we report the dependence of $E_g^i$ on the number of bands $N$ as well as the evolution of the QP corrections to the Kohn-Sham DFT eigenvalues at $\Gamma$
for the conduction band minimum (CBM, TM-d$^0$) and valence band maximum (VBM, O-$p$). For the sake of comparison the corresponding data for SrTiO$_3$
are also included. As a general result we found that by applying Eq.~\ref{BSC_KC} well converged values of $E_g$  [$E_{g}$($N_k$, $N_\infty$)] are obtained by setting
$n_k$=2$\times$2$\times$2, $N_k$=6$\times$6$\times$6 (but also $N_k$=4$\times$4$\times$4 leads to accurate results); these data are reported in
Tab.~\ref{tab:convergence}.

The most important result that one notices is that the difference between NC and US data is substantially reduced
for SrZrO$_3$ and SrHfO$_3$, as compared to SrTiO$_3$ (see Fig.~\ref{fig:06}, and note the different scalefor $E^i_g$ plots). As already mentioned,  the difference in $\Delta_N$ between NC and US energies in SrTiO$_3$
is about 0.5~eV, whereas for SrZrO$_3$ and SrHfO$_3$ it is almost zero: US and NC PAWs deliver roughly the same $\Delta_N$
for both materials, $\approx$ 0.15 eV (see Tab.~\ref{tab:convergence}). This result can be readily explained by the much lower norm-violation
$\delta_d$ in 4$d$ Zr (0.02) and 5$d$ Hf (0.03) as compared to 3$d$ Ti (0.2), which originates from the smoothness of the 4$d$ and 5$d$  orbitals
as compared to the more localized nature of 3$d$ orbitals (see Fig.~\ref{fig:01}). This conclusion correlates well with the
behavior of the QP corrections shown in Fig.~\ref{fig:06}, in particular, by looking at the
differences between the US and NC QP corrections at the CBM and VBM at $\Gamma$, defined as
$\Delta{E}_{QP}^{V} = |E_{QP-VBM}^{NC} - E_{QP-VBM}^{US}|$ and $\Delta{E}_{QP}^{C} = |E_{QP-CBM}^{NC} - E_{QP-CBM}^{US}|$:
for the Ti-3$d$ empty states $\Delta{E}_{QP}^{C}$ is 0.68 eV, whereas for Zr and Hf $d^0$ states as well as for the highest occupied O-$p$
states $\Delta{E}_{QP}^{C}$ and $\Delta{E}_{QP}^{V}$ are in the range 0.04-0.1 eV. The exact values are listed in Tab.~\ref{tab:convergence}.

In terms of band gaps, US and NC PAWs lead to similar values, in particular for 5$d$ SrHfO$_3$ (5.36 eV and 5.43 eV, respectively) in satisfactory agreement with the available experimental
estimates (see Tab.~\ref{tab:convergence}).

To conclude this part, we have shown that the type of convergence scheme, the type of potential employed in the calculations and the type of TM $d$-orbital
affect the QP energies and therefore the final 'converged' value of the band gap.
Overall, we have tested four different procedures to compute the gap:
conventional scheme (no extrapolation, labeled 'nE' in Tab.~\ref{tab:convergence}) and basis-set extrapolation
(labeled 'E') using US or NC PAWs. As mentioned before, nE-NC calculations were only done for SrTiO$_3$.
The main conclusion is that extrapolated-NC values agree better with the experimentally measured data, in particular for 3$d$ SrTiO$_3$
for which the large norm-violation underestimates the electronic correlation contribution to the Ti 3$d$ CBM states for the US-PAW potentials, and thus a too large band gap~\cite{Klimes2014}. For 4$d$ SrZrO$_3$ and 5$d$ SrHfO$_3$ the difference between NC and US PAWs
is strongly attenuated and the final extrapolated values of the gap are almost identical.
Also, our results suggest that there are not pronounced differences in the gap between the two schemes
for a specific type of potential: the two schemes yield very similar gaps for SrTiO$_3$, 3.55 eV (see Tab.~\ref{tab:convergence}).
Qualitatively similar results are obtained for the larger and magnetically order 3$d$, 4$d$, and 5$d$ systems, as discussed below.

\subsubsection{Large magnetic systems}
We show here the convergence tests for the basis-set extrapolation scheme applied to the $t_{2g}^3$ series
SrMnO$_3$, SrTcO$_3$, and NaOsO$_3$. For the other compounds included in our dataset as well as for the data obtained using the
conventional scheme, we will only discuss the converged values of the gap and compare them to available experimental measurements.
Further details and graphs can be found in the Supplemental Materials (SM)~\cite{SM} and in Ref.~\onlinecite{PhD}.

The unit cells used to model SrMnO$_3$, SrTcO$_3$, and NaOsO$_3$ contain four formula units (20 atoms),
which are necessary to model the internal structural distortions and the antiferromagnetic ordering (see Tab.~\ref{tab1:structures}). This leads to
an increase of the number of basis functions and, therefore, to more substantial memory requirements and computing times. As a result, the calculations become technically heavier
and almost prohibitive for NC-based calculations. Due to this computational limitation, in some cases, we have performed the NC-based extrapolation using only 2 or 3 points (see Fig.~\ref{fig:07} and SM).

The trends for the QP energies and gaps for this series is plotted in Fig.~\ref{fig:07}.
For US-PAW calculations we have inspected the $N$ range from $\sim$5000 up to $\sim$10000 in about 10 steps (a denser mesh has been used for
the largest $N$ in order to improve the extrapolation for $N \rightarrow \infty$); however, for NC calculations due to the computational restrictions
mentioned above, we could scrutinize a smaller $N$ range, between 10000 and 12000.

The violation of the norm is much larger for 3$d$ Mn (0.4) compared to 4$d$ Tc and 5$d$ Os (0.1), which
explains the bigger difference between NC and US results in SrMnO$_3$ as compared to SrTcO$_3$ and NaOsO$_3$, particularly evident for the QP 
correction in Fig.~\ref{fig:07}(d) but also the gap [Fig.~\ref{fig:07}(a)].
Unlike $d^0$ cubic perovskites, for this $t_{2g}^3$ series the difference between NC and US PAW is not limited to the bottom of the conduction band,
but is also manifested at the top of the valence band that has a strong $d$ character. This is shown in the bottom panels of Fig.~\ref{fig:07}, that displays
the QP corrections at the US and NC level for the VBM and CBM.
The energy shifts  $\Delta{E}_{QP}^{V}$ and $\Delta{E}_{QP}^{C}$ that measure the differences between the US and NC QP corrections at the CBM and VBM tabulated in Tab.~\ref{tab:convergence},
shows that in  SrMnO$_3$ the difference is larger for the CBM  than VBM (0.43 eV and 0.27 eV, respectively), whereas in SrTcO$_3$ and NaOsO$_3$ the deviation is about the same for
filled and empty states, $\approx$ 0.3 eV. This behavior can be explained by the amount of $d$ states present in the CBM and VBM, which is also listed in Tab.~\ref{tab:convergence}
(see the additional column for SrMnO$_3$, SrTcO$_3$ and NaOsO$_3$): in SrTcO$_3$ and NaOsO$_3$ the CBM and VBM possess about the same amount of $d$ character, $\approx$ 0.8,
but in SrMnO$_3$ the CBM is almost completely formed by Mn-$d$ states, 93\%, twice larger than the $d$-population at the valence band, 54\%.

As expected from the above considerations, the final values of the extrapolated gap for this sub-series vary less than for 
Sr(Ti, Zr, Hf)O$_3$. For SrMnO$_3$ the NC potential lowers the gap by 0.2 eV, for the other materials there is hardly  any difference between the predicted gaps. 
This outcome is qualitatively similar to the situation discussed for the
3$d$, 4$d$, and 5$d$ cubic non-magnetic perovskites (see Tab.~\ref{tab:convergence}).

\begin{table}[ht!]
\caption{Compilation of the calculated (G$_0$W$_0$) and experimental band gaps for the perovskites dataset studied in this paper for both type of convergence schemes (extrapolated, E, and non-extrapolated, nE) using US- and when available,
NC-based data. Both, the calculated direct gap at $\Gamma$ ($E_g^\Gamma$) and the indirect gap ($E_g^i$, only if smaller than $E_g^\Gamma$) are listed. As explained in the text NC-nE results are only given for SrTiO$_3$.
The results are also compared with other previously calculated GW results. 
For NaOsO$_3$ the calculaitons were done including SOC.
Due to the large computational cost, it was not possible to obtain NC data for Ca$_2$RuO$_4$.
The experimental techniques used to extract the gap are also reported: photoemission spectroscopy (Refs.~\onlinecite{Takizawa2009,Tezuka1994,Saitoh1995}), electron energy loss spectroscopy (Ref.~\onlinecite{VanBenthem2001a}),
optical spectroscopy (Refs.~\onlinecite{Lee2002, Sousa2007, Arima1993}), spectroscopic ellipsometry (Ref.~\onlinecite{Jellison2006}), X-ray absorption spectroscopy (Ref.~\onlinecite{Fatuzzo2015}), and
optical conductivity (Ref.~\onlinecite{Vecchio2013}). All energies are expressed in eV.}
\begin{ruledtabular}
\begin{tabular}{lccccccc}
Compound         & ~PAW~ & \multicolumn{2}{c}{$E^{\Gamma}_{g}$}    & \multicolumn{2}{c}{$E^{i}_{g}$}    & $E^{Expt}_{g}$                                                               & Other GW \\ 
                 &     & ~(E)~                &  ~(nE)~              &  ~(E)~            & ~(nE)~             &                                                                             &                                             \\
\hline
SrTiO$_3$        &  US & 4.45               &      4.39          &    4.08           & 4.06             & 3.3~\cite{Takizawa2009,Tezuka1994,VanBenthem2001a}                          & 3.82~\cite{Kang2015a,Friedrich2010}         \\
                 &  NC & 3.94               &      3.99          &    3.55           & 3.55             &                                                                             &                                             \\
SrZrO$_3$        &  US & 5.73               &      5.64          &    5.36           & 5.29             & 5.6~\cite{Lee2002}                                                          &                                             \\
                 &  NC & 5.80               &        -           &    5.43           &      -           &                                                                             &                                             \\
SrHfO$_3$        &  US & 6.17               &      6.01          &    5.76           & 5.69             & 6.1~\cite{Sousa2007}                                                        &                                             \\
                 &  NC & 6.21               &        -           &    5.81           &       -          &                                                                             &                                             \\
KTaO$_3$         &  US & 4.40               &      4.31          &    3.67           & 3.59             & 3.6~\cite{Jellison2006}                                                    &  3.57~\cite{Sousa2007} 3.51~\cite{Wang2011} \\
                 &  NC & {            4.39} &        -           &{            3.64} &       -          &                                                                             &                                             \\
LaScO$_3$        &  US & 4.87               &      4.56          &                   &                  & 6.0~\cite{Arima1993}                                                        &                                             \\
                 &  NC & {            4.93} &        -           &                   &       -          &                                                                             &                                     \\
LaTiO$_3$        &  US & 1.12               &      1.00          &    0.63           & 0.49             & 0.1~\cite{Arima1993}                                                        & 0.77~\cite{Nohara2009}              \\
                 &  NC &{            1.17}  &        -           &{            0.54} &                  &                                                                             &                                     \\
LaVO$_3$         &  US & 1.73               &      1.74          &    1.19           & 1.14             & 1.1~\cite{Arima1993}                                                        & 2.47~\cite{Nohara2009}              \\
                 &  NC & {            1.71} &        -           &{            1.14} &                  &                                                                             &                                     \\
LaCrO$_3$        &  US & 2.98               &      2.95          &                   &                  & 3.3~\cite{Arima1993}                                                        & 3.25~\cite{Nohara2009}              \\
                 &  NC & {            2.77} &        -           &                   &                  &                                                                             &                                     \\
LaMnO$_3$        &  US & 1.33               &      1.34          &    0.96           & 0.97             & 1.1~\cite{Arima1993}                                                        & 1.63~\cite{Nohara2009}              \\
                 &  NC & {            1.30} &        -           & {            0.87}&     -            &                                                                             &                                     \\
LaFeO$_3$        &  US & 2.61               &      2.65          &    1.95           & 1.91             & 2.1~\cite{Arima1993}                                                        & 1.76~\cite{Nohara2009}              \\
                 &  NC & {            2.46} &        -           &{            1.73} &       -          &                                                                             &                                     \\
SrMnO$_3$        &  US & 1.66               &      1.75          &                   &                  &                                                                             &                                     \\
                 &  NC & {            1.46} &                    &                   &                  &                                                                             &                                     \\
SrTcO$_3$        &  US & 1.62               &      1.62          &    1.18           & 1.14             &                                                                             &                                     \\
                 &  NC & {            1.58} &        -           &{            1.20} &   -              &                                                                             &                                     \\
Ca$_2$RuO$_4$    &  US & 0.96               &      0.98          &    0.53           & 0.50             & 0.3-0.5~\cite{Fatuzzo2015}                                                  &                                     \\
NaOsO$_3$        &  US & 0.79               &      0.82          &    0.28           & 0.27             &      0.1~\cite{Vecchio2013}                                                 &                                     \\
                 &  NC & 0.92               &        -            &    0.27           &        -          &                                                                             &                                     \\
\end{tabular}
\end{ruledtabular}
\label{bandgaps}
\end{table}

\begin{figure}[ht!]
\includegraphics[width=0.49\textwidth]{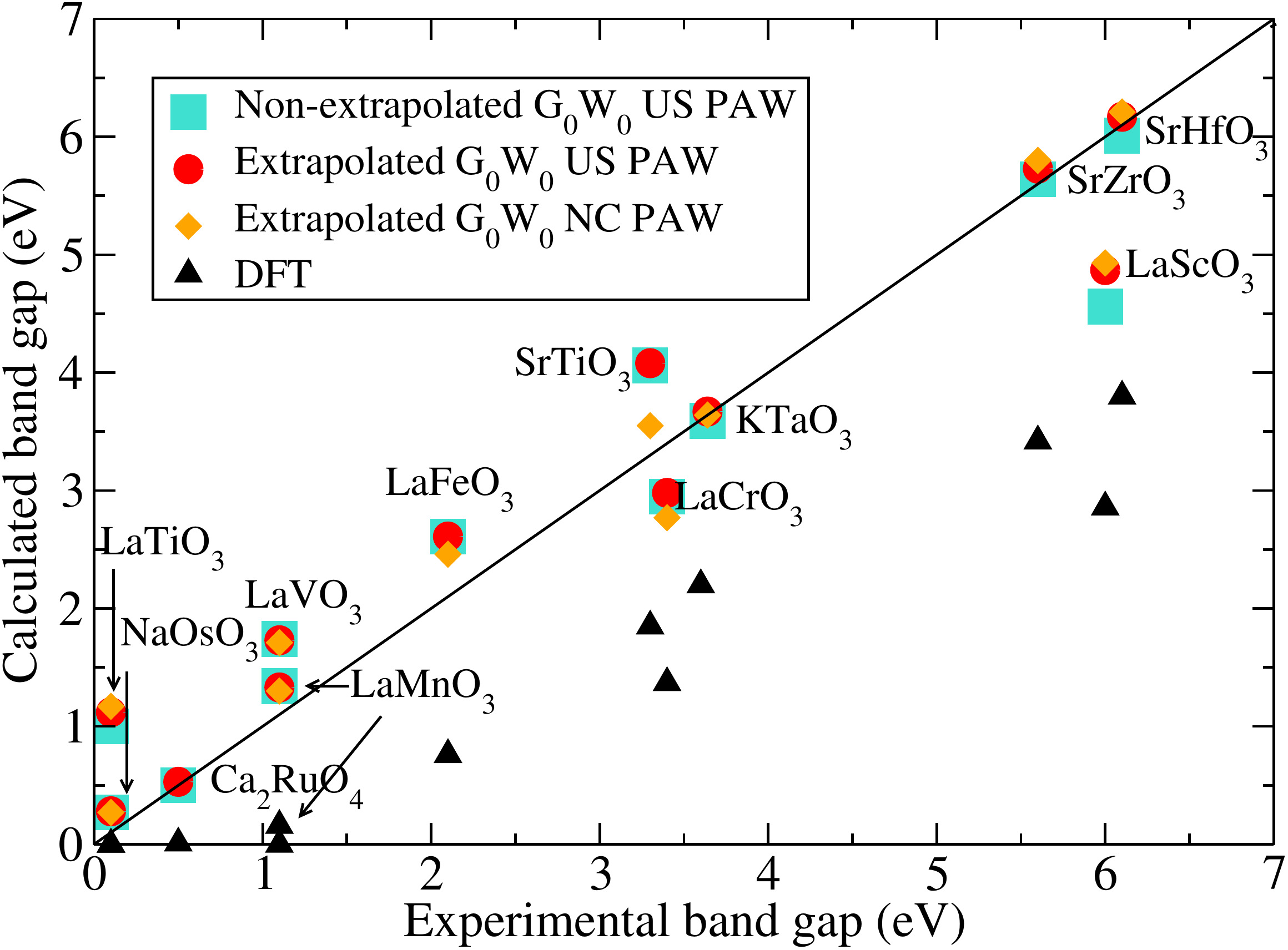}
\caption{(Color online) Comparison between the DFT and G$_0$W$_0$ band gaps and available experimental measurements (reference given in Tab.~\ref{bandgaps}).
The G$_0$W$_0$ refer to three different sets: non-extrapolated results obtained using US potentials, and
 basis-set corrected values using both type of potentials, US and NC (not for Ca$_2$RuO$_4$).}
\label{fig:08}
\end{figure}

The complete collection of bandgaps for the entire series of perovskites considered in our study is compiled in Tab.~\ref{bandgaps}.
A graphical summary of the comparison between the computed (GW and DFT) and
available photoemission and inverse photoemission spectroscopy measurements is provided in Fig.~\ref{fig:08}.
The technical parameters (energy cutoff, number of \textbf{k}-points and number of bands $N$)
that guarantee well converged QP energies (accuracy $\approx$ 100 meV) within the conventional non-extrapolated scheme
are listed in Tab.~\ref{tab:bands}.

Clearly, GW outperforms DFT, which underestimates the gaps by more than 50\% and, in some cases (NaOsO$_3$, LaTiO$_3$, and LaVO$_3$) finds a metallic solution. This is a well-known behavior that has been widely discussed in literature~\cite{VanSchilfgaarde2006b, Hybertsen1985, Hybertsen1986, aulbur, GWKresse, Nohara2009}.
Regardless of the specific convergence scheme and type of potential, the GW gaps are in overall good agreement with experiment. However, we should note that the experimental data must be treated with care because perovskite materials can
often exhibit oxygen deficiencies that unavoidably alter the value of the gap. In addition to this, we should also mention that the GW gap refers to the fundamental gap, meaning that excitonic effects are not taken into
account. The calculation and the experimental estimation of electron-hole interactions are not an easy task~\cite{VanSchilfgaarde2006b,PhysRevLett.99.246403} and remain a largely unexplored issue in TM perovskites.

\begin{table}[ht!]
\caption{Set of parameters (energy cutoff $E_{pw}$, \textbf{k}-point mesh and number of bands $N$) used for the calculation of the band structures and optical spectra at G$_0$W$_0$ level within the non-extrapolated scheme.
This setup guarantees well-converged QP energies within a accuracy of typically 100 meV. All of the parameters are for US PAWs except for SrTiO$_3$ where NC PAW is used.}
\begin{tabular}{lccc}
\hline\hline
Compound        &      $E_{pw}$                     &  \textbf{k}-points mesh            &   $N$           \\
\hline
SrTiO$_3$       &  600                              &  4$\times$4$\times$4               &  512            \\
SrZrO$_3$       &  650                              &  4$\times$4$\times$4               &  1791           \\
SrHfO$_3$       &  650                              &  4$\times$4$\times$4               &  2304           \\
KTaO$_3$        &  500                              &  4$\times$4$\times$4               &  896            \\
SrMnO$_3$       &  500	                            &  4$\times$4$\times$2               &  400            \\
SrTcO$_3$       &  500                              &  5$\times$3$\times$5               &  512            \\
Ca$_2$RuO$_4$   &  500                              &  4$\times$4$\times$2               &  512            \\
NaOsO$_3$       &  500                              &  5$\times$3$\times$5               &  400            \\
LaScO$_3$       &  500                              &  5$\times$5$\times$3               &  1280           \\
LaTiO$_3$       &  500                              &  5$\times$3$\times$5               &  400            \\
LaVO$_3$        &  500                              &  5$\times$3$\times$5               &  400            \\
LaCrO$_3$       &  500                              &  5$\times$3$\times$5               &  400            \\
LaMnO$_3$       &  500                              &  5$\times$3$\times$5               &  400            \\
LaFeO$_3$       &  500                              &  5$\times$3$\times$5               &  400            \\
\hline
\end{tabular}
\label{tab:bands}
\end{table}

US and NC data are generally very similar, apart from the 3$d$ systems, in particular, titanates SrTiO$_3$ and LaTiO$_3$, for which the US gaps are larger by about 15\% compared to the NC values and, to a lesser extent,
LaCrO$_3$, LaFeO$_3$, LaMnO$_3$ where the  difference reduces to $\approx$ 10\%: clearly, the discrepancy is correlated with the difference $\delta_d$ between the all-electron and pseudized norm of the 3$d$ orbitals, which is
larger for $d$ elements (as discussed previously, see Tab.~\ref{tab:pp}), and the character of the VBM and CBM.
Finally, the comparison between extrapolated and non-extrapolated schemes (here inspected for US PAW only) confirms that these two  methods lead to similar results for the entire TM perovskite dataset, with differences of about 0.1 eV. The only exception is LaScO$_3$ for which the non-extrapolated value of the gap is 0.3 eV smaller than the extrapolated one.
We reaffirm, however, that only the basis-set corrected scheme is founded on a solid mathematical basis.  Especially for US PAWs, it is also computational more efficient than the non-extrapolated scheme as it reduces the number of
calculations to be performed with a large number of bands and \textbf{k}-points.

\subsection{Statistical correlations}

\begin{figure}[ht!]
\includegraphics[width=0.5\textwidth]{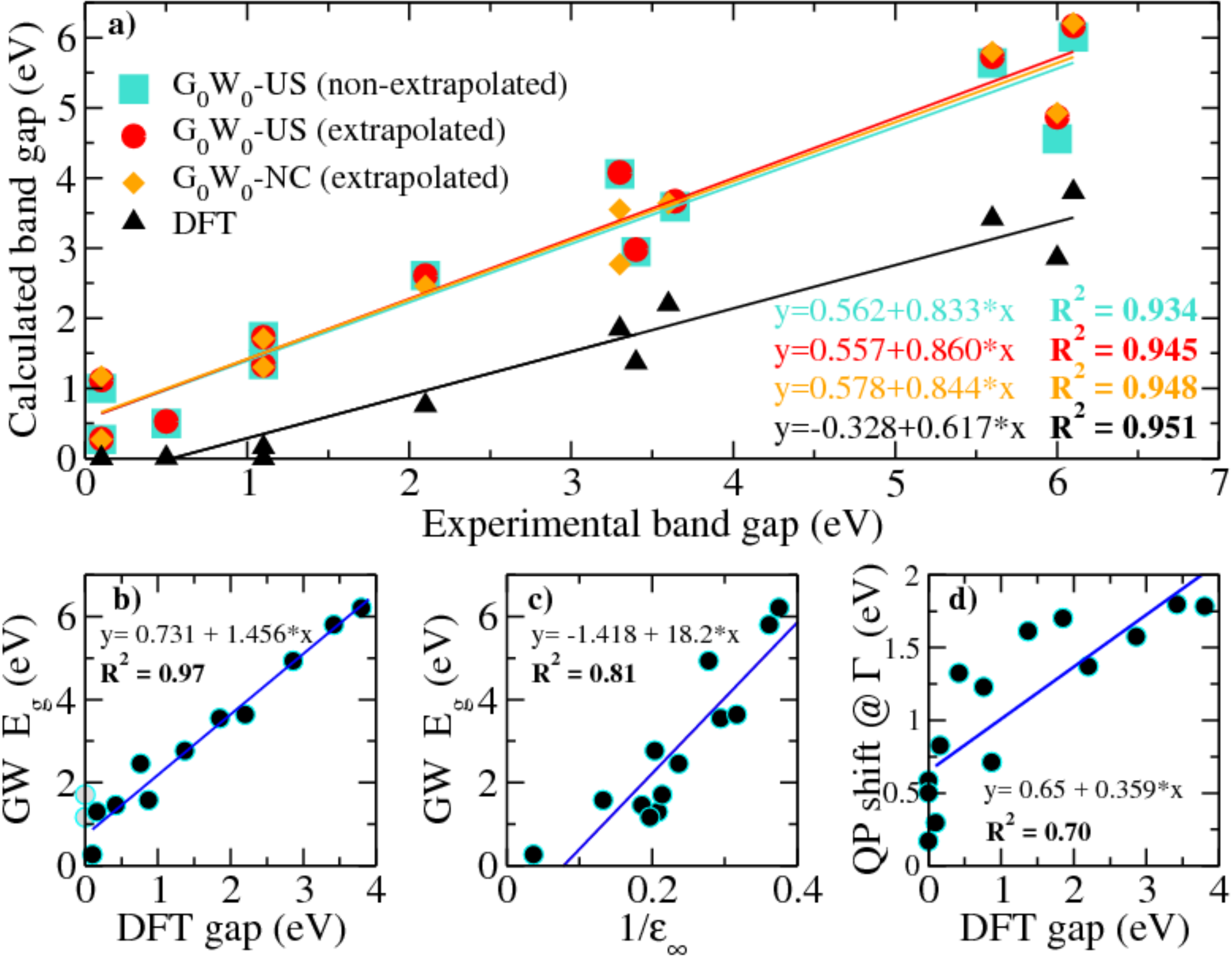}
\caption{(Color online) Statistical interpretation of the GW data by means of a linear regression.
(a) Comparison between the calculated and experimental gaps. The calculated values include G$_0$W$_0$ results obtained following the different schemes discussed in the main text: conventional non-extrapolated scheme using US PAW and the basis-set correction procedure using both NC and US PAW. (b) Correlation between the  G$_0$W$_0$ and the DFT gap. (c) Correlation between the G$_0$W$_0$  gap and the calculated static dielectric constant $\epsilon_\infty$. (d) Comparison between the DFT gap and the QP shift at $\Gamma$.
In each panel the linear relation and the R$^2$ factors are given in the insets.}
\label{fig:09}
\end{figure}

Even though the material data set under scrutiny is limited to 14  compounds, a minimal statistical analysis of the results is useful, in particular considering the complexity of the systems, the degree of accuracy of the method adopted and the increasing interest in automatizing
first-principles calculations within a high-throughput framework~\cite{Curtarolo}.  To this end, we have inspected possible correlations between different types of identifiers: QP gap, QP shift, DFT gap and the static dielectric constant $\epsilon_\infty$. The results are summarized in Fig.~\ref{fig:09}. 
First, we note that there is a relatively strong correlation, $\sim$0.97, between the calculated and experimental gaps, regardless of the specific GW flavor (NC-PAW extrapolated data are slightly better than the others) and the correlation is essentially identical for GW and DFT 
[see Fig.~\ref{fig:09}(a)]. This result is in line with the very recent results of van Setten and coworkers~\cite{Setten},
who found a correlation of $R^2=$0.962 (GW) and 0.957 (DFT) for a larger set of 77 materials including monoatomic and binary semiconductors.
In the insets of Fig.~\ref{fig:09}(a) we provide the linear relations to reproduce the experimental gap starting from the calculated band gaps.

In agreement with the conclusions of Ref.~\onlinecite{Setten} we confirm that also for TMO perovskites the correlation between the GW and DFT gap, 0.97, [see Fig.~\ref{fig:09}(b)] is larger than the correlation between the calculated and experimental gap
[see Fig.~\ref{fig:09}(a)], meaning that it is more accurate (smaller average error) to reproduce the GW gap starting from PBE gap than to approximate the experimental gap based on GW data. A standard  linear regression procedure leads to an 
$E_g^{GW} = 0.731 + 1.456E_g^{DFT}$ relation, which should be compared with the corresponding relation found by van Setten, namely
$E_g^{GW} = 0.51 + 1.32E_g^{DFT}$~\cite{Setten}. 

Interestingly, we found that the static dielectric constant $\epsilon_\infty$ (the average of the diagonal part of the static dielectric tensor, see next section for more details) is another  identifier that can be used in high-throughput automatic GW calculations. In this case  the correlation between the  G$_0$W$_0$ gap and $\epsilon_\infty$ is quantified by the relation 
$E_g^{GW} = -1.418 + 18.2 \frac{1}{\epsilon_\infty}$, with an associated correlation of 0.81 [see Fig.~\ref{fig:09}(c)]. Even though the linear relation between the GW gap and $\epsilon_\infty$  can be useful, the accurate calculation or measurement of $\epsilon_\infty$ is not an easy task~\cite{paier, jghe}. Finally, we have also inspected the relation between the DFT gap and the QP shift @ $\Gamma$ but we found a rather low correlation of 0.70 (see Fig.~\ref{fig:09}d).

\begin{figure}[ht!]
\includegraphics[width=0.49\textwidth]{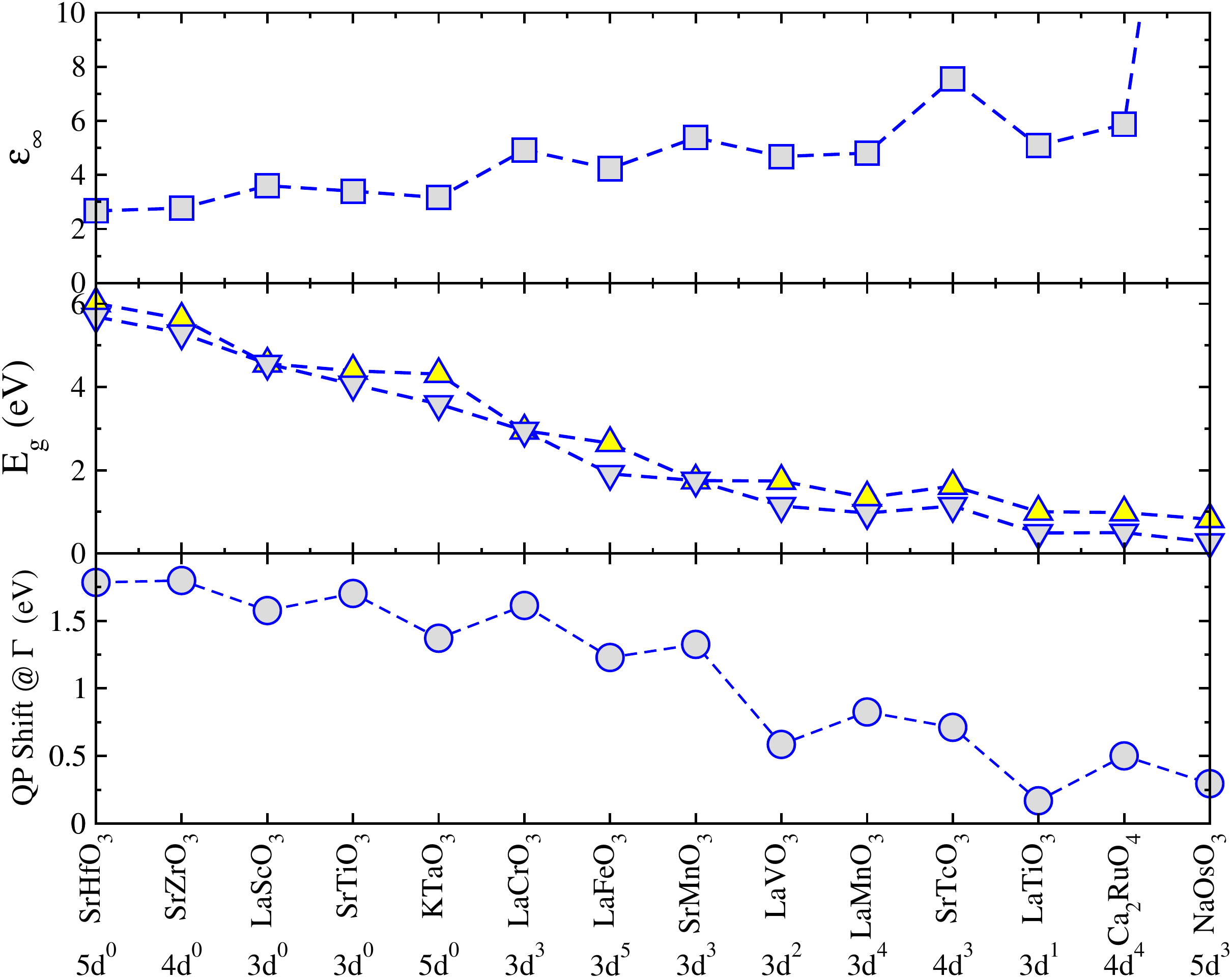}
\caption{(Color online) Comparison between calculated QP shift, band gap (indirect--down-triangles, and direct at $\Gamma$--up-triangles), and 
avarage of the diagonal component of the static dielectric tensor using the conventional non-extrapolated  scheme and 
adopting US PAWs. The QP shift is defined as the difference between the G$_0$W$_0$ QP energy and DFT eigenvalue.}
\label{fig:10}
\end{figure}

To conclude this part we collect in Fig.~\ref{fig:10} the various identifiers used for the statistical analysis:
band gap, QP shift and $\epsilon_\infty$. The cubic $d^0$ compounds exhibit the largest gaps (4-6 eV) and the largest QP shifts ($\approx$~1.5 eV). In the 3$d$ series, the gap $E_g^\Gamma$ decreases progressively depending on the filling of the $d$ orbitals (larger, 1.7-3 eV, for half-filled systems) and so does the QP shift (1.5-0.6 eV). Finally for 4$d$ and 5$d$ compounds with partially filled $d$ bands the direct gap at $\Gamma$ is $\approx$~1~eV, and the corresponding QP shift is about 0.5 eV. 
Summing up, for less correlated $d^0$ and 4-5$d$ materials the QP shift is roughly 25\% of the direct gap, whereas
in the more correlated 3$d$ perovskites the QP shift increases to 50-60\% of the gap.
As expected, the dielectric constant follows an opposite trend. It increases with decreasing gap size and approaches a metallic-like limit for NaOsO$_3$ (which is on the verge of a Lifshitz insulator-to-metal transition~\cite{Lifshitz}) for which $\epsilon_\infty \approx 27$. 

\subsection{Band Structures and Optical Spectra}

After having analyzed in detail the convergence of the QP energies in the G$_0$W$_0$ method, we turn now to the calculation of the electronic band structure and optical spectra for the considered TM perovskite dataset.
To this end,  we have used US PAWs and the non-extrapolated scheme according to the technical set-up given in Tab.~\ref{tab:bands}.
In fact, for the calculation of the band structure the basis-set correction scheme is unpractical because
it would be necessary to apply the extrapolation procedure to each QP energy using a sufficiently large
\textbf{k}-point mesh (required for the Wannier interpolation, see below). This would clearly result in a cumbersome procedure, and the need to use many  \textbf{k}-points would wipe out the advantages of the \textbf{k}-points correction scheme.

Due to technical reasons related to the \textbf{k}-point sampling, it is presently not possible to calculate
the QP energies for non-uniform \textbf{k}-point meshes in the GW method. A common alternative is the interpolation of the QP energies obtained for a uniform mesh using maximally localized Wannier functions (MLWF); in VASP this is done by using the VASP2WANNIER90 interface~\cite{CF1} which connects VASP with the Wannier90 suite~\cite{wannier90}. We have followed this approach for the calculation of the band structures,
and used as an orbital basis for the Wannier projections the full $d$ manifold of the TM ion ($e_g$ and $t_{2g}$) and the O-$p$ states.
This choice is adequate to accurately describe the electronic bands in a few eV windows around the Fermi energy, as for all materials
the top part of the valence band has mixed O-$p$/TM-$d$ character and the bottom portion of the conduction band is generally dominated
by empty TM-$d$ states (see SM~\cite{SM}).

\begin{table}[h!]
\centering
\caption{Percentage of O-$p$ and TM-$d$ states at $\Gamma$ at the VBM and CBM from the projected orbitals.
For Ca$_2$RuO$_4$ the data in brackets refer to the values taken for the VBM and CBM at (0.25,0,0).}
\begin{tabular}{ccccc}
\hline\hline
Compound            &   ${VBM}$ O-$p$     &    ${VBM}$ TM-$d$       & ${CBM}$ O-$p$                 &  ${CBM}$ TM-$d$          \\
\hline
SrTiO$_3$           &    100  $\%$        &         0 $\%$          &     0   $\%$                  &     100 $\%$          \\
SrZrO$_3$           &    100  $\%$        &         0 $\%$          &     0   $\%$                  &     100 $\%$          \\
SrHfO$_3$           &    100  $\%$        &         0 $\%$          &     0   $\%$                  &     100 $\%$          \\
KTaO$_3$            &    100  $\%$        &         0 $\%$          &     0   $\%$                  &     100 $\%$          \\
SrMnO$_3$           &    45   $\%$        &        54 $\%$          &     3   $\%$                  &      93 $\%$           \\
LaScO$_3$           &    98   $\%$        &         1 $\%$          &     8   $\%$                  &      76 $\%$           \\
LaTiO$_3$           &     6   $\%$        &        90 $\%$          &     2   $\%$                  &      93 $\%$           \\
LaVO$_3$            &    10   $\%$        &        89 $\%$          &    14   $\%$                  &      81 $\%$           \\
LaCrO$_3$           &    20   $\%$        &        78 $\%$          &     4   $\%$                  &      92 $\%$          \\
LaMnO$_3$           &    24   $\%$        &        72 $\%$          &    11   $\%$                  &      85 $\%$            \\
LaFeO$_3$           &    33   $\%$        &        61 $\%$          &     7   $\%$                  &      90  $\%$            \\
SrTcO$_3$           &    14   $\%$        &        86 $\%$          &    13    $\%$                 &      87  $\%$         \\
Ca$_2$RuO$_4$       &    14   $\%$        &        84 $\%$          &    24   $\%$                  &      74  $\%$        \\
                    &   (24   $\%$)       &       (75  $\%$)        &     (24   $\%$)               &     (75  $\%$)        \\
NaOsO$_3$           &    23   $\%$        &        73 $\%$          &    18    $\%$                 &      78 $\%$         \\
\hline\hline
\end{tabular}
\label{tab:occupation_gamma}
\end{table}

\begin{figure*}[ht!]
\includegraphics[width=0.9\textwidth]{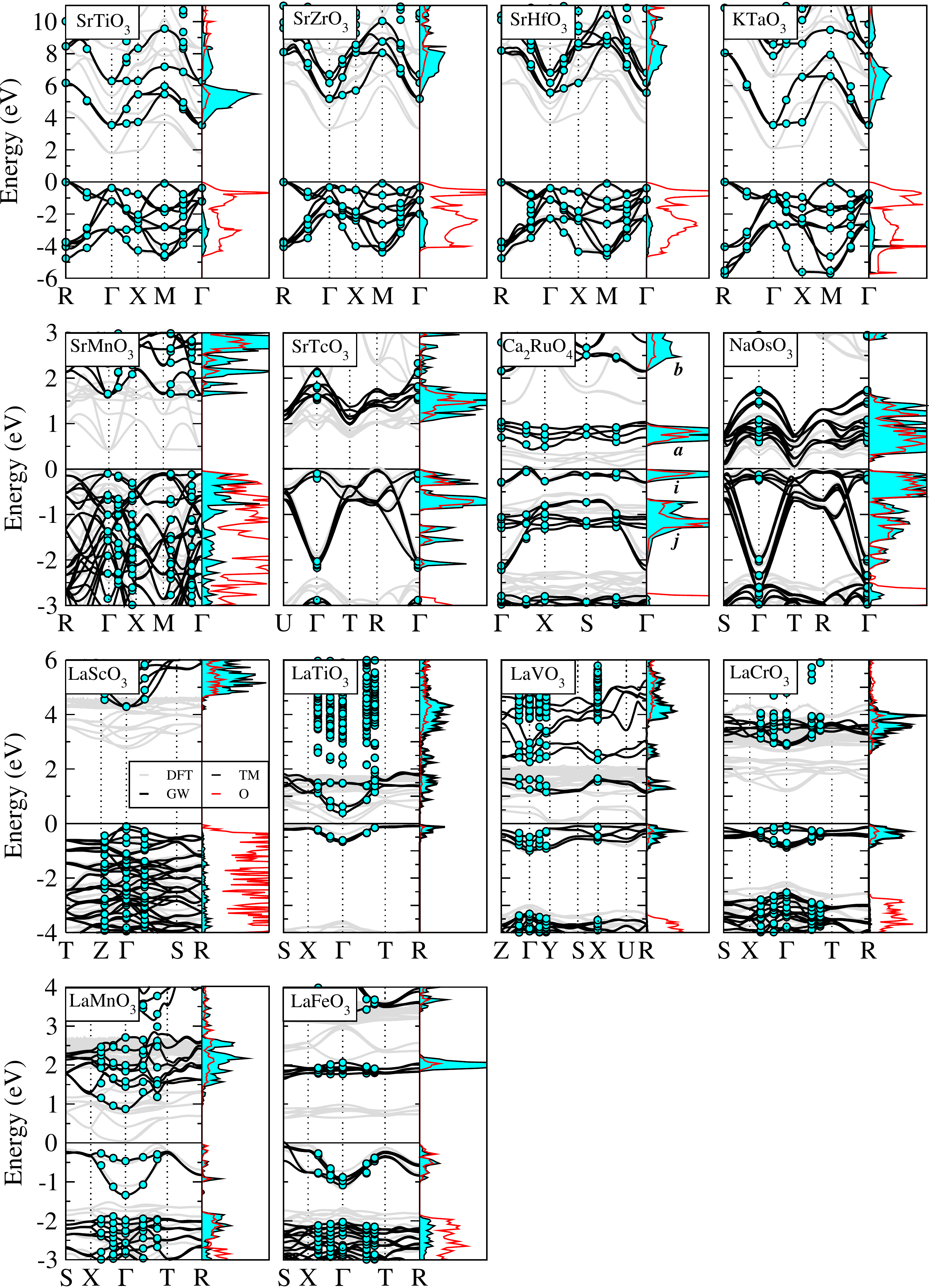}
\caption{(Color online) Collection of the calculated  band structures for DFT (gray lines) and GW (black) together with the GW density of TM-$d$ (shadow, cyan line) and O-$p$ (full line, red) states. The filled circles indicate the calculated GW QP energies  (used for the Wannier interpolation). As mentioned in the main text, the DFT calculations for LaTiO$_3$ and LaVO$_3$ were performed with the addition of a small
effective $U$.}
\label{fig:11}
\end{figure*}

The band structures are compiled in Fig.~\ref{fig:11}, where we show a comparison between PBE and GW-derived bands, along with the computed GW density of states (DOS).  First, we note that the quality of the Wannier interpolation is generally very good, as established by the excellent
match between the interpolated bands and the actual GW QP energies  used for the interpolation procedure (shown as filled circles) and by the
smoothness of the electron dispersions.

\begin{figure*}[ht!]
\includegraphics[width=0.98\textwidth]{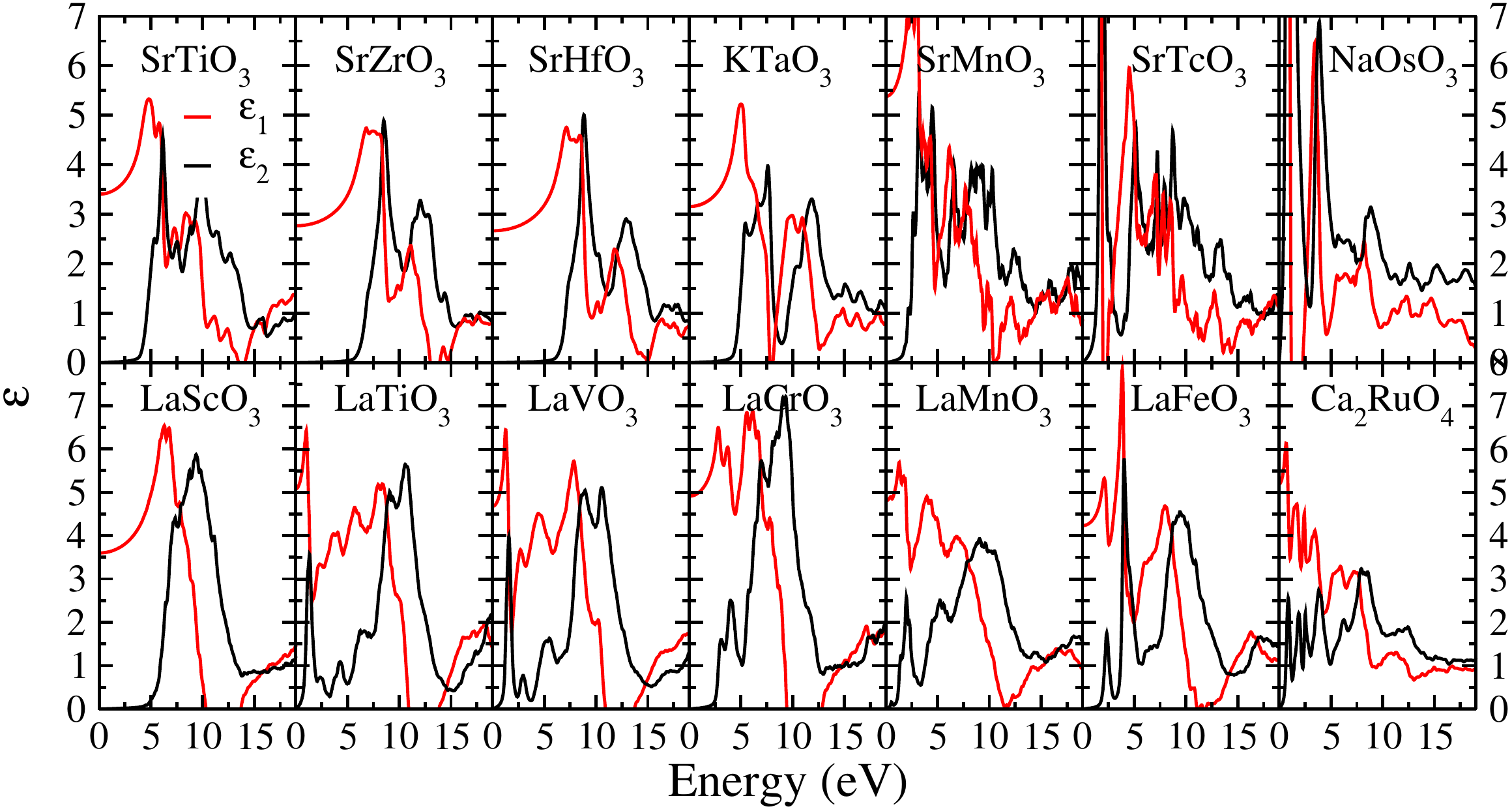}
\caption{(Color online) Average of diagonal component of the real ($\epsilon_1$) and imaginary ($\epsilon_2$) part of the dielectric function obtained by G$_0$W$_0$ for the
entire materials dataset studied in this work.
The values of the static ion-clamped dielectric function $\epsilon_{\infty}$
are listed in Tab.~\ref{epsilon}.
}
\label{fig:12}
\end{figure*}

\begin{figure*}[ht!]
\includegraphics[width=0.98\textwidth]{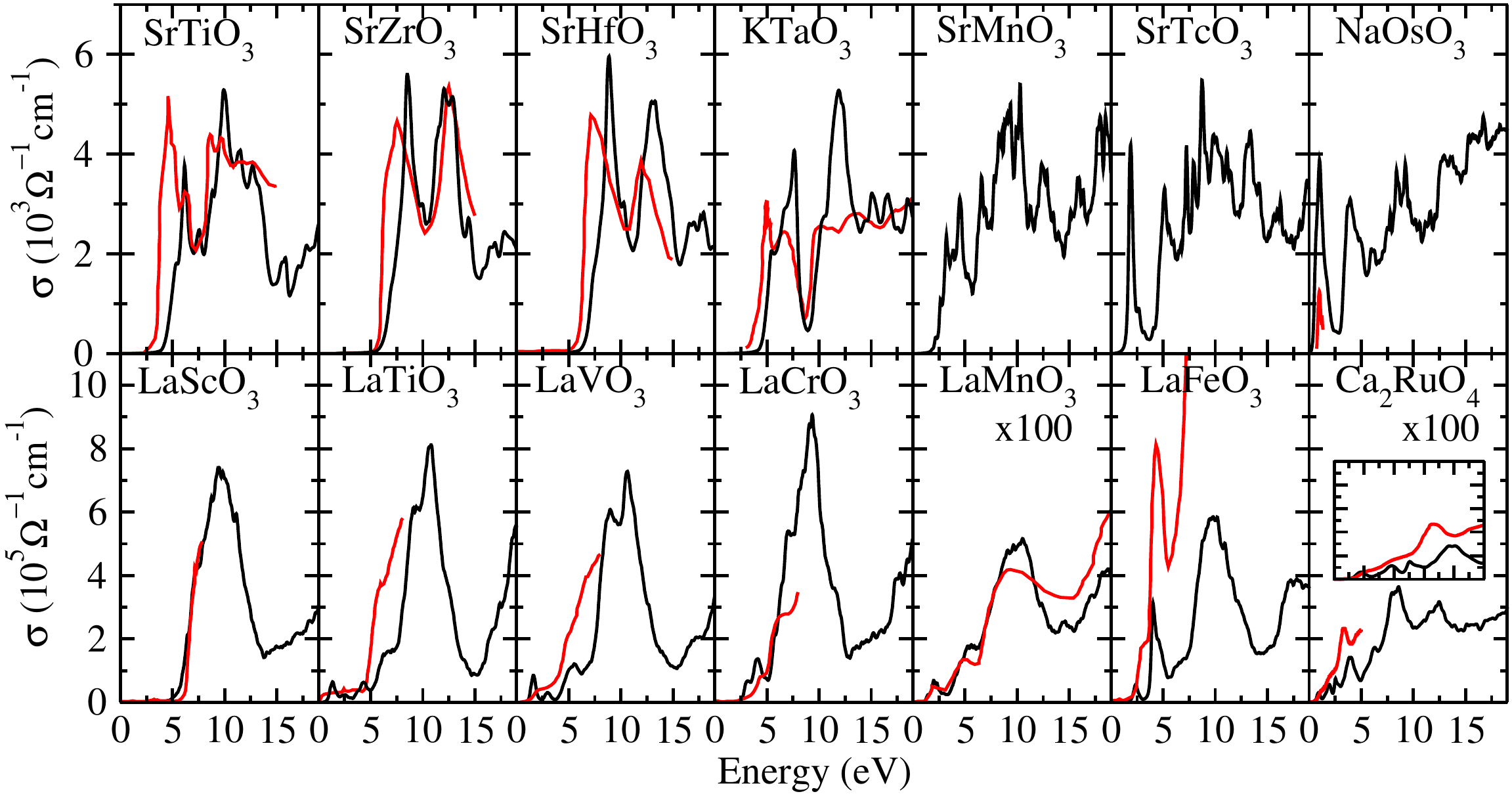}
\caption{(Color online) Collection of calculated (black line) optical spectra along with available experimental curves (gray/red line); for
SrTiO$_3$, SrZrO$_3$ and SrHfO$_3$ the experimental data are taken from
Ref.~\onlinecite{Lee2010}, for KTaO$_3$ from Ref.~\onlinecite{KTO},
for NaOsO$_3$ from Ref.~\onlinecite{Vecchio2013},
for the La series from Ref.~\onlinecite{Arima1993} (LaMnO$_3$ from Ref.~\onlinecite{PhysRevB.55.15489},
and for Ca$_2$RuO$_4$ from Ref.~\onlinecite{PhysRevLett.91.056403}.
For Ca$_2$RuO$_4$ the inset shows a zoom of the low-energy region.
The calculated dielectric functions from which the optical conductivity spectra have been derived are shown in Fig.~\ref{fig:12}.}
\label{fig:13}
\end{figure*}

By combining the information included in the band structures and DOS with the quantitative analysis of the orbital character at
the CBM and VBM at the $\Gamma$ point (see Tab.~\ref{tab:occupation_gamma}) it is possible to draw some conclusions on the type of band gap.
The $d^0$ cubic systems SrTiO$_3$, SrZrO$_3$, SrHfO$_3$, and KTaO$_3$ are band insulators characterized by a $p$-$d$ fundamental gap
that is also well visible as first excitation peak in the calculated and experimental optical spectra shown in Fig.~\ref{fig:13}.
Also LaScO$_3$ falls in the category of band insulators, even though the conduction band has a sizable amount of  O-$p$ states, which causes a
broadening of the first excitation peak (see Fig.~\ref{fig:13}). The other compounds have a predominant $d$-$d$ fundamental gap, with some distinctions:
LaTiO$_3$  exhibits a clear Mott gap with only marginal (about 10 \%) O-$p$ states at the valence band;
LaVO$_3$, LaCrO$_3$, LaMnO$_3$ and SrTcO$_3$ appear to have a predominant Mott character too, but it is known that
the gap in LaMnO$_3$ originates also from the Jahn-Teller instability~\cite{PhysRevLett.104.086402, PhysRevB.85.195135};
SrTcO$_3$ was reported to possess a substantial itinerant character, which places it on the verge of a Mott transition~\cite{PhysRevB.83.220402, PhysRevLett.108.197202}. 
The data suggest that SrMnO$_3$ and LaFeO$_3$ can be assimilated to an intermediate Mott/charge-transfer nature, as they have a strong intermix of O-$p$ and TM-$d$ states at the valence bands,
whilst the conduction bands are largely formed by empty $d$ orbitals.
Ca$_2$RuO$_4$ exhibits a $pd$-$pd$ gap, with 25\% of O-$p$ states at both VBM and CBM.
NaOsO$_3$ is a peculiar case, characterized by electron and spin itinerancy, a relatively
strong SOC and a weak electron-electron correlation~\cite{Lifshitz, Vecchio2013, NOO}. In this case the valence and conduction bands are formed by a
strong mixture of O-$p$ and Os-$d$ states.

Finally, we used the GW QP energies to calculated the dielecric function in the independent particle approximation.
The real ($\epsilon{_1}$) and imaginary ($\epsilon{_2}$) parts of dielectric functions $\epsilon=\epsilon_1+i \epsilon_2$ (shown in Fig.~\ref{fig:12}) were used to compute the
optical conductivity spectra as:
\begin{equation}
\sigma(\omega)= -i \omega \epsilon_0 [\epsilon(\omega)-1],
\end{equation}
where $\varepsilon_0$ is the vacuum dielectric constant.
The results are displayed in Fig.~\ref{fig:13} and include a comparison with the measured spectra from Ref.~\onlinecite{Lee2010}
(SrTiO$_3$, SrZrO$_3$, SrHfO$_3$), Ref.~\onlinecite{KTO} and Ref.~\onlinecite{Arima1993} (Lanthanum series). For SrMnO$_3$ and
SrTcO$_3$ we could not find any experimental reports in the literature. The agreement between the calculated and measured values are
generally good. Unfortunately for the La-series, the experimental data are limited to a small frequency window, which only allows for
a comparison with the onset of the optical excitations. The cubic systems exhibit two prominent structures. According to the electronic
structure properties discussed above, the first peak corresponds to the interband transition from O-2$p$ to TM-$t_{2g}$ states,
and the second one is associated with the transition from O-2$p$ to TM-$e_{g}$ orbitals.

The energy separation between the two main peaks is predominantly determined by the crystal field splitting. 
Using the GW QP energies, the independent particle approximation underestimates the energy seperation by 15-20\%: we obtained 1.5 eV (expt: 1.9 eV), 4.0 eV (expt: 5.1 eV) and  4.3 eV (expt: 4.9 eV) in SrTiO$_3$, SrZrO$_3$, SrHfO$_3$, respectively. We believe that this error is mainly related to the independent particle
approximation which places the t$_{2g}$ at too high energies. Including excitonic effects would lower
these t$_{2g}$ states [preliminary calculations on SrTiO$_3$ based on the Bethe-Salpeter equation (BSE) 
confirm this conclusion, in agreement with recent BSE data~\cite{PhysRevB.87.235102}].
The spectrum of KTaO$_3$ exhibits a less pronounced separation into two main peaks. This is due
to the larger band-width of both the valence and conduction bands (see bandstructure in Fig.~\ref{fig:11}), which allows for
more broadened optical transitions and the appearance of shoulders close to the main peaks.

For NaOsO$_3$ the calculated optical
conductivity follows well the measurements of Lo Vecchio~\cite{Vecchio2013} \emph{et al.}: the main absorption edge is mainly associated to
charge-transfer excitations among Os 5$d$ and O 2$p$ states (see bandstructure in Fig.~\ref{fig:11}).

The variation of the optical
properties and of the band gaps in the La-series has been discussed  in the seminal paper of Arima and coworkers.~\cite{Arima1993}.
Also for this set of compounds, the agreement with the measured data is satisfactory, with the exception of LaVO$_3$ for which
GW predicts a substantial blue-shift of the strongest excitation peak at about 2.5 eV. This discrepancy can again be related
to the neglect of excitonic effects, which are strong for $p\to d$ transitions.
However, the onset of the optical spectrum
at 1.2 eV is well reproduced by theory and corresponds to the characteristic $d$-$d$ transition, which is weak due to the low
density of empty states at the bottom of the conduction band (see Fig.~\ref{fig:11}). LaScO$_3$ is a clear band insulator with the first
(charge-transfer) optical excitation arising from the O-$p$ to Sc-$d$ transition; the experimental spectrum does not clearly show the tail
at the bottom of the spectrum well visible in GW. We believe that the GW bandstructure is reliable in this respect, 
and that the onset of optical adsorption is not easily dedected in the experiment. We therefore trust
that our  predicted band gap of 4.9~eV should be more reliable than the experimental estimate of 6.0 eV (see Tab.~\ref{bandgaps}, and Fig.~\ref{fig:11}).
As the 3$d$ states start to become occupied (LaTiO$_3$, 3$d^1$) a Mott peak shows up in the low-energy region of the optical conductivity,
but the overall spectra are still dominated by the intense charge-transfer peak located at about 9-10 eV (depending on the specific
system). In LaFeO$_3$ a third relatively intense feature appears at 4.1 eV between the lowest Mott peak (2.3 eV) and the charge
transfer peak (9.8 eV), which can be assigned to the transition from the VBM to the group of bands centered at 4 eV above the Fermi
level which have a mixed O-$p$ and Fe-$d$ character (see Fig.~\ref{fig:11}).

Finally, Ca$_2$RuO$_4$ displays 4 main peaks in the lowest part of the optical conductivity (i.e., in the energy window up to
5 eV for which experimental reports are available). On the basis of the electronic properties (bands and DOS) and following the
labelling given in Fig.~\ref{fig:11} we can tentatively assign the first peak at 1 eV to the $i \rightarrow a$ transition,
the second two peaks at 2 and 2.5 eV can be interpreted as $j \rightarrow a$ and $i \rightarrow b$  excitations, and finally
the broad and intense peak at about 4 eV should correspond to transition from the valence bands $i,j$ to the main $d$ bands $b$.
While the more intense peak should have a clear charge-transfer character, the other transition might involve $d$-$d$ Mott-like
transitions~\cite{PhysRevLett.91.056403}. However, a more quantitative and certain analysis of the specific type of transition would
require beyond GW approaches such as the solution of the Bethe-Salpeter equation, which will be the topic of a future work. This comment
applies to some degree to the interpretation of all spectra.

\begin{table}
\caption{Diagonal part of the static ($\omega=0$, ion-clamped) dielectric matrix $\epsilon^{\alpha\beta}_\infty$ calculated by means of the G$_0$W$_0$ approximation.}
\begin{tabular}{lccccccc}
\hline\hline
Compound           & $\epsilon^{XX}_\infty$     &  $\epsilon^{YY}_\infty$     &    $\epsilon^{ZZ}_\infty$               \\
\hline
SrTiO$_3$          &   3.40                          &  3.40                           &   3.40                                        \\
SrZrO$_3$          &   2.77                          &  2.77                           &   2.77                                       \\
SrHfO$_3$          &   2.67                          &  2.67                           &   2.67                                        \\
KTaO$_3$           &   3.16			     &  3.16		 	       &   3.16 		         		\\
SrMnO$_3$          &   5.26                          &  5.26                           &   5.61                             		 \\
LaScO$_3$          &   3.46                          &  3.67                           &   3.66                                         \\
LaTiO$_3$          &   5.31                          &  5.05                           &   4.89                           		\\
LaVO$_3$           &   3.83                          &  6.51		               &   3.70                          		\\
LaCrO$_3$          &   4.90                          &  4.89                           &   4.96                                         \\
LaMnO$_3$          &   4.83                          &  4.58                           &   5.02                                         \\
LaFeO$_3$          &   4.23                          &  4.19                           &   4.28                      			\\
SrTcO$_3$          &   7.53                          &  7.46                           &   7.66                                         \\
Ca$_2$RuO$_4$      &   5.80                          &  5.90                           &   3.90                                         \\
NaOsO$_3$          &  26.02                          &  28.63                          &   27.05                                       \\
\hline\hline
\end{tabular}
\label{epsilon}
\end{table}

\section{Conclusions}

In summary, in this study, we have assessed the performance and accuracy of the single shot G$_0$W$_0$ approximation for the calculation of
converged QP energies for transition-metal perovskites using two different schemes (the basis-set correction procedure and the conventional
non-extrapolated method) and inspected the dependence of the results on the type of PAW used in the computation (ultrasoft vs. norm-conserving).
In order to draw general conclusions valid for different physical environments, we have performed a series of calculations on a TM perovskites
dataset comprising 14 compounds representative of the variety of properties characteristic of this class of materials: magnetic and non-magnetic systems,
with and without structural distortions, with different occupancies ($d^0$ $\rightarrow$ $d^5$) and spatial extension [3$d$, 4$d$ and 5$d$]
of the outermost $d$ shell, with band-gap ranging from 0.1 eV (NaOsO$_3$) up to 6.1 eV (SrHfO$_3$) and different types of main optical excitations
(Mott-Hubbard, charge-transfer, relativistic and band insulators).

We reassert that the formally (mathematically) corrected procedure to obtain accurate QP energies requires a basis-set as well as a \textbf{k}-points
correction. However, these corrections, in particular, the basis-set extrapolation, can become computationally prohibitive when combined with NC PAWs because
norm-conserving pseudopotentials are generally constructed with a much larger number of plane-waves compared to ultrasoft potentials (this is the case for Ca$_2$RuO$_4$, 
for which we could not perform NC-based calculations).
On the other side, the use of US PAWs makes the basis-set corrections scheme
computationally more advantageous than the non-extrapolated scheme because well-converged results can be achieved with few k-points and the k-point correction requires only
a small number of bands (and energy cutoffs).
Even though the reliability of the conventional scheme, based on a progressive increase of the most important technical parameters influencing the convergence of the results
(cutoff energy, number of bands, and number of \textbf{k}-points) is not supported by a mathematical demonstration, our numerical
results indicate that in most cases this scheme leads to reasonably converged QP energies very similar to those achieved by means of the basis set correction scheme.
This conclusion is of great practical importance as this allows to easily compute energy dispersion relations (band structures) and optical spectra, which cannot
be computed using the basis-set correction method. The so obtained optical spectra, based on the calculation of the frequency dependent dielectric function
(without the inclusion of excitonic effects), are in good agreement with experimentally available measurements and provide useful insight on the
characterization of the most important optical transitions.

Concerning the difference between NC and US PAWs, the main source of inaccuracy in US-based calculations is the degree of norm-violation for the $d$ shell, which
can be as large as 0.4-0.45 electrons for Mn and Fe. For most of the compounds considered in the present study, this inaccuracy is somehow compensated by using
a robust technical setup in US-based calculations [typically $E_{pw}$=500-600~eV, N=500 (in some cases up to 2000), and a 4$\times$4$\times$4 \textbf{k}-point mesh].
In this respect, the most problematic compounds turned out to be SrTiO$_3$, for which US PAWs deliver a band gap 0.5 eV larger than the corresponding
NC value. In general, particular care is required for 3$d$ charge-transfer insulators ($p \rightarrow d$ transition). Also, late transition metals are more difficult 
than early transisiotn metals, since the $d$ electrons become more localized towards the end of the $d$ series.

Finally, a basic statistical analysis of our results indicates a strong correlation between the calculated and experimental band gap ($R^2$=0.94-0.95) as well as a very robust correlation
between the GW and DFT gaps ($R^2$=0.98). Also, we found a correlation between the GW gap and the static dielectric function which could be particularly useful for automatic
high-throughput calculations.  As expected, the QP shift decreases 
with decreasing band gap, and this trend is characterized by two distinct behaviors: for the less electronically correlated $d^0$ $p$-$d$ insulators and 4-5$d$ compounds the
QP shift is about 25\% of the value of the gap, whereas for the more correlated 3$d$ materials the QP shifts increase up to 50-60\%.

We hope that the results and conclusions of our work will serve as useful references for future GW calculations on complex transition metal oxides.

\begin{acknowledgments}
This work was supported by the joint FWF (Austrian Science Fund) and Indian Department of Science and Technology (DST) project INDOX (I1490-N19),
and by the FWF-SFB ViCoM (Grant No. F41).
Computing time at the Vienna Scientific Cluster is greatly acknowledged.
\end{acknowledgments}

\appendix

\section{Details on the employed PAW potentials}

\begin{table}[ht!]
\centering
\caption{List of radial cutoff parameters (core radii, in atomic units) $r_s$, $r_p$, $r_d$, $r_f$ for each angular quantum number and default $E_{pw}$ for all type of potentials used in the present work.}
\begin{tabular}{lcccccc}
\hline\hline
Element                    & PAW      &  ~~$r_s$~~  &  ~~$r_p$~~  & ~~$r_d$~~  &   ~~$r_f$~~  &     ~~$E_{pw}$~~     \\
\hline
 O                         & US      &  1.2    & 1.5     & 1.6    &  1.4     &     434  \\
                           & NC      &  1.0    & 1.1     & 1.1    &   -      &     765   \\
 Na                        & US      &  1.4    & 2.2     & 2.2    &   -      &     373         \\
                           & NC      &  1.2    & 2.2     & 2.2    &   -      &     467           \\
 K                         & US      &  1.7    & 2.0     & 2.5    &   -      &     249    \\
                           & NC      &  1.0    & 1.8     & 2.1    &  2.1     &     633    \\
 Ca                        & US      &  1.6    & 1.9     & 2.2    &   -      &     281        \\
                           & NC      &  0.9    & 1.7     & 1.9    &   2.1    &     771    \\
 Sc                        & US      &  1.7    & 1.7     & 1.9    &   2.0    &     379  \\
                           & NC      &  0.9    & 1.5     & 1.9    &   1.9    &     778   \\ 
 Ti                        & US      &  1.7    & 1.7     & 2.0    &   2.0    &     384  \\
                           & NC      &  0.9    & 1.4     & 1.9    &   1.9    &     785   \\
 V                         & US      &  1.8    & 1.7     & 1.9    &   2.0    &     384   \\
                           & NC      &  0.8    & 1.3     & 1.9    &   1.9    &     800  \\
 Cr                        & US      &  2.8    & 2.5     & 2.5    &   2.8    &     219  \\
                           & NC      &  0.8    & 1.2     & 1.9    &   1.9    &     819  \\
 Mn                        & US      &  1.6    & 1.7     & 1.9    &   1.9    &     385   \\
                           & NC      &  1.6    & 1.3     & 1.9    &   1.9    &     781   \\
 Fe                        & US      &  1.5    & 1.7     & 1.9    &   2.0    &     388 \\
                           & NC      &  1.5    & 1.7     & 1.9    &   1.9    &     786 \\
 Sr                        & US      &  1.7    & 2.1     & 2.5    &  2.5     &     225       \\
                           & NC      &  1.1    & 2.0     & 2.3    &  2.1     &     543       \\
 Zr                        & US      &  1.3    & 1.8     & 2.0    &   2.1    &     346   \\
                           & NC      &  1.0    & 1.9     & 2.1    &   1.9    &     637  \\
 Tc                        & US      &  2.0    & 1.8     & 2.0    &   2.1    &     351   \\
                           & NC      &  0.9    & 1.9     & 2.1    &   1.9    &     639   \\
 Ru                        & US      &  2.0    & 1.8     & 2.0    &   2.1    &     348   \\
                           & NC      &  0.9    & 1.9     & 2.1    &   1.9    &     660   \\
 La                        & US      &  1.6    & 1.8     & 2.2    &   2.5    &     314            \\
 Hf                        & US      &  1.5    & 1.9     & 2.2    &   2.5    &     283   \\
                           & NC      &  2.2    & 2.0     & 2.2    &   1.9    &     576 \\
 Ta                        & US      &  1.5    & 1.9     & 2.2    &   2.5    &     286  \\
                           & NC      &  2.1    & 2.0     & 2.2    &   1.9    &     584  \\
 Os                        & US      &  1.5    & 1.8     & 2.2    &   2.3    &     319   \\
                           & NC      &  2.0    & 2.0     & 2.1    &   1.9    &     647   \\

 \hline
 \end{tabular}
 \label{corerad}
 \end{table}

\bibliography{reference}

\begin{thebibliography}{104}%
\makeatletter
\providecommand \@ifxundefined [1]{%
 \@ifx{#1\undefined}
}%
\providecommand \@ifnum [1]{%
 \ifnum #1\expandafter \@firstoftwo
 \else \expandafter \@secondoftwo
 \fi
}%
\providecommand \@ifx [1]{%
 \ifx #1\expandafter \@firstoftwo
 \else \expandafter \@secondoftwo
 \fi
}%
\providecommand \natexlab [1]{#1}%
\providecommand \enquote  [1]{``#1''}%
\providecommand \bibnamefont  [1]{#1}%
\providecommand \bibfnamefont [1]{#1}%
\providecommand \citenamefont [1]{#1}%
\providecommand \href@noop [0]{\@secondoftwo}%
\providecommand \href [0]{\begingroup \@sanitize@url \@href}%
\providecommand \@href[1]{\@@startlink{#1}\@@href}%
\providecommand \@@href[1]{\endgroup#1\@@endlink}%
\providecommand \@sanitize@url [0]{\catcode `\\12\catcode `\$12\catcode
  `\&12\catcode `\#12\catcode `\^12\catcode `\_12\catcode `\%12\relax}%
\providecommand \@@startlink[1]{}%
\providecommand \@@endlink[0]{}%
\providecommand \url  [0]{\begingroup\@sanitize@url \@url }%
\providecommand \@url [1]{\endgroup\@href {#1}{\urlprefix }}%
\providecommand \urlprefix  [0]{URL }%
\providecommand \Eprint [0]{\href }%
\providecommand \doibase [0]{http://dx.doi.org/}%
\providecommand \selectlanguage [0]{\@gobble}%
\providecommand \bibinfo  [0]{\@secondoftwo}%
\providecommand \bibfield  [0]{\@secondoftwo}%
\providecommand \translation [1]{[#1]}%
\providecommand \BibitemOpen [0]{}%
\providecommand \bibitemStop [0]{}%
\providecommand \bibitemNoStop [0]{.\EOS\space}%
\providecommand \EOS [0]{\spacefactor3000\relax}%
\providecommand \BibitemShut  [1]{\csname bibitem#1\endcsname}%
\let\auto@bib@innerbib\@empty
\bibitem [{\citenamefont {von Helmolt}\ \emph {et~al.}(1993)\citenamefont {von
  Helmolt}, \citenamefont {Wecker}, \citenamefont {Holzapfel}, \citenamefont
  {Schultz},\ and\ \citenamefont {Samwer}}]{PhysRevLett.71.2331}%
  \BibitemOpen
  \bibfield  {author} {\bibinfo {author} {\bibfnamefont {R.}~\bibnamefont {von
  Helmolt}}, \bibinfo {author} {\bibfnamefont {J.}~\bibnamefont {Wecker}},
  \bibinfo {author} {\bibfnamefont {B.}~\bibnamefont {Holzapfel}}, \bibinfo
  {author} {\bibfnamefont {L.}~\bibnamefont {Schultz}}, \ and\ \bibinfo
  {author} {\bibfnamefont {K.}~\bibnamefont {Samwer}},\ }\href {\doibase
  10.1103/PhysRevLett.71.2331} {\bibfield  {journal} {\bibinfo  {journal}
  {Phys. Rev. Lett.}\ }\textbf {\bibinfo {volume} {71}},\ \bibinfo {pages}
  {2331} (\bibinfo {year} {1993})}\BibitemShut {NoStop}%
\bibitem [{\citenamefont {Salamon}\ and\ \citenamefont
  {Jaime}(2001)}]{RevModPhys.73.583}%
  \BibitemOpen
  \bibfield  {author} {\bibinfo {author} {\bibfnamefont {M.~B.}\ \bibnamefont
  {Salamon}}\ and\ \bibinfo {author} {\bibfnamefont {M.}~\bibnamefont
  {Jaime}},\ }\href {\doibase 10.1103/RevModPhys.73.583} {\bibfield  {journal}
  {\bibinfo  {journal} {Rev. Mod. Phys.}\ }\textbf {\bibinfo {volume} {73}},\
  \bibinfo {pages} {583} (\bibinfo {year} {2001})}\BibitemShut {NoStop}%
\bibitem [{\citenamefont {Imada}\ \emph {et~al.}(1998)\citenamefont {Imada},
  \citenamefont {Fujimori},\ and\ \citenamefont {Tokura}}]{Imada1998}%
  \BibitemOpen
  \bibfield  {author} {\bibinfo {author} {\bibfnamefont {M.}~\bibnamefont
  {Imada}}, \bibinfo {author} {\bibfnamefont {A.}~\bibnamefont {Fujimori}}, \
  and\ \bibinfo {author} {\bibfnamefont {Y.}~\bibnamefont {Tokura}},\ }\href
  {\doibase 10.1103/RevModPhys.70.1039} {\bibfield  {journal} {\bibinfo
  {journal} {Rev. Mod. Phys.}\ }\textbf {\bibinfo {volume} {70}},\ \bibinfo
  {pages} {1039} (\bibinfo {year} {1998})}\BibitemShut {NoStop}%
\bibitem [{\citenamefont {Bednorz}\ and\ \citenamefont
  {M{\"u}ller}(1986)}]{Bednorz1986}%
  \BibitemOpen
  \bibfield  {author} {\bibinfo {author} {\bibfnamefont {J.~G.}\ \bibnamefont
  {Bednorz}}\ and\ \bibinfo {author} {\bibfnamefont {K.~A.}\ \bibnamefont
  {M{\"u}ller}},\ }\href {\doibase 10.1007/BF01303701} {\bibfield  {journal}
  {\bibinfo  {journal} {Z. Phys. B Condens. Matter}\ }\textbf {\bibinfo
  {volume} {64}},\ \bibinfo {pages} {189} (\bibinfo {year} {1986})}\BibitemShut
  {NoStop}%
\bibitem [{\citenamefont {Tokura}(2003)}]{Tokura2003}%
  \BibitemOpen
  \bibfield  {author} {\bibinfo {author} {\bibfnamefont {Y.}~\bibnamefont
  {Tokura}},\ }\href {\doibase 10.1063/1.1603080} {\bibfield  {journal}
  {\bibinfo  {journal} {Phys. Today}\ }\textbf {\bibinfo {volume} {56}},\
  \bibinfo {pages} {50} (\bibinfo {year} {2003})}\BibitemShut {NoStop}%
\bibitem [{\citenamefont {Ohtomo}\ and\ \citenamefont
  {Hwang}(2004)}]{Ohtomo2004}%
  \BibitemOpen
  \bibfield  {author} {\bibinfo {author} {\bibfnamefont {A.}~\bibnamefont
  {Ohtomo}}\ and\ \bibinfo {author} {\bibfnamefont {H.~Y.}\ \bibnamefont
  {Hwang}},\ }\href {http://dx.doi.org/10.1038/nature02308} {\bibfield
  {journal} {\bibinfo  {journal} {Nature}\ }\textbf {\bibinfo {volume} {427}},\
  \bibinfo {pages} {423} (\bibinfo {year} {2004})}\BibitemShut {NoStop}%
\bibitem [{\citenamefont {Wang}\ \emph {et~al.}(2009)\citenamefont {Wang},
  \citenamefont {Liu},\ and\ \citenamefont {Ren}}]{Wang09}%
  \BibitemOpen
  \bibfield  {author} {\bibinfo {author} {\bibfnamefont {K.}~\bibnamefont
  {Wang}}, \bibinfo {author} {\bibfnamefont {J.-M.}\ \bibnamefont {Liu}}, \
  and\ \bibinfo {author} {\bibfnamefont {Z.}~\bibnamefont {Ren}},\ }\href
  {http://dx.doi.org/10.1080/00018730902920554} {\bibfield  {journal} {\bibinfo
   {journal} {Adv. Phys.}\ }\textbf {\bibinfo {volume} {58}},\ \bibinfo {pages}
  {321} (\bibinfo {year} {2009})}\BibitemShut {NoStop}%
\bibitem [{\citenamefont {Rao}(2000)}]{Rao}%
  \BibitemOpen
  \bibfield  {author} {\bibinfo {author} {\bibfnamefont {C.}~\bibnamefont
  {Rao}},\ }\href {http://dx.doi.org/10.1021/jp0004866} {\bibfield  {journal}
  {\bibinfo  {journal} {J. Phys. Chem. B}\ }\textbf {\bibinfo {volume} {104}},\
  \bibinfo {pages} {5877} (\bibinfo {year} {2000})}\BibitemShut {NoStop}%
\bibitem [{\citenamefont {Arima}\ \emph
  {et~al.}(1993{\natexlab{a}})\citenamefont {Arima}, \citenamefont {Tokura},\
  and\ \citenamefont {Torrance}}]{arima}%
  \BibitemOpen
  \bibfield  {author} {\bibinfo {author} {\bibfnamefont {T.}~\bibnamefont
  {Arima}}, \bibinfo {author} {\bibfnamefont {Y.}~\bibnamefont {Tokura}}, \
  and\ \bibinfo {author} {\bibfnamefont {J.~B.}\ \bibnamefont {Torrance}},\
  }\href {\doibase 10.1103/PhysRevB.48.17006} {\bibfield  {journal} {\bibinfo
  {journal} {Phys. Rev. B}\ }\textbf {\bibinfo {volume} {48}},\ \bibinfo
  {pages} {17006} (\bibinfo {year} {1993}{\natexlab{a}})}\BibitemShut {NoStop}%
\bibitem [{\citenamefont {Zhu}\ \emph {et~al.}(2014)\citenamefont {Zhu},
  \citenamefont {Li}, \citenamefont {Zhong}, \citenamefont {Xiao},
  \citenamefont {Xu}, \citenamefont {Yang}, \citenamefont {Zhao},\ and\
  \citenamefont {Li}}]{Zhu}%
  \BibitemOpen
  \bibfield  {author} {\bibinfo {author} {\bibfnamefont {J.}~\bibnamefont
  {Zhu}}, \bibinfo {author} {\bibfnamefont {H.}~\bibnamefont {Li}}, \bibinfo
  {author} {\bibfnamefont {L.}~\bibnamefont {Zhong}}, \bibinfo {author}
  {\bibfnamefont {P.}~\bibnamefont {Xiao}}, \bibinfo {author} {\bibfnamefont
  {X.}~\bibnamefont {Xu}}, \bibinfo {author} {\bibfnamefont {X.}~\bibnamefont
  {Yang}}, \bibinfo {author} {\bibfnamefont {Z.}~\bibnamefont {Zhao}}, \ and\
  \bibinfo {author} {\bibfnamefont {J.}~\bibnamefont {Li}},\ }\href
  {http://dx.doi.org/10.1021/cs500606g} {\bibfield  {journal} {\bibinfo
  {journal} {ACS Catalysis}\ }\textbf {\bibinfo {volume} {4}},\ \bibinfo
  {pages} {2917} (\bibinfo {year} {2014})}\BibitemShut {NoStop}%
\bibitem [{\citenamefont {Dogan}\ \emph {et~al.}(2015)\citenamefont {Dogan},
  \citenamefont {Lin}, \citenamefont {Guilloux-Viry},\ and\ \citenamefont
  {Peña}}]{applications}%
  \BibitemOpen
  \bibfield  {author} {\bibinfo {author} {\bibfnamefont {F.}~\bibnamefont
  {Dogan}}, \bibinfo {author} {\bibfnamefont {H.}~\bibnamefont {Lin}}, \bibinfo
  {author} {\bibfnamefont {M.}~\bibnamefont {Guilloux-Viry}}, \ and\ \bibinfo
  {author} {\bibfnamefont {O.}~\bibnamefont {Peña}},\ }\href
  {http://stacks.iop.org/1468-6996/16/i=2/a=020301} {\bibfield  {journal}
  {\bibinfo  {journal} {Sci. Technol. Adv. Mat.}\ }\textbf {\bibinfo {volume}
  {16}},\ \bibinfo {pages} {020301} (\bibinfo {year} {2015})}\BibitemShut
  {NoStop}%
\bibitem [{\citenamefont {Kim}\ \emph {et~al.}(2008)\citenamefont {Kim},
  \citenamefont {Jin}, \citenamefont {Moon}, \citenamefont {Kim}, \citenamefont
  {Park}, \citenamefont {Leem}, \citenamefont {Yu}, \citenamefont {Noh},
  \citenamefont {Kim}, \citenamefont {Oh}, \citenamefont {Park}, \citenamefont
  {Durairaj}, \citenamefont {Cao},\ and\ \citenamefont {Rotenberg}}]{diracsoc}%
  \BibitemOpen
  \bibfield  {author} {\bibinfo {author} {\bibfnamefont {B.~J.}\ \bibnamefont
  {Kim}}, \bibinfo {author} {\bibfnamefont {H.}~\bibnamefont {Jin}}, \bibinfo
  {author} {\bibfnamefont {S.~J.}\ \bibnamefont {Moon}}, \bibinfo {author}
  {\bibfnamefont {J.-Y.}\ \bibnamefont {Kim}}, \bibinfo {author} {\bibfnamefont
  {B.-G.}\ \bibnamefont {Park}}, \bibinfo {author} {\bibfnamefont {C.~S.}\
  \bibnamefont {Leem}}, \bibinfo {author} {\bibfnamefont {J.}~\bibnamefont
  {Yu}}, \bibinfo {author} {\bibfnamefont {T.~W.}\ \bibnamefont {Noh}},
  \bibinfo {author} {\bibfnamefont {C.}~\bibnamefont {Kim}}, \bibinfo {author}
  {\bibfnamefont {S.-J.}\ \bibnamefont {Oh}}, \bibinfo {author} {\bibfnamefont
  {J.-H.}\ \bibnamefont {Park}}, \bibinfo {author} {\bibfnamefont
  {V.}~\bibnamefont {Durairaj}}, \bibinfo {author} {\bibfnamefont
  {G.}~\bibnamefont {Cao}}, \ and\ \bibinfo {author} {\bibfnamefont
  {E.}~\bibnamefont {Rotenberg}},\ }\href {\doibase
  10.1103/PhysRevLett.101.076402} {\bibfield  {journal} {\bibinfo  {journal}
  {Phys. Rev. Lett.}\ }\textbf {\bibinfo {volume} {101}},\ \bibinfo {pages}
  {076402} (\bibinfo {year} {2008})}\BibitemShut {NoStop}%
\bibitem [{\citenamefont {Kim}\ \emph {et~al.}(2016)\citenamefont {Kim},
  \citenamefont {Liu}, \citenamefont {Erg\"onenc}, \citenamefont {Toschi},
  \citenamefont {Khmelevskyi},\ and\ \citenamefont {Franchini}}]{Lifshitz}%
  \BibitemOpen
  \bibfield  {author} {\bibinfo {author} {\bibfnamefont {B.}~\bibnamefont
  {Kim}}, \bibinfo {author} {\bibfnamefont {P.}~\bibnamefont {Liu}}, \bibinfo
  {author} {\bibfnamefont {Z.}~\bibnamefont {Erg\"onenc}}, \bibinfo {author}
  {\bibfnamefont {A.}~\bibnamefont {Toschi}}, \bibinfo {author} {\bibfnamefont
  {S.}~\bibnamefont {Khmelevskyi}}, \ and\ \bibinfo {author} {\bibfnamefont
  {C.}~\bibnamefont {Franchini}},\ }\href {\doibase 10.1103/PhysRevB.94.241113}
  {\bibfield  {journal} {\bibinfo  {journal} {Phys. Rev. B}\ }\textbf {\bibinfo
  {volume} {94}},\ \bibinfo {pages} {241113} (\bibinfo {year}
  {2016})}\BibitemShut {NoStop}%
\bibitem [{\citenamefont {Jackeli}\ and\ \citenamefont
  {Khaliullin}(2009)}]{Jackeli}%
  \BibitemOpen
  \bibfield  {author} {\bibinfo {author} {\bibfnamefont {G.}~\bibnamefont
  {Jackeli}}\ and\ \bibinfo {author} {\bibfnamefont {G.}~\bibnamefont
  {Khaliullin}},\ }\href {\doibase 10.1103/PhysRevLett.102.017205} {\bibfield
  {journal} {\bibinfo  {journal} {Phys. Rev. Lett.}\ }\textbf {\bibinfo
  {volume} {102}},\ \bibinfo {pages} {017205} (\bibinfo {year}
  {2009})}\BibitemShut {NoStop}%
\bibitem [{\citenamefont {Liu}\ \emph {et~al.}(2015)\citenamefont {Liu},
  \citenamefont {Khmelevskyi}, \citenamefont {Kim}, \citenamefont {Marsman},
  \citenamefont {Li}, \citenamefont {Chen}, \citenamefont {Sarma},
  \citenamefont {Kresse},\ and\ \citenamefont {Franchini}}]{DM}%
  \BibitemOpen
  \bibfield  {author} {\bibinfo {author} {\bibfnamefont {P.}~\bibnamefont
  {Liu}}, \bibinfo {author} {\bibfnamefont {S.}~\bibnamefont {Khmelevskyi}},
  \bibinfo {author} {\bibfnamefont {B.}~\bibnamefont {Kim}}, \bibinfo {author}
  {\bibfnamefont {M.}~\bibnamefont {Marsman}}, \bibinfo {author} {\bibfnamefont
  {D.}~\bibnamefont {Li}}, \bibinfo {author} {\bibfnamefont {X.-Q.}\
  \bibnamefont {Chen}}, \bibinfo {author} {\bibfnamefont {D.~D.}\ \bibnamefont
  {Sarma}}, \bibinfo {author} {\bibfnamefont {G.}~\bibnamefont {Kresse}}, \
  and\ \bibinfo {author} {\bibfnamefont {C.}~\bibnamefont {Franchini}},\ }\href
  {\doibase 10.1103/PhysRevB.92.054428} {\bibfield  {journal} {\bibinfo
  {journal} {Phys. Rev. B}\ }\textbf {\bibinfo {volume} {92}},\ \bibinfo
  {pages} {054428} (\bibinfo {year} {2015})}\BibitemShut {NoStop}%
\bibitem [{\citenamefont {Kim}\ \emph {et~al.}(2017{\natexlab{a}})\citenamefont
  {Kim}, \citenamefont {Khmelevskyi}, \citenamefont {Mazin}, \citenamefont
  {Agterberg},\ and\ \citenamefont {Franchini}}]{Kim2017ER}%
  \BibitemOpen
  \bibfield  {author} {\bibinfo {author} {\bibfnamefont {B.}~\bibnamefont
  {Kim}}, \bibinfo {author} {\bibfnamefont {S.}~\bibnamefont {Khmelevskyi}},
  \bibinfo {author} {\bibfnamefont {I.~I.}\ \bibnamefont {Mazin}}, \bibinfo
  {author} {\bibfnamefont {D.~F.}\ \bibnamefont {Agterberg}}, \ and\ \bibinfo
  {author} {\bibfnamefont {C.}~\bibnamefont {Franchini}},\ }\href {\doibase
  10.1038/s41535-017-0041-8} {\bibfield  {journal} {\bibinfo  {journal} {npj
  Quantum Materials}\ }\textbf {\bibinfo {volume} {2}},\ \bibinfo {pages} {37}
  (\bibinfo {year} {2017}{\natexlab{a}})}\BibitemShut {NoStop}%
\bibitem [{\citenamefont {Hohenberg}\ and\ \citenamefont {{W.
  Kohn}}(1964)}]{kohn1964}%
  \BibitemOpen
  \bibfield  {author} {\bibinfo {author} {\bibfnamefont {P.}~\bibnamefont
  {Hohenberg}}\ and\ \bibinfo {author} {\bibnamefont {{W. Kohn}}},\ }\href
  {\doibase 10.1103/PhysRevB.7.1912} {\bibfield  {journal} {\bibinfo  {journal}
  {Phys. Rev. B}\ }\textbf {\bibinfo {volume} {136}},\ \bibinfo {pages} {B864}
  (\bibinfo {year} {1964})}\BibitemShut {NoStop}%
\bibitem [{\citenamefont {{Van Schilfgaarde}}\ \emph
  {et~al.}(2006)\citenamefont {{Van Schilfgaarde}}, \citenamefont {Kotani},\
  and\ \citenamefont {Faleev}}]{VanSchilfgaarde2006b}%
  \BibitemOpen
  \bibfield  {author} {\bibinfo {author} {\bibfnamefont {M.}~\bibnamefont {{Van
  Schilfgaarde}}}, \bibinfo {author} {\bibfnamefont {T.}~\bibnamefont
  {Kotani}}, \ and\ \bibinfo {author} {\bibfnamefont {S.}~\bibnamefont
  {Faleev}},\ }\href@noop {} {\bibfield  {journal} {\bibinfo  {journal} {Phys.
  Rev. Lett.}\ } (\bibinfo {year} {2006})}\BibitemShut {NoStop}%
\bibitem [{\citenamefont {Hedin}(1965)}]{Hedin}%
  \BibitemOpen
  \bibfield  {author} {\bibinfo {author} {\bibfnamefont {L.}~\bibnamefont
  {Hedin}},\ }\href@noop {} {\bibfield  {journal} {\bibinfo  {journal} {Phys.
  Rev.}\ }\textbf {\bibinfo {volume} {3A}},\ \bibinfo {pages} {A796} (\bibinfo
  {year} {1965})}\BibitemShut {NoStop}%
\bibitem [{\citenamefont {Strinati}\ \emph {et~al.}(1980)\citenamefont
  {Strinati}, \citenamefont {Mattausch},\ and\ \citenamefont
  {Hanke}}]{PhysRevLett.45.290}%
  \BibitemOpen
  \bibfield  {author} {\bibinfo {author} {\bibfnamefont {G.}~\bibnamefont
  {Strinati}}, \bibinfo {author} {\bibfnamefont {H.~J.}\ \bibnamefont
  {Mattausch}}, \ and\ \bibinfo {author} {\bibfnamefont {W.}~\bibnamefont
  {Hanke}},\ }\href {\doibase 10.1103/PhysRevLett.45.290} {\bibfield  {journal}
  {\bibinfo  {journal} {Phys. Rev. Lett.}\ }\textbf {\bibinfo {volume} {45}},\
  \bibinfo {pages} {290} (\bibinfo {year} {1980})}\BibitemShut {NoStop}%
\bibitem [{\citenamefont {Strinati}\ \emph {et~al.}(1982)\citenamefont
  {Strinati}, \citenamefont {Mattausch},\ and\ \citenamefont
  {Hanke}}]{PhysRevB.25.2867}%
  \BibitemOpen
  \bibfield  {author} {\bibinfo {author} {\bibfnamefont {G.}~\bibnamefont
  {Strinati}}, \bibinfo {author} {\bibfnamefont {H.~J.}\ \bibnamefont
  {Mattausch}}, \ and\ \bibinfo {author} {\bibfnamefont {W.}~\bibnamefont
  {Hanke}},\ }\href {\doibase 10.1103/PhysRevB.25.2867} {\bibfield  {journal}
  {\bibinfo  {journal} {Phys. Rev. B}\ }\textbf {\bibinfo {volume} {25}},\
  \bibinfo {pages} {2867} (\bibinfo {year} {1982})}\BibitemShut {NoStop}%
\bibitem [{\citenamefont {Hybertsen}\ and\ \citenamefont
  {Louie}(1985)}]{Hybertsen1985}%
  \BibitemOpen
  \bibfield  {author} {\bibinfo {author} {\bibfnamefont {M.~S.}\ \bibnamefont
  {Hybertsen}}\ and\ \bibinfo {author} {\bibfnamefont {S.~G.}\ \bibnamefont
  {Louie}},\ }\href@noop {} {\bibfield  {journal} {\bibinfo  {journal} {Phys.
  Rev. Lett.}\ }\textbf {\bibinfo {volume} {55}},\ \bibinfo {pages} {1418}
  (\bibinfo {year} {1985})}\BibitemShut {NoStop}%
\bibitem [{\citenamefont {Hybertsen}\ and\ \citenamefont
  {Louie}(1986)}]{Hybertsen1986}%
  \BibitemOpen
  \bibfield  {author} {\bibinfo {author} {\bibfnamefont {M.~S.}\ \bibnamefont
  {Hybertsen}}\ and\ \bibinfo {author} {\bibfnamefont {S.~G.}\ \bibnamefont
  {Louie}},\ }\href@noop {} {\bibfield  {journal} {\bibinfo  {journal} {Phys.
  Rev. B}\ }\textbf {\bibinfo {volume} {34}},\ \bibinfo {pages} {5390}
  (\bibinfo {year} {1986})}\BibitemShut {NoStop}%
\bibitem [{\citenamefont {Aulbur}\ \emph {et~al.}(1999)\citenamefont {Aulbur},
  \citenamefont {Jonsson},\ and\ \citenamefont {Wilkins}}]{aulbur}%
  \BibitemOpen
  \bibfield  {author} {\bibinfo {author} {\bibfnamefont {W.~G.}\ \bibnamefont
  {Aulbur}}, \bibinfo {author} {\bibfnamefont {L.}~\bibnamefont {Jonsson}}, \
  and\ \bibinfo {author} {\bibfnamefont {J.~W.}\ \bibnamefont {Wilkins}},\
  }\href {\doibase 10.1016/S0081-1947(08)60248-9} {\bibfield  {journal}
  {\bibinfo  {journal} {Solid State Physics - Advances in Research and
  Applications}\ }\textbf {\bibinfo {volume} {54}},\ \bibinfo {pages} {1}
  (\bibinfo {year} {1999})}\BibitemShut {NoStop}%
\bibitem [{\citenamefont {Shishkin}\ and\ \citenamefont
  {Kresse}(2007)}]{GWKresse}%
  \BibitemOpen
  \bibfield  {author} {\bibinfo {author} {\bibfnamefont {M.}~\bibnamefont
  {Shishkin}}\ and\ \bibinfo {author} {\bibfnamefont {G.}~\bibnamefont
  {Kresse}},\ }\href {\doibase 10.1103/PhysRevB.75.235102} {\bibfield
  {journal} {\bibinfo  {journal} {Phys. Rev. B}\ }\textbf {\bibinfo {volume}
  {75}},\ \bibinfo {pages} {235102} (\bibinfo {year} {2007})}\BibitemShut
  {NoStop}%
\bibitem [{\citenamefont {Fuchs}\ \emph {et~al.}(2007)\citenamefont {Fuchs},
  \citenamefont {Furthm\"uller}, \citenamefont {Bechstedt}, \citenamefont
  {Shishkin},\ and\ \citenamefont {Kresse}}]{GWGK1}%
  \BibitemOpen
  \bibfield  {author} {\bibinfo {author} {\bibfnamefont {F.}~\bibnamefont
  {Fuchs}}, \bibinfo {author} {\bibfnamefont {J.}~\bibnamefont
  {Furthm\"uller}}, \bibinfo {author} {\bibfnamefont {F.}~\bibnamefont
  {Bechstedt}}, \bibinfo {author} {\bibfnamefont {M.}~\bibnamefont {Shishkin}},
  \ and\ \bibinfo {author} {\bibfnamefont {G.}~\bibnamefont {Kresse}},\ }\href
  {\doibase 10.1103/PhysRevB.76.115109} {\bibfield  {journal} {\bibinfo
  {journal} {Phys. Rev. B}\ }\textbf {\bibinfo {volume} {76}},\ \bibinfo
  {pages} {115109} (\bibinfo {year} {2007})}\BibitemShut {NoStop}%
\bibitem [{\citenamefont {Paier}\ \emph
  {et~al.}(2008{\natexlab{a}})\citenamefont {Paier}, \citenamefont {Marsman},\
  and\ \citenamefont {Kresse}}]{GWGK2}%
  \BibitemOpen
  \bibfield  {author} {\bibinfo {author} {\bibfnamefont {J.}~\bibnamefont
  {Paier}}, \bibinfo {author} {\bibfnamefont {M.}~\bibnamefont {Marsman}}, \
  and\ \bibinfo {author} {\bibfnamefont {G.}~\bibnamefont {Kresse}},\ }\href
  {\doibase 10.1103/PhysRevB.78.121201} {\bibfield  {journal} {\bibinfo
  {journal} {Phys. Rev. B}\ }\textbf {\bibinfo {volume} {78}},\ \bibinfo
  {pages} {121201} (\bibinfo {year} {2008}{\natexlab{a}})}\BibitemShut
  {NoStop}%
\bibitem [{\citenamefont {Shih}\ \emph {et~al.}(2010)\citenamefont {Shih},
  \citenamefont {Xue}, \citenamefont {Zhang}, \citenamefont {Cohen},\ and\
  \citenamefont {Louie}}]{Shih2010}%
  \BibitemOpen
  \bibfield  {author} {\bibinfo {author} {\bibfnamefont {B.-C.}\ \bibnamefont
  {Shih}}, \bibinfo {author} {\bibfnamefont {Y.}~\bibnamefont {Xue}}, \bibinfo
  {author} {\bibfnamefont {P.}~\bibnamefont {Zhang}}, \bibinfo {author}
  {\bibfnamefont {M.~L.}\ \bibnamefont {Cohen}}, \ and\ \bibinfo {author}
  {\bibfnamefont {S.~G.}\ \bibnamefont {Louie}},\ }\href {\doibase
  10.1103/PhysRevLett.105.146401} {\bibfield  {journal} {\bibinfo  {journal}
  {Phys. Rev. Lett.}\ }\textbf {\bibinfo {volume} {105}},\ \bibinfo {pages}
  {146401} (\bibinfo {year} {2010})}\BibitemShut {NoStop}%
\bibitem [{\citenamefont {Klime{\v{s}}}\ \emph {et~al.}(2014)\citenamefont
  {Klime{\v{s}}}, \citenamefont {Kaltak},\ and\ \citenamefont
  {Kresse}}]{Klimes2014}%
  \BibitemOpen
  \bibfield  {author} {\bibinfo {author} {\bibfnamefont {J.}~\bibnamefont
  {Klime{\v{s}}}}, \bibinfo {author} {\bibfnamefont {M.}~\bibnamefont
  {Kaltak}}, \ and\ \bibinfo {author} {\bibfnamefont {G.}~\bibnamefont
  {Kresse}},\ }\href {\doibase 10.1103/PhysRevB.90.075125} {\bibfield
  {journal} {\bibinfo  {journal} {Phys. Rev. B}\ }\textbf {\bibinfo {volume}
  {90}},\ \bibinfo {pages} {075125} (\bibinfo {year} {2014})}\BibitemShut
  {NoStop}%
\bibitem [{\citenamefont {Harl}\ and\ \citenamefont {Kresse}(2008)}]{Harl2008}%
  \BibitemOpen
  \bibfield  {author} {\bibinfo {author} {\bibfnamefont {J.}~\bibnamefont
  {Harl}}\ and\ \bibinfo {author} {\bibfnamefont {G.}~\bibnamefont {Kresse}},\
  }\href {\doibase 10.1103/PhysRevB.77.045136} {\bibfield  {journal} {\bibinfo
  {journal} {Phys. Rev. B}\ }\textbf {\bibinfo {volume} {77}},\ \bibinfo
  {pages} {045136} (\bibinfo {year} {2008})}\BibitemShut {NoStop}%
\bibitem [{\citenamefont {Bj\"orkman}\ \emph {et~al.}(2012)\citenamefont
  {Bj\"orkman}, \citenamefont {Gulans}, \citenamefont {Krasheninnikov},\ and\
  \citenamefont {Nieminen}}]{Bjorkman2012}%
  \BibitemOpen
  \bibfield  {author} {\bibinfo {author} {\bibfnamefont {T.}~\bibnamefont
  {Bj\"orkman}}, \bibinfo {author} {\bibfnamefont {A.}~\bibnamefont {Gulans}},
  \bibinfo {author} {\bibfnamefont {A.~V.}\ \bibnamefont {Krasheninnikov}}, \
  and\ \bibinfo {author} {\bibfnamefont {R.~M.}\ \bibnamefont {Nieminen}},\
  }\href {\doibase 10.1103/PhysRevLett.108.235502} {\bibfield  {journal}
  {\bibinfo  {journal} {Phys. Rev. Lett.}\ }\textbf {\bibinfo {volume} {108}},\
  \bibinfo {pages} {235502} (\bibinfo {year} {2012})}\BibitemShut {NoStop}%
\bibitem [{\citenamefont {Schindlmayr}(2013)}]{PhysRevB.87.075104}%
  \BibitemOpen
  \bibfield  {author} {\bibinfo {author} {\bibfnamefont {A.}~\bibnamefont
  {Schindlmayr}},\ }\href {\doibase 10.1103/PhysRevB.87.075104} {\bibfield
  {journal} {\bibinfo  {journal} {Phys. Rev. B}\ }\textbf {\bibinfo {volume}
  {87}},\ \bibinfo {pages} {075104} (\bibinfo {year} {2013})}\BibitemShut
  {NoStop}%
\bibitem [{\citenamefont {Liu}\ \emph {et~al.}(2016)\citenamefont {Liu},
  \citenamefont {Kaltak}, \citenamefont {Klime\v{s}},\ and\ \citenamefont
  {Kresse}}]{PhysRevB.94.165109}%
  \BibitemOpen
  \bibfield  {author} {\bibinfo {author} {\bibfnamefont {P.}~\bibnamefont
  {Liu}}, \bibinfo {author} {\bibfnamefont {M.}~\bibnamefont {Kaltak}},
  \bibinfo {author} {\bibfnamefont {J.}~\bibnamefont {Klime\v{s}}}, \ and\
  \bibinfo {author} {\bibfnamefont {G.}~\bibnamefont {Kresse}},\ }\href
  {\doibase 10.1103/PhysRevB.94.165109} {\bibfield  {journal} {\bibinfo
  {journal} {Phys. Rev. B}\ }\textbf {\bibinfo {volume} {94}},\ \bibinfo
  {pages} {165109} (\bibinfo {year} {2016})}\BibitemShut {NoStop}%
\bibitem [{\citenamefont {Kim}\ \emph {et~al.}(2017{\natexlab{b}})\citenamefont
  {Kim}, \citenamefont {Martyna},\ and\ \citenamefont
  {Ismail-Beigi}}]{arXiv:1707.06752}%
  \BibitemOpen
  \bibfield  {author} {\bibinfo {author} {\bibfnamefont {M.}~\bibnamefont
  {Kim}}, \bibinfo {author} {\bibfnamefont {G.~J.}\ \bibnamefont {Martyna}}, \
  and\ \bibinfo {author} {\bibfnamefont {S.}~\bibnamefont {Ismail-Beigi}},\
  }\href@noop {} {\bibfield  {journal} {\bibinfo  {journal} {arXiv:1707.06752}\
  } (\bibinfo {year} {2017}{\natexlab{b}})}\BibitemShut {NoStop}%
\bibitem [{\citenamefont {Umari}\ \emph {et~al.}(2015)\citenamefont {Umari},
  \citenamefont {Mosconi},\ and\ \citenamefont {{De Angelis}}}]{Umari2014}%
  \BibitemOpen
  \bibfield  {author} {\bibinfo {author} {\bibfnamefont {P.}~\bibnamefont
  {Umari}}, \bibinfo {author} {\bibfnamefont {E.}~\bibnamefont {Mosconi}}, \
  and\ \bibinfo {author} {\bibfnamefont {F.}~\bibnamefont {{De Angelis}}},\
  }\href {\doibase 10.1038/srep04467} {\bibfield  {journal} {\bibinfo
  {journal} {Sci. Rep.}\ }\textbf {\bibinfo {volume} {4}},\ \bibinfo {pages}
  {4467} (\bibinfo {year} {2015})},\ \Eprint
  {http://arxiv.org/abs/arXiv:1402.1923v1} {arXiv:arXiv:1402.1923v1}
  \BibitemShut {NoStop}%
\bibitem [{\citenamefont {Filip}\ and\ \citenamefont
  {Giustino}(2014)}]{PhysRevB.90.245145}%
  \BibitemOpen
  \bibfield  {author} {\bibinfo {author} {\bibfnamefont {M.~R.}\ \bibnamefont
  {Filip}}\ and\ \bibinfo {author} {\bibfnamefont {F.}~\bibnamefont
  {Giustino}},\ }\href {\doibase 10.1103/PhysRevB.90.245145} {\bibfield
  {journal} {\bibinfo  {journal} {Phys. Rev. B}\ }\textbf {\bibinfo {volume}
  {90}},\ \bibinfo {pages} {245145} (\bibinfo {year} {2014})}\BibitemShut
  {NoStop}%
\bibitem [{\citenamefont {Castelli}\ \emph {et~al.}(2014)\citenamefont
  {Castelli}, \citenamefont {García-Lastra}, \citenamefont {Thygesen},\ and\
  \citenamefont {Jacobsen}}]{IvanoAPL2014}%
  \BibitemOpen
  \bibfield  {author} {\bibinfo {author} {\bibfnamefont {I.~E.}\ \bibnamefont
  {Castelli}}, \bibinfo {author} {\bibfnamefont {J.~M.}\ \bibnamefont
  {García-Lastra}}, \bibinfo {author} {\bibfnamefont {K.~S.}\ \bibnamefont
  {Thygesen}}, \ and\ \bibinfo {author} {\bibfnamefont {K.~W.}\ \bibnamefont
  {Jacobsen}},\ }\href {http://dx.doi.org/10.1063/1.4893495} {\bibfield
  {journal} {\bibinfo  {journal} {APL Materials}\ }\textbf {\bibinfo {volume}
  {2}},\ \bibinfo {pages} {081514} (\bibinfo {year} {2014})}\BibitemShut
  {NoStop}%
\bibitem [{\citenamefont {Bokdam}\ \emph {et~al.}(2016)\citenamefont {Bokdam},
  \citenamefont {Sander}, \citenamefont {Stroppa}, \citenamefont {Picozzi},
  \citenamefont {Sarma}, \citenamefont {Franchini},\ and\ \citenamefont
  {Kresse}}]{Bokdam2016}%
  \BibitemOpen
  \bibfield  {author} {\bibinfo {author} {\bibfnamefont {M.}~\bibnamefont
  {Bokdam}}, \bibinfo {author} {\bibfnamefont {T.}~\bibnamefont {Sander}},
  \bibinfo {author} {\bibfnamefont {A.}~\bibnamefont {Stroppa}}, \bibinfo
  {author} {\bibfnamefont {S.}~\bibnamefont {Picozzi}}, \bibinfo {author}
  {\bibfnamefont {D.~D.}\ \bibnamefont {Sarma}}, \bibinfo {author}
  {\bibfnamefont {C.}~\bibnamefont {Franchini}}, \ and\ \bibinfo {author}
  {\bibfnamefont {G.}~\bibnamefont {Kresse}},\ }\href
  {http://dx.doi.org/10.1038/srep28618} {\ \textbf {\bibinfo {volume} {6}},\
  \bibinfo {pages} {28618 EP } (\bibinfo {year} {2016})},\ \bibinfo {note}
  {article}\BibitemShut {NoStop}%
\bibitem [{\citenamefont {Friedrich}\ \emph {et~al.}(2010)\citenamefont
  {Friedrich}, \citenamefont {Bl{\"{u}}gel},\ and\ \citenamefont
  {Schindlmayr}}]{Friedrich2010}%
  \BibitemOpen
  \bibfield  {author} {\bibinfo {author} {\bibfnamefont {C.}~\bibnamefont
  {Friedrich}}, \bibinfo {author} {\bibfnamefont {S.}~\bibnamefont
  {Bl{\"{u}}gel}}, \ and\ \bibinfo {author} {\bibfnamefont {A.}~\bibnamefont
  {Schindlmayr}},\ }\href {\doibase 10.1103/PhysRevB.81.125102} {\bibfield
  {journal} {\bibinfo  {journal} {Phys. Rev. B Condens. Matter Mater. Phys.}\
  }\textbf {\bibinfo {volume} {81}},\ \bibinfo {pages} {1} (\bibinfo {year}
  {2010})}\BibitemShut {NoStop}%
\bibitem [{\citenamefont {Nohara}\ \emph {et~al.}(2009)\citenamefont {Nohara},
  \citenamefont {Yamamoto},\ and\ \citenamefont {Fujiwara}}]{Nohara2009}%
  \BibitemOpen
  \bibfield  {author} {\bibinfo {author} {\bibfnamefont {Y.}~\bibnamefont
  {Nohara}}, \bibinfo {author} {\bibfnamefont {S.}~\bibnamefont {Yamamoto}}, \
  and\ \bibinfo {author} {\bibfnamefont {T.}~\bibnamefont {Fujiwara}},\ }\href
  {\doibase 10.1103/PhysRevB.79.195110} {\bibfield  {journal} {\bibinfo
  {journal} {Phys. Rev. B}\ }\textbf {\bibinfo {volume} {79}},\ \bibinfo
  {pages} {195110} (\bibinfo {year} {2009})}\BibitemShut {NoStop}%
\bibitem [{\citenamefont {Franchini}\ \emph {et~al.}(2010)\citenamefont
  {Franchini}, \citenamefont {Sanna}, \citenamefont {Marsman},\ and\
  \citenamefont {Kresse}}]{BBO}%
  \BibitemOpen
  \bibfield  {author} {\bibinfo {author} {\bibfnamefont {C.}~\bibnamefont
  {Franchini}}, \bibinfo {author} {\bibfnamefont {A.}~\bibnamefont {Sanna}},
  \bibinfo {author} {\bibfnamefont {M.}~\bibnamefont {Marsman}}, \ and\
  \bibinfo {author} {\bibfnamefont {G.}~\bibnamefont {Kresse}},\ }\href
  {\doibase 10.1103/PhysRevB.81.085213} {\bibfield  {journal} {\bibinfo
  {journal} {Phys. Rev. B}\ }\textbf {\bibinfo {volume} {81}},\ \bibinfo
  {pages} {085213} (\bibinfo {year} {2010})}\BibitemShut {NoStop}%
\bibitem [{\citenamefont {Franchini}\ \emph {et~al.}(2012)\citenamefont
  {Franchini}, \citenamefont {Kovacik}, \citenamefont {Marsman}, \citenamefont
  {Murthy}, \citenamefont {He}, \citenamefont {Ederer},\ and\ \citenamefont
  {Kresse}}]{CF1}%
  \BibitemOpen
  \bibfield  {author} {\bibinfo {author} {\bibfnamefont {C.}~\bibnamefont
  {Franchini}}, \bibinfo {author} {\bibfnamefont {R.}~\bibnamefont {Kovacik}},
  \bibinfo {author} {\bibfnamefont {M.}~\bibnamefont {Marsman}}, \bibinfo
  {author} {\bibfnamefont {S.~S.}\ \bibnamefont {Murthy}}, \bibinfo {author}
  {\bibfnamefont {J.}~\bibnamefont {He}}, \bibinfo {author} {\bibfnamefont
  {C.}~\bibnamefont {Ederer}}, \ and\ \bibinfo {author} {\bibfnamefont
  {G.}~\bibnamefont {Kresse}},\ }\href
  {http://stacks.iop.org/0953-8984/24/i=23/a=235602} {\bibfield  {journal}
  {\bibinfo  {journal} {‎J. Phys. Condens. Matter}\ }\textbf {\bibinfo
  {volume} {24}},\ \bibinfo {pages} {235602} (\bibinfo {year}
  {2012})}\BibitemShut {NoStop}%
\bibitem [{\citenamefont {Kang}\ \emph {et~al.}(2015)\citenamefont {Kang},
  \citenamefont {Kang},\ and\ \citenamefont {Han}}]{Kang2015a}%
  \BibitemOpen
  \bibfield  {author} {\bibinfo {author} {\bibfnamefont {G.}~\bibnamefont
  {Kang}}, \bibinfo {author} {\bibfnamefont {Y.}~\bibnamefont {Kang}}, \ and\
  \bibinfo {author} {\bibfnamefont {S.}~\bibnamefont {Han}},\ }\href@noop {}
  {\bibfield  {journal} {\bibinfo  {journal} {Phys. Rev. B Condens. Matter
  Mater. Phys.}\ }\textbf {\bibinfo {volume} {91}},\ \bibinfo {pages} {155141}
  (\bibinfo {year} {2015})}\BibitemShut {NoStop}%
\bibitem [{\citenamefont {Kov\'a\ifmmode~\check{c}\else \v{c}\fi{}ik}\ \emph
  {et~al.}(2016)\citenamefont {Kov\'a\ifmmode~\check{c}\else \v{c}\fi{}ik},
  \citenamefont {Murthy}, \citenamefont {Quiroga}, \citenamefont {Ederer},\
  and\ \citenamefont {Franchini}}]{CF2}%
  \BibitemOpen
  \bibfield  {author} {\bibinfo {author} {\bibfnamefont {R.}~\bibnamefont
  {Kov\'a\ifmmode~\check{c}\else \v{c}\fi{}ik}}, \bibinfo {author}
  {\bibfnamefont {S.~S.}\ \bibnamefont {Murthy}}, \bibinfo {author}
  {\bibfnamefont {C.~E.}\ \bibnamefont {Quiroga}}, \bibinfo {author}
  {\bibfnamefont {C.}~\bibnamefont {Ederer}}, \ and\ \bibinfo {author}
  {\bibfnamefont {C.}~\bibnamefont {Franchini}},\ }\href {\doibase
  10.1103/PhysRevB.93.075139} {\bibfield  {journal} {\bibinfo  {journal} {Phys.
  Rev. B}\ }\textbf {\bibinfo {volume} {93}},\ \bibinfo {pages} {075139}
  (\bibinfo {year} {2016})}\BibitemShut {NoStop}%
\bibitem [{\citenamefont {Lany}(2013)}]{Lany}%
  \BibitemOpen
  \bibfield  {author} {\bibinfo {author} {\bibfnamefont {S.}~\bibnamefont
  {Lany}},\ }\href {\doibase 10.1103/PhysRevB.87.085112} {\bibfield  {journal}
  {\bibinfo  {journal} {Phys. Rev. B}\ }\textbf {\bibinfo {volume} {87}},\
  \bibinfo {pages} {085112} (\bibinfo {year} {2013})}\BibitemShut {NoStop}%
\bibitem [{\citenamefont {Sousa}\ \emph {et~al.}(2007)\citenamefont {Sousa},
  \citenamefont {Rossel}, \citenamefont {Marchiori}, \citenamefont {Siegwart},
  \citenamefont {Caimi}, \citenamefont {Locquet}, \citenamefont {Webb},
  \citenamefont {Germann}, \citenamefont {Fompeyrine}, \citenamefont {Babich},
  \citenamefont {Seo},\ and\ \citenamefont {Dieker}}]{Sousa2007}%
  \BibitemOpen
  \bibfield  {author} {\bibinfo {author} {\bibfnamefont {M.}~\bibnamefont
  {Sousa}}, \bibinfo {author} {\bibfnamefont {C.}~\bibnamefont {Rossel}},
  \bibinfo {author} {\bibfnamefont {C.}~\bibnamefont {Marchiori}}, \bibinfo
  {author} {\bibfnamefont {H.}~\bibnamefont {Siegwart}}, \bibinfo {author}
  {\bibfnamefont {D.}~\bibnamefont {Caimi}}, \bibinfo {author} {\bibfnamefont
  {J.-P.}\ \bibnamefont {Locquet}}, \bibinfo {author} {\bibfnamefont {D.~J.}\
  \bibnamefont {Webb}}, \bibinfo {author} {\bibfnamefont {R.}~\bibnamefont
  {Germann}}, \bibinfo {author} {\bibfnamefont {J.}~\bibnamefont {Fompeyrine}},
  \bibinfo {author} {\bibfnamefont {K.}~\bibnamefont {Babich}}, \bibinfo
  {author} {\bibfnamefont {J.~W.}\ \bibnamefont {Seo}}, \ and\ \bibinfo
  {author} {\bibfnamefont {C.}~\bibnamefont {Dieker}},\ }\href {\doibase
  10.1063/1.2812425} {\bibfield  {journal} {\bibinfo  {journal} {J. Appl.
  Phys.}\ }\textbf {\bibinfo {volume} {102}},\ \bibinfo {pages} {104103}
  (\bibinfo {year} {2007})}\BibitemShut {NoStop}%
\bibitem [{\citenamefont {Ryee}\ \emph {et~al.}(2016)\citenamefont {Ryee},
  \citenamefont {Jang}, \citenamefont {Kino}, \citenamefont {Kotani},\ and\
  \citenamefont {Han}}]{PhysRevB.93.075125}%
  \BibitemOpen
  \bibfield  {author} {\bibinfo {author} {\bibfnamefont {S.}~\bibnamefont
  {Ryee}}, \bibinfo {author} {\bibfnamefont {S.~W.}\ \bibnamefont {Jang}},
  \bibinfo {author} {\bibfnamefont {H.}~\bibnamefont {Kino}}, \bibinfo {author}
  {\bibfnamefont {T.}~\bibnamefont {Kotani}}, \ and\ \bibinfo {author}
  {\bibfnamefont {M.~J.}\ \bibnamefont {Han}},\ }\href {\doibase
  10.1103/PhysRevB.93.075125} {\bibfield  {journal} {\bibinfo  {journal} {Phys.
  Rev. B}\ }\textbf {\bibinfo {volume} {93}},\ \bibinfo {pages} {075125}
  (\bibinfo {year} {2016})}\BibitemShut {NoStop}%
\bibitem [{\citenamefont {He}\ and\ \citenamefont
  {Franchini}(2014{\natexlab{a}})}]{He2014}%
  \BibitemOpen
  \bibfield  {author} {\bibinfo {author} {\bibfnamefont {J.}~\bibnamefont
  {He}}\ and\ \bibinfo {author} {\bibfnamefont {C.}~\bibnamefont {Franchini}},\
  }\href {\doibase 10.1103/PhysRevB.89.045104} {\bibfield  {journal} {\bibinfo
  {journal} {Phys. Rev. B}\ }\textbf {\bibinfo {volume} {89}},\ \bibinfo
  {pages} {045104} (\bibinfo {year} {2014}{\natexlab{a}})}\BibitemShut
  {NoStop}%
\bibitem [{\citenamefont {van Setten}\ \emph {et~al.}(2017)\citenamefont {van
  Setten}, \citenamefont {Giantomassi}, \citenamefont {Gonze}, \citenamefont
  {Rignanese},\ and\ \citenamefont {Hautier}}]{Setten}%
  \BibitemOpen
  \bibfield  {author} {\bibinfo {author} {\bibfnamefont {M.~J.}\ \bibnamefont
  {van Setten}}, \bibinfo {author} {\bibfnamefont {M.}~\bibnamefont
  {Giantomassi}}, \bibinfo {author} {\bibfnamefont {X.}~\bibnamefont {Gonze}},
  \bibinfo {author} {\bibfnamefont {G.-M.}\ \bibnamefont {Rignanese}}, \ and\
  \bibinfo {author} {\bibfnamefont {G.}~\bibnamefont {Hautier}},\ }\href
  {\doibase 10.1103/PhysRevB.96.155207} {\bibfield  {journal} {\bibinfo
  {journal} {Phys. Rev. B}\ }\textbf {\bibinfo {volume} {96}},\ \bibinfo
  {pages} {155207} (\bibinfo {year} {2017})}\BibitemShut {NoStop}%
\bibitem [{\citenamefont {Gulans}(2014)}]{Gulans}%
  \BibitemOpen
  \bibfield  {author} {\bibinfo {author} {\bibfnamefont {A.}~\bibnamefont
  {Gulans}},\ }\href {\doibase 10.1063/1.4900447} {\bibfield  {journal}
  {\bibinfo  {journal} {The Journal of Chemical Physics}\ }\textbf {\bibinfo
  {volume} {141}},\ \bibinfo {pages} {164127} (\bibinfo {year} {2014})},\
  \Eprint {http://arxiv.org/abs/https://doi.org/10.1063/1.4900447}
  {https://doi.org/10.1063/1.4900447} \BibitemShut {NoStop}%
\bibitem [{\citenamefont {He}\ and\ \citenamefont
  {Franchini}(2012)}]{PhysRevB.86.235117}%
  \BibitemOpen
  \bibfield  {author} {\bibinfo {author} {\bibfnamefont {J.}~\bibnamefont
  {He}}\ and\ \bibinfo {author} {\bibfnamefont {C.}~\bibnamefont {Franchini}},\
  }\href {\doibase 10.1103/PhysRevB.86.235117} {\bibfield  {journal} {\bibinfo
  {journal} {Phys. Rev. B}\ }\textbf {\bibinfo {volume} {86}},\ \bibinfo
  {pages} {235117} (\bibinfo {year} {2012})}\BibitemShut {NoStop}%
\bibitem [{\citenamefont {He}\ and\ \citenamefont
  {Franchini}(2014{\natexlab{b}})}]{PhysRevB.90.039907}%
  \BibitemOpen
  \bibfield  {author} {\bibinfo {author} {\bibfnamefont {J.}~\bibnamefont
  {He}}\ and\ \bibinfo {author} {\bibfnamefont {C.}~\bibnamefont {Franchini}},\
  }\href {\doibase 10.1103/PhysRevB.90.039907} {\bibfield  {journal} {\bibinfo
  {journal} {Phys. Rev. B}\ }\textbf {\bibinfo {volume} {90}},\ \bibinfo
  {pages} {039907} (\bibinfo {year} {2014}{\natexlab{b}})}\BibitemShut
  {NoStop}%
\bibitem [{\citenamefont {Swanson}\ \emph {et~al.}(1954)\citenamefont
  {Swanson}, \citenamefont {Ugrinic}, \citenamefont {Fuyat},\ and\
  \citenamefont {States}}]{Swanson1954}%
  \BibitemOpen
  \bibfield  {author} {\bibinfo {author} {\bibfnamefont {H.~E.}\ \bibnamefont
  {Swanson}}, \bibinfo {author} {\bibfnamefont {G.~M.}\ \bibnamefont
  {Ugrinic}}, \bibinfo {author} {\bibfnamefont {R.~K.}\ \bibnamefont {Fuyat}},
  \ and\ \bibinfo {author} {\bibfnamefont {U.}~\bibnamefont {States}},\ }\href
  {file://catalog.hathitrust.org/Record/007291113
  http://hdl.handle.net/2027/mdp.39015077578519} {\ \bibinfo {series} {National
  Bureau of Standards circular539, v. 3},\ \bibinfo {pages} {II, 73 p.}
  (\bibinfo {year} {1954})}\BibitemShut {NoStop}%
\bibitem [{\citenamefont {Smith}\ and\ \citenamefont
  {Welch}(1960)}]{Smith1960}%
  \BibitemOpen
  \bibfield  {author} {\bibinfo {author} {\bibfnamefont {A.~J.}\ \bibnamefont
  {Smith}}\ and\ \bibinfo {author} {\bibfnamefont {A.~J.~E.}\ \bibnamefont
  {Welch}},\ }\href {\doibase 10.1107/S0365110X60001540} {\bibfield  {journal}
  {\bibinfo  {journal} {Acta Crystallogr.}\ }\textbf {\bibinfo {volume} {13}},\
  \bibinfo {pages} {653} (\bibinfo {year} {1960})}\BibitemShut {NoStop}%
\bibitem [{\citenamefont {Kennedy}\ \emph {et~al.}(1999)\citenamefont
  {Kennedy}, \citenamefont {Howard},\ and\ \citenamefont
  {Chakoumakos}}]{Kennedy1999}%
  \BibitemOpen
  \bibfield  {author} {\bibinfo {author} {\bibfnamefont {B.~J.}\ \bibnamefont
  {Kennedy}}, \bibinfo {author} {\bibfnamefont {C.~J.}\ \bibnamefont {Howard}},
  \ and\ \bibinfo {author} {\bibfnamefont {B.~C.}\ \bibnamefont
  {Chakoumakos}},\ }\href {\doibase 10.1103/PhysRevB.60.2972} {\bibfield
  {journal} {\bibinfo  {journal} {Phys. Rev. B}\ }\textbf {\bibinfo {volume}
  {60}},\ \bibinfo {pages} {2972} (\bibinfo {year} {1999})}\BibitemShut
  {NoStop}%
\bibitem [{\citenamefont {Sigman}\ \emph {et~al.}(2002)\citenamefont {Sigman},
  \citenamefont {Norton}, \citenamefont {Christen}, \citenamefont {Fleming},\
  and\ \citenamefont {Boatner}}]{Sigman2002}%
  \BibitemOpen
  \bibfield  {author} {\bibinfo {author} {\bibfnamefont {J.}~\bibnamefont
  {Sigman}}, \bibinfo {author} {\bibfnamefont {D.~P.}\ \bibnamefont {Norton}},
  \bibinfo {author} {\bibfnamefont {H.~M.}\ \bibnamefont {Christen}}, \bibinfo
  {author} {\bibfnamefont {P.~H.}\ \bibnamefont {Fleming}}, \ and\ \bibinfo
  {author} {\bibfnamefont {L.~A.}\ \bibnamefont {Boatner}},\ }\href {\doibase
  10.1103/PhysRevLett.88.097601} {\bibfield  {journal} {\bibinfo  {journal}
  {Phys. Rev. Lett.}\ }\textbf {\bibinfo {volume} {88}},\ \bibinfo {pages}
  {97601} (\bibinfo {year} {2002})}\BibitemShut {NoStop}%
\bibitem [{\citenamefont {S\o{}nden\aa{}}\ \emph {et~al.}(2006)\citenamefont
  {S\o{}nden\aa{}}, \citenamefont {Ravindran}, \citenamefont {St\o{}len},
  \citenamefont {Grande},\ and\ \citenamefont {Hanfland}}]{PhysRevB.74.144102}%
  \BibitemOpen
  \bibfield  {author} {\bibinfo {author} {\bibfnamefont {R.}~\bibnamefont
  {S\o{}nden\aa{}}}, \bibinfo {author} {\bibfnamefont {P.}~\bibnamefont
  {Ravindran}}, \bibinfo {author} {\bibfnamefont {S.}~\bibnamefont
  {St\o{}len}}, \bibinfo {author} {\bibfnamefont {T.}~\bibnamefont {Grande}}, \
  and\ \bibinfo {author} {\bibfnamefont {M.}~\bibnamefont {Hanfland}},\ }\href
  {\doibase 10.1103/PhysRevB.74.144102} {\bibfield  {journal} {\bibinfo
  {journal} {Phys. Rev. B}\ }\textbf {\bibinfo {volume} {74}},\ \bibinfo
  {pages} {144102} (\bibinfo {year} {2006})}\BibitemShut {NoStop}%
\bibitem [{\citenamefont {Geller}(1957)}]{Geller1957}%
  \BibitemOpen
  \bibfield  {author} {\bibinfo {author} {\bibfnamefont {S.}~\bibnamefont
  {Geller}},\ }\href {\doibase 10.1107/S0365110X57000778} {\bibfield  {journal}
  {\bibinfo  {journal} {Acta Crystallogr.}\ }\textbf {\bibinfo {volume} {10}},\
  \bibinfo {pages} {243} (\bibinfo {year} {1957})}\BibitemShut {NoStop}%
\bibitem [{\citenamefont {Cwik}\ \emph {et~al.}(2003)\citenamefont {Cwik},
  \citenamefont {Lorenz}, \citenamefont {Baier}, \citenamefont {M{\"{u}}ller},
  \citenamefont {Andr{\'{e}}}, \citenamefont {Bour{\'{e}}e}, \citenamefont
  {Lichtenberg}, \citenamefont {Freimuth}, \citenamefont {Schmitz},
  \citenamefont {M{\"{u}}ller-Hartmann},\ and\ \citenamefont
  {Braden}}]{Cwik2003}%
  \BibitemOpen
  \bibfield  {author} {\bibinfo {author} {\bibfnamefont {M.}~\bibnamefont
  {Cwik}}, \bibinfo {author} {\bibfnamefont {T.}~\bibnamefont {Lorenz}},
  \bibinfo {author} {\bibfnamefont {J.}~\bibnamefont {Baier}}, \bibinfo
  {author} {\bibfnamefont {R.}~\bibnamefont {M{\"{u}}ller}}, \bibinfo {author}
  {\bibfnamefont {G.}~\bibnamefont {Andr{\'{e}}}}, \bibinfo {author}
  {\bibfnamefont {F.}~\bibnamefont {Bour{\'{e}}e}}, \bibinfo {author}
  {\bibfnamefont {F.}~\bibnamefont {Lichtenberg}}, \bibinfo {author}
  {\bibfnamefont {A.}~\bibnamefont {Freimuth}}, \bibinfo {author}
  {\bibfnamefont {R.}~\bibnamefont {Schmitz}}, \bibinfo {author} {\bibfnamefont
  {E.}~\bibnamefont {M{\"{u}}ller-Hartmann}}, \ and\ \bibinfo {author}
  {\bibfnamefont {M.}~\bibnamefont {Braden}},\ }\href {\doibase
  10.1103/PhysRevB.68.060401} {\bibfield  {journal} {\bibinfo  {journal} {Phys.
  Rev. B}\ }\textbf {\bibinfo {volume} {68}},\ \bibinfo {pages} {060401}
  (\bibinfo {year} {2003})}\BibitemShut {NoStop}%
\bibitem [{\citenamefont {Bordet}\ \emph {et~al.}(1993)\citenamefont {Bordet},
  \citenamefont {Chaillout}, \citenamefont {Marezio}, \citenamefont {Huang},
  \citenamefont {Santoro}, \citenamefont {Cheong}, \citenamefont {Takagi},
  \citenamefont {Oglesby},\ and\ \citenamefont {Batlogg}}]{Bordet1993}%
  \BibitemOpen
  \bibfield  {author} {\bibinfo {author} {\bibfnamefont {P.}~\bibnamefont
  {Bordet}}, \bibinfo {author} {\bibfnamefont {C.}~\bibnamefont {Chaillout}},
  \bibinfo {author} {\bibfnamefont {M.}~\bibnamefont {Marezio}}, \bibinfo
  {author} {\bibfnamefont {Q.}~\bibnamefont {Huang}}, \bibinfo {author}
  {\bibfnamefont {A.}~\bibnamefont {Santoro}}, \bibinfo {author} {\bibfnamefont
  {S.-W.}\ \bibnamefont {Cheong}}, \bibinfo {author} {\bibfnamefont
  {H.}~\bibnamefont {Takagi}}, \bibinfo {author} {\bibfnamefont
  {C.}~\bibnamefont {Oglesby}}, \ and\ \bibinfo {author} {\bibfnamefont
  {B.}~\bibnamefont {Batlogg}},\ }\href {\doibase 10.1006/jssc.1993.1285}
  {\bibfield  {journal} {\bibinfo  {journal} {J. Solid State Chem}\ }\textbf
  {\bibinfo {volume} {106}},\ \bibinfo {pages} {253} (\bibinfo {year}
  {1993})}\BibitemShut {NoStop}%
\bibitem [{\citenamefont {States}\ \emph {et~al.}(1977)\citenamefont {States},
  \citenamefont {Khattak}, \citenamefont {Cox},\ and\ \citenamefont
  {February}}]{States1977}%
  \BibitemOpen
  \bibfield  {author} {\bibinfo {author} {\bibfnamefont {U.}~\bibnamefont
  {States}}, \bibinfo {author} {\bibfnamefont {C.~P.}\ \bibnamefont {Khattak}},
  \bibinfo {author} {\bibfnamefont {D.~E.}\ \bibnamefont {Cox}}, \ and\
  \bibinfo {author} {\bibfnamefont {R.}~\bibnamefont {February}},\ }\href
  {\doibase http://dx.doi.org/10.1016/0025-5408(77)90111-8} {\bibfield
  {journal} {\bibinfo  {journal} {Mater. Res. Bull.}\ }\textbf {\bibinfo
  {volume} {12}},\ \bibinfo {pages} {463} (\bibinfo {year} {1977})}\BibitemShut
  {NoStop}%
\bibitem [{\citenamefont {Elemans}\ \emph {et~al.}(1971)\citenamefont
  {Elemans}, \citenamefont {{Van Laar}}, \citenamefont {{Van Der Veen}},\ and\
  \citenamefont {Loopstra}}]{Elemans1971}%
  \BibitemOpen
  \bibfield  {author} {\bibinfo {author} {\bibfnamefont {J.~B. A.~A.}\
  \bibnamefont {Elemans}}, \bibinfo {author} {\bibfnamefont {B.}~\bibnamefont
  {{Van Laar}}}, \bibinfo {author} {\bibfnamefont {K.~R.}\ \bibnamefont {{Van
  Der Veen}}}, \ and\ \bibinfo {author} {\bibfnamefont {B.~O.}\ \bibnamefont
  {Loopstra}},\ }\href {\doibase 10.1016/0022-4596(71)90034-X} {\bibfield
  {journal} {\bibinfo  {journal} {‎J. Solid State Chem}\ }\textbf {\bibinfo
  {volume} {3}},\ \bibinfo {pages} {238} (\bibinfo {year} {1971})}\BibitemShut
  {NoStop}%
\bibitem [{\citenamefont {Dann}\ \emph {et~al.}(1994)\citenamefont {Dann},
  \citenamefont {Currie}, \citenamefont {Weller}, \citenamefont {Thomas},\ and\
  \citenamefont {a.D. Al-Rawwas}}]{Dann1994}%
  \BibitemOpen
  \bibfield  {author} {\bibinfo {author} {\bibfnamefont {S.}~\bibnamefont
  {Dann}}, \bibinfo {author} {\bibfnamefont {D.}~\bibnamefont {Currie}},
  \bibinfo {author} {\bibfnamefont {M.}~\bibnamefont {Weller}}, \bibinfo
  {author} {\bibfnamefont {M.}~\bibnamefont {Thomas}}, \ and\ \bibinfo {author}
  {\bibnamefont {a.D. Al-Rawwas}},\ }\bibfield  {booktitle} {\emph {\bibinfo
  {booktitle} {J. Solid State Chem}},\ }\href {\doibase 10.1006/jssc.1994.1083}
  {\ \textbf {\bibinfo {volume} {109}},\ \bibinfo {pages} {134} (\bibinfo
  {year} {1994})}\BibitemShut {NoStop}%
\bibitem [{\citenamefont {Rodriguez}\ \emph {et~al.}(2011)\citenamefont
  {Rodriguez}, \citenamefont {Poineau}, \citenamefont {Llobet}, \citenamefont
  {Kennedy}, \citenamefont {Avdeev}, \citenamefont {Thorogood}, \citenamefont
  {Carter}, \citenamefont {Seshadri}, \citenamefont {Singh},\ and\
  \citenamefont {Cheetham}}]{Rodriguez2011}%
  \BibitemOpen
  \bibfield  {author} {\bibinfo {author} {\bibfnamefont {E.~E.}\ \bibnamefont
  {Rodriguez}}, \bibinfo {author} {\bibfnamefont {F.}~\bibnamefont {Poineau}},
  \bibinfo {author} {\bibfnamefont {A.}~\bibnamefont {Llobet}}, \bibinfo
  {author} {\bibfnamefont {B.~J.}\ \bibnamefont {Kennedy}}, \bibinfo {author}
  {\bibfnamefont {M.}~\bibnamefont {Avdeev}}, \bibinfo {author} {\bibfnamefont
  {G.~J.}\ \bibnamefont {Thorogood}}, \bibinfo {author} {\bibfnamefont {M.~L.}\
  \bibnamefont {Carter}}, \bibinfo {author} {\bibfnamefont {R.}~\bibnamefont
  {Seshadri}}, \bibinfo {author} {\bibfnamefont {D.~J.}\ \bibnamefont {Singh}},
  \ and\ \bibinfo {author} {\bibfnamefont {A.~K.}\ \bibnamefont {Cheetham}},\
  }\href {\doibase 10.1103/PhysRevLett.106.067201} {\bibfield  {journal}
  {\bibinfo  {journal} {Phys. Rev. Lett.}\ }\textbf {\bibinfo {volume} {106}},\
  \bibinfo {pages} {1} (\bibinfo {year} {2011})}\BibitemShut {NoStop}%
\bibitem [{\citenamefont {Braden}\ \emph {et~al.}(1998)\citenamefont {Braden},
  \citenamefont {Andr{\'{e}}}, \citenamefont {Nakatsuji},\ and\ \citenamefont
  {Maeno}}]{Braden1998}%
  \BibitemOpen
  \bibfield  {author} {\bibinfo {author} {\bibfnamefont {M.}~\bibnamefont
  {Braden}}, \bibinfo {author} {\bibfnamefont {G.}~\bibnamefont {Andr{\'{e}}}},
  \bibinfo {author} {\bibfnamefont {S.}~\bibnamefont {Nakatsuji}}, \ and\
  \bibinfo {author} {\bibfnamefont {Y.}~\bibnamefont {Maeno}},\ }\href
  {\doibase 10.1103/PhysRevB.58.847} {\bibfield  {journal} {\bibinfo  {journal}
  {Phys. Rev. B}\ }\textbf {\bibinfo {volume} {58}},\ \bibinfo {pages} {847}
  (\bibinfo {year} {1998})}\BibitemShut {NoStop}%
\bibitem [{\citenamefont {Shi}\ \emph {et~al.}(2009)\citenamefont {Shi},
  \citenamefont {Guo}, \citenamefont {Yu}, \citenamefont {Arai}, \citenamefont
  {Belik}, \citenamefont {Sato}, \citenamefont {Yamaura}, \citenamefont
  {Takayama-Muromachi}, \citenamefont {Tian}, \citenamefont {Yang},
  \citenamefont {Li}, \citenamefont {Varga}, \citenamefont {Mitchell},\ and\
  \citenamefont {Okamoto}}]{Shi2009}%
  \BibitemOpen
  \bibfield  {author} {\bibinfo {author} {\bibfnamefont {Y.~G.}\ \bibnamefont
  {Shi}}, \bibinfo {author} {\bibfnamefont {Y.~F.}\ \bibnamefont {Guo}},
  \bibinfo {author} {\bibfnamefont {S.}~\bibnamefont {Yu}}, \bibinfo {author}
  {\bibfnamefont {M.}~\bibnamefont {Arai}}, \bibinfo {author} {\bibfnamefont
  {A.~A.}\ \bibnamefont {Belik}}, \bibinfo {author} {\bibfnamefont
  {A.}~\bibnamefont {Sato}}, \bibinfo {author} {\bibfnamefont {K.}~\bibnamefont
  {Yamaura}}, \bibinfo {author} {\bibfnamefont {E.}~\bibnamefont
  {Takayama-Muromachi}}, \bibinfo {author} {\bibfnamefont {H.~F.}\ \bibnamefont
  {Tian}}, \bibinfo {author} {\bibfnamefont {H.~X.}\ \bibnamefont {Yang}},
  \bibinfo {author} {\bibfnamefont {J.~Q.}\ \bibnamefont {Li}}, \bibinfo
  {author} {\bibfnamefont {T.}~\bibnamefont {Varga}}, \bibinfo {author}
  {\bibfnamefont {J.~F.}\ \bibnamefont {Mitchell}}, \ and\ \bibinfo {author}
  {\bibfnamefont {S.}~\bibnamefont {Okamoto}},\ }\href@noop {} {\bibfield
  {journal} {\bibinfo  {journal} {Phys. Rev. B Condens. Matter Mater. Phys.}\
  }\textbf {\bibinfo {volume} {80}},\ \bibinfo {pages} {1} (\bibinfo {year}
  {2009})}\BibitemShut {NoStop}%
\bibitem [{\citenamefont {Takeda}\ and\ \citenamefont
  {Ōhara}(1974)}]{doi:10.1143/JPSJ.37.275}%
  \BibitemOpen
  \bibfield  {author} {\bibinfo {author} {\bibfnamefont {T.}~\bibnamefont
  {Takeda}}\ and\ \bibinfo {author} {\bibfnamefont {S.}~\bibnamefont
  {Ōhara}},\ }\href {\doibase 10.1143/JPSJ.37.275} {\bibfield  {journal}
  {\bibinfo  {journal} {J. Phys. Soc. Jpn}\ }\textbf {\bibinfo {volume} {37}},\
  \bibinfo {pages} {275} (\bibinfo {year} {1974})}\BibitemShut {NoStop}%
\bibitem [{\citenamefont {Kresse}\ and\ \citenamefont
  {Hafner}(1993)}]{Kresse1993}%
  \BibitemOpen
  \bibfield  {author} {\bibinfo {author} {\bibfnamefont {G.}~\bibnamefont
  {Kresse}}\ and\ \bibinfo {author} {\bibfnamefont {J.}~\bibnamefont
  {Hafner}},\ }\href {\doibase 10.1103/PhysRevB.47.558} {\bibfield  {journal}
  {\bibinfo  {journal} {Phys. Rev. B}\ }\textbf {\bibinfo {volume} {47}},\
  \bibinfo {pages} {558} (\bibinfo {year} {1993})}\BibitemShut {NoStop}%
\bibitem [{\citenamefont {Kresse}\ and\ \citenamefont
  {Furthm{\"{u}}ller}(1996)}]{Kresse1996}%
  \BibitemOpen
  \bibfield  {author} {\bibinfo {author} {\bibfnamefont {G.}~\bibnamefont
  {Kresse}}\ and\ \bibinfo {author} {\bibfnamefont {J.}~\bibnamefont
  {Furthm{\"{u}}ller}},\ }\href {\doibase 10.1103/PhysRevB.54.11169} {\bibfield
   {journal} {\bibinfo  {journal} {Phys. Rev. B}\ }\textbf {\bibinfo {volume}
  {54}},\ \bibinfo {pages} {11169} (\bibinfo {year} {1996})}\BibitemShut
  {NoStop}%
\bibitem [{Blo(1994)}]{Blochl1994}%
  \BibitemOpen
  \href {\doibase 10.1103/PhysRevB.50.17953} {\bibfield  {journal} {\bibinfo
  {journal} {Phys. Rev. B}\ }\textbf {\bibinfo {volume} {50}},\ \bibinfo
  {pages} {17953} (\bibinfo {year} {1994})}\BibitemShut {NoStop}%
\bibitem [{\citenamefont {Perdew}\ \emph {et~al.}(1997)\citenamefont {Perdew},
  \citenamefont {Burke},\ and\ \citenamefont
  {Ernzerhof}}]{PhysRevLett.78.1396}%
  \BibitemOpen
  \bibfield  {author} {\bibinfo {author} {\bibfnamefont {J.~P.}\ \bibnamefont
  {Perdew}}, \bibinfo {author} {\bibfnamefont {K.}~\bibnamefont {Burke}}, \
  and\ \bibinfo {author} {\bibfnamefont {M.}~\bibnamefont {Ernzerhof}},\ }\href
  {\doibase 10.1103/PhysRevLett.78.1396} {\bibfield  {journal} {\bibinfo
  {journal} {Phys. Rev. Lett.}\ }\textbf {\bibinfo {volume} {78}},\ \bibinfo
  {pages} {1396} (\bibinfo {year} {1997})}\BibitemShut {NoStop}%
\bibitem [{\citenamefont {Dudarev}\ \emph {et~al.}(1998)\citenamefont
  {Dudarev}, \citenamefont {Botton}, \citenamefont {Savrasov}, \citenamefont
  {Humphreys},\ and\ \citenamefont {Sutton}}]{Dudarev1998}%
  \BibitemOpen
  \bibfield  {author} {\bibinfo {author} {\bibfnamefont {S.~L.}\ \bibnamefont
  {Dudarev}}, \bibinfo {author} {\bibfnamefont {G.~A.}\ \bibnamefont {Botton}},
  \bibinfo {author} {\bibfnamefont {S.~Y.}\ \bibnamefont {Savrasov}}, \bibinfo
  {author} {\bibfnamefont {C.~J.}\ \bibnamefont {Humphreys}}, \ and\ \bibinfo
  {author} {\bibfnamefont {A.~P.}\ \bibnamefont {Sutton}},\ }\href {\doibase
  10.1103/PhysRevB.57.1505} {\bibfield  {journal} {\bibinfo  {journal} {Phys.
  Rev. B}\ }\textbf {\bibinfo {volume} {57}},\ \bibinfo {pages} {1505}
  (\bibinfo {year} {1998})}\BibitemShut {NoStop}%
\bibitem [{\citenamefont {Shishkin}\ and\ \citenamefont
  {Kresse}(2006)}]{PhysRevB.74.035101}%
  \BibitemOpen
  \bibfield  {author} {\bibinfo {author} {\bibfnamefont {M.}~\bibnamefont
  {Shishkin}}\ and\ \bibinfo {author} {\bibfnamefont {G.}~\bibnamefont
  {Kresse}},\ }\href {\doibase 10.1103/PhysRevB.74.035101} {\bibfield
  {journal} {\bibinfo  {journal} {Phys. Rev. B}\ }\textbf {\bibinfo {volume}
  {74}},\ \bibinfo {pages} {035101} (\bibinfo {year} {2006})}\BibitemShut
  {NoStop}%
\bibitem [{\citenamefont {Kang}\ and\ \citenamefont
  {Hybertsen}(2010)}]{Kang2000}%
  \BibitemOpen
  \bibfield  {author} {\bibinfo {author} {\bibfnamefont {W.}~\bibnamefont
  {Kang}}\ and\ \bibinfo {author} {\bibfnamefont {M.~S.}\ \bibnamefont
  {Hybertsen}},\ }\href {\doibase 10.1103/PhysRevB.82.085203} {\bibfield
  {journal} {\bibinfo  {journal} {Phys. Rev. B}\ }\textbf {\bibinfo {volume}
  {82}},\ \bibinfo {pages} {085203} (\bibinfo {year} {2010})}\BibitemShut
  {NoStop}%
\bibitem [{\citenamefont {Kato}(1957)}]{Kato}%
  \BibitemOpen
  \bibfield  {author} {\bibinfo {author} {\bibfnamefont {T.}~\bibnamefont
  {Kato}},\ }\href {\doibase 10.1002/cpa.3160100201} {\bibfield  {journal}
  {\bibinfo  {journal} {Comm. Pure Appl. Math}\ }\textbf {\bibinfo {volume}
  {10}},\ \bibinfo {pages} {151} (\bibinfo {year} {1957})}\BibitemShut
  {NoStop}%
\bibitem [{enc()}]{encutgw}%
  \BibitemOpen
  \href@noop {} {}\bibinfo {note} {In {\tt{VASP}} the energy cutoff for the
  plane wave expansion of the orbitals $E_{pw}$ is controlled by the flag
  {\tt{ENCUT}}, whereas the energy cutoff for the response function
  $E_{pw}^\chi$ is defined by the flag {\tt{ENCUTGW}} and controls how many
  ${\mathbf{G}}$ vectors are included in the response function
  $\chi_{{\mathbf{q}}}^0 ({\mathbf{G}}, {\mathbf{G}}', \omega)$. {\tt{ENCUTGW}}
  controls the basis set for the response functions in exactly the same manner
  as ENCUT does for the orbitals, in Hartree units: $N_{\rm{pw}} \approx
  \frac{\Omega}{2\pi^2}(2E_{\rm{pw}})^{3/2}$. In {\tt{VASP}} the total number
  of plane waves $N_{\rm{pw}}$ is given by the maximum number of plane-waves
  per irreducible \textbf{k}-point.}\BibitemShut {Stop}%
\bibitem [{\citenamefont {Nabok}\ \emph {et~al.}(2016)\citenamefont {Nabok},
  \citenamefont {Gulans},\ and\ \citenamefont {Draxl}}]{PhysRevB.94.035118}%
  \BibitemOpen
  \bibfield  {author} {\bibinfo {author} {\bibfnamefont {D.}~\bibnamefont
  {Nabok}}, \bibinfo {author} {\bibfnamefont {A.}~\bibnamefont {Gulans}}, \
  and\ \bibinfo {author} {\bibfnamefont {C.}~\bibnamefont {Draxl}},\ }\href
  {\doibase 10.1103/PhysRevB.94.035118} {\bibfield  {journal} {\bibinfo
  {journal} {Phys. Rev. B}\ }\textbf {\bibinfo {volume} {94}},\ \bibinfo
  {pages} {035118} (\bibinfo {year} {2016})}\BibitemShut {NoStop}%
\bibitem [{SM()}]{SM}%
  \BibitemOpen
  \href {http://prb...} {\enquote {\bibinfo {title} {See supplemental material
  for further details.}}\ }\BibitemShut {NoStop}%
\bibitem [{\citenamefont {Takizawa}\ \emph {et~al.}(2009)\citenamefont
  {Takizawa}, \citenamefont {Maekawa}, \citenamefont {Wadati}, \citenamefont
  {Yoshida}, \citenamefont {Fujimori}, \citenamefont {Kumigashira},\ and\
  \citenamefont {Oshima}}]{Takizawa2009}%
  \BibitemOpen
  \bibfield  {author} {\bibinfo {author} {\bibfnamefont {M.}~\bibnamefont
  {Takizawa}}, \bibinfo {author} {\bibfnamefont {K.}~\bibnamefont {Maekawa}},
  \bibinfo {author} {\bibfnamefont {H.}~\bibnamefont {Wadati}}, \bibinfo
  {author} {\bibfnamefont {T.}~\bibnamefont {Yoshida}}, \bibinfo {author}
  {\bibfnamefont {A.}~\bibnamefont {Fujimori}}, \bibinfo {author}
  {\bibfnamefont {H.}~\bibnamefont {Kumigashira}}, \ and\ \bibinfo {author}
  {\bibfnamefont {M.}~\bibnamefont {Oshima}},\ }\href {\doibase
  10.1103/PhysRevB.79.113103} {\bibfield  {journal} {\bibinfo  {journal} {Phys.
  Rev. B}\ }\textbf {\bibinfo {volume} {79}},\ \bibinfo {pages} {113103}
  (\bibinfo {year} {2009})}\BibitemShut {NoStop}%
\bibitem [{\citenamefont {Tezuka}\ \emph {et~al.}(1994)\citenamefont {Tezuka},
  \citenamefont {Shin}, \citenamefont {Ishii}, \citenamefont {Ejima},
  \citenamefont {Suzuki},\ and\ \citenamefont {Sato}}]{Tezuka1994}%
  \BibitemOpen
  \bibfield  {author} {\bibinfo {author} {\bibfnamefont {Y.}~\bibnamefont
  {Tezuka}}, \bibinfo {author} {\bibfnamefont {S.}~\bibnamefont {Shin}},
  \bibinfo {author} {\bibfnamefont {T.}~\bibnamefont {Ishii}}, \bibinfo
  {author} {\bibfnamefont {T.}~\bibnamefont {Ejima}}, \bibinfo {author}
  {\bibfnamefont {S.}~\bibnamefont {Suzuki}}, \ and\ \bibinfo {author}
  {\bibfnamefont {S.}~\bibnamefont {Sato}},\ }\href {\doibase
  10.1143/JPSJ.63.347} {\bibfield  {journal} {\bibinfo  {journal} {J. Phys.
  Soc. Jpn}\ }\textbf {\bibinfo {volume} {63}},\ \bibinfo {pages} {347}
  (\bibinfo {year} {1994})}\BibitemShut {NoStop}%
\bibitem [{\citenamefont {{Van Benthem}}\ \emph {et~al.}(2001)\citenamefont
  {{Van Benthem}}, \citenamefont {Els{\"{a}}sser},\ and\ \citenamefont
  {French}}]{VanBenthem2001a}%
  \BibitemOpen
  \bibfield  {author} {\bibinfo {author} {\bibfnamefont {K.}~\bibnamefont {{Van
  Benthem}}}, \bibinfo {author} {\bibfnamefont {C.}~\bibnamefont
  {Els{\"{a}}sser}}, \ and\ \bibinfo {author} {\bibfnamefont {R.~H.}\
  \bibnamefont {French}},\ }\href {\doibase 10.1063/1.1415766} {\bibfield
  {journal} {\bibinfo  {journal} {J. Appl. Phys.}\ }\textbf {\bibinfo {volume}
  {90}},\ \bibinfo {pages} {6156} (\bibinfo {year} {2001})}\BibitemShut
  {NoStop}%
\bibitem [{\citenamefont {Lee}\ \emph {et~al.}(2003)\citenamefont {Lee},
  \citenamefont {Lee}, \citenamefont {Noh}, \citenamefont {Byun}, \citenamefont
  {Yoo}, \citenamefont {Yamaura},\ and\ \citenamefont
  {Takayama-Muromachi}}]{Lee2002}%
  \BibitemOpen
  \bibfield  {author} {\bibinfo {author} {\bibfnamefont {Y.~S.}\ \bibnamefont
  {Lee}}, \bibinfo {author} {\bibfnamefont {J.~S.}\ \bibnamefont {Lee}},
  \bibinfo {author} {\bibfnamefont {T.~W.}\ \bibnamefont {Noh}}, \bibinfo
  {author} {\bibfnamefont {D.-Y.}\ \bibnamefont {Byun}}, \bibinfo {author}
  {\bibfnamefont {K.~S.}\ \bibnamefont {Yoo}}, \bibinfo {author} {\bibfnamefont
  {K.}~\bibnamefont {Yamaura}}, \ and\ \bibinfo {author} {\bibfnamefont
  {E.}~\bibnamefont {Takayama-Muromachi}},\ }\href {\doibase
  10.1103/PhysRevB.67.113101} {\ \textbf {\bibinfo {volume} {67}},\ \bibinfo
  {pages} {113101} (\bibinfo {year} {2003})}\BibitemShut {NoStop}%
\bibitem [{\citenamefont {Vecchio}\ \emph {et~al.}(2013)\citenamefont
  {Vecchio}, \citenamefont {Perucchi}, \citenamefont {{Di Pietro}},
  \citenamefont {Limaj}, \citenamefont {Schade}, \citenamefont {Sun},
  \citenamefont {Arai}, \citenamefont {Yamaura},\ and\ \citenamefont
  {Lupi}}]{Vecchio2013}%
  \BibitemOpen
  \bibfield  {author} {\bibinfo {author} {\bibfnamefont {I.~L.}\ \bibnamefont
  {Vecchio}}, \bibinfo {author} {\bibfnamefont {A.}~\bibnamefont {Perucchi}},
  \bibinfo {author} {\bibfnamefont {P.}~\bibnamefont {{Di Pietro}}}, \bibinfo
  {author} {\bibfnamefont {O.}~\bibnamefont {Limaj}}, \bibinfo {author}
  {\bibfnamefont {U.}~\bibnamefont {Schade}}, \bibinfo {author} {\bibfnamefont
  {Y.}~\bibnamefont {Sun}}, \bibinfo {author} {\bibfnamefont {M.}~\bibnamefont
  {Arai}}, \bibinfo {author} {\bibfnamefont {K.}~\bibnamefont {Yamaura}}, \
  and\ \bibinfo {author} {\bibfnamefont {S.}~\bibnamefont {Lupi}},\ }\href
  {\doibase 10.1038/srep02990} {\bibfield  {journal} {\bibinfo  {journal} {Sci.
  Rep.}\ }\textbf {\bibinfo {volume} {3}},\ \bibinfo {pages} {2990} (\bibinfo
  {year} {2013})}\BibitemShut {NoStop}%
\bibitem [{\citenamefont {Erg\"onenc}(2017)}]{PhD}%
  \BibitemOpen
  \bibfield  {author} {\bibinfo {author} {\bibfnamefont {Z.}~\bibnamefont
  {Erg\"onenc}},\ }\href@noop {} {\emph {\bibinfo {title} {GW quasiparticle
  energies for transition metal oxide perovskites}}}\ (\bibinfo  {publisher}
  {PhD Thesis, University of Vienna},\ \bibinfo {year} {2017})\BibitemShut
  {NoStop}%
\bibitem [{\citenamefont {Saitoh}\ \emph {et~al.}(1995)\citenamefont {Saitoh},
  \citenamefont {Bocquet}, \citenamefont {Mizokawa}, \citenamefont {Namatame},
  \citenamefont {Fujimori}, \citenamefont {Abbate}, \citenamefont {Takeda},\
  and\ \citenamefont {Takano}}]{Saitoh1995}%
  \BibitemOpen
  \bibfield  {author} {\bibinfo {author} {\bibfnamefont {T.}~\bibnamefont
  {Saitoh}}, \bibinfo {author} {\bibfnamefont {A.~E.}\ \bibnamefont {Bocquet}},
  \bibinfo {author} {\bibfnamefont {T.}~\bibnamefont {Mizokawa}}, \bibinfo
  {author} {\bibfnamefont {H.}~\bibnamefont {Namatame}}, \bibinfo {author}
  {\bibfnamefont {A.}~\bibnamefont {Fujimori}}, \bibinfo {author}
  {\bibfnamefont {M.}~\bibnamefont {Abbate}}, \bibinfo {author} {\bibfnamefont
  {Y.}~\bibnamefont {Takeda}}, \ and\ \bibinfo {author} {\bibfnamefont
  {M.}~\bibnamefont {Takano}},\ }\href {\doibase 10.1103/PhysRevB.51.13942}
  {\bibfield  {journal} {\bibinfo  {journal} {Phys. Rev. B}\ }\textbf {\bibinfo
  {volume} {51}},\ \bibinfo {pages} {13942} (\bibinfo {year}
  {1995})}\BibitemShut {NoStop}%
\bibitem [{\citenamefont {Arima}\ \emph
  {et~al.}(1993{\natexlab{b}})\citenamefont {Arima}, \citenamefont {Tokura},\
  and\ \citenamefont {Torrance}}]{Arima1993}%
  \BibitemOpen
  \bibfield  {author} {\bibinfo {author} {\bibfnamefont {T.}~\bibnamefont
  {Arima}}, \bibinfo {author} {\bibfnamefont {Y.}~\bibnamefont {Tokura}}, \
  and\ \bibinfo {author} {\bibfnamefont {J.~B.}\ \bibnamefont {Torrance}},\
  }\href {\doibase 10.1103/PhysRevB.48.17006} {\bibfield  {journal} {\bibinfo
  {journal} {Phys. Rev. B}\ }\textbf {\bibinfo {volume} {48}},\ \bibinfo
  {pages} {17006} (\bibinfo {year} {1993}{\natexlab{b}})}\BibitemShut {NoStop}%
\bibitem [{\citenamefont {Jellison}\ \emph {et~al.}(2006)\citenamefont
  {Jellison}, \citenamefont {Paulauskas}, \citenamefont {Boatner},\ and\
  \citenamefont {Singh}}]{Jellison2006}%
  \BibitemOpen
  \bibfield  {author} {\bibinfo {author} {\bibfnamefont {G.~E.}\ \bibnamefont
  {Jellison}}, \bibinfo {author} {\bibfnamefont {I.}~\bibnamefont
  {Paulauskas}}, \bibinfo {author} {\bibfnamefont {L.~A.}\ \bibnamefont
  {Boatner}}, \ and\ \bibinfo {author} {\bibfnamefont {D.~J.}\ \bibnamefont
  {Singh}},\ }\href {\doibase 10.1103/PhysRevB.74.155130} {\bibfield  {journal}
  {\bibinfo  {journal} {Phys. Rev. B Condens. Matter Mater. Phys.}\ }\textbf
  {\bibinfo {volume} {74}},\ \bibinfo {pages} {1} (\bibinfo {year}
  {2006})}\BibitemShut {NoStop}%
\bibitem [{\citenamefont {Fatuzzo}\ \emph {et~al.}(2015)\citenamefont
  {Fatuzzo}, \citenamefont {Dantz}, \citenamefont {Fatale}, \citenamefont
  {Olalde-Velasco}, \citenamefont {Shaik}, \citenamefont {{Dalla Piazza}},
  \citenamefont {Toth}, \citenamefont {Pelliciari}, \citenamefont {Fittipaldi},
  \citenamefont {Vecchione}, \citenamefont {Kikugawa}, \citenamefont {Brooks},
  \citenamefont {R$\o{}$nnow}, \citenamefont {Grioni}, \citenamefont {R\"uegg},
  \citenamefont {Schmitt},\ and\ \citenamefont {Chang}}]{Fatuzzo2015}%
  \BibitemOpen
  \bibfield  {author} {\bibinfo {author} {\bibfnamefont {C.~G.}\ \bibnamefont
  {Fatuzzo}}, \bibinfo {author} {\bibfnamefont {M.}~\bibnamefont {Dantz}},
  \bibinfo {author} {\bibfnamefont {S.}~\bibnamefont {Fatale}}, \bibinfo
  {author} {\bibfnamefont {P.}~\bibnamefont {Olalde-Velasco}}, \bibinfo
  {author} {\bibfnamefont {N.~E.}\ \bibnamefont {Shaik}}, \bibinfo {author}
  {\bibfnamefont {B.}~\bibnamefont {{Dalla Piazza}}}, \bibinfo {author}
  {\bibfnamefont {S.}~\bibnamefont {Toth}}, \bibinfo {author} {\bibfnamefont
  {J.}~\bibnamefont {Pelliciari}}, \bibinfo {author} {\bibfnamefont
  {R.}~\bibnamefont {Fittipaldi}}, \bibinfo {author} {\bibfnamefont
  {A.}~\bibnamefont {Vecchione}}, \bibinfo {author} {\bibfnamefont
  {N.}~\bibnamefont {Kikugawa}}, \bibinfo {author} {\bibfnamefont {J.~S.}\
  \bibnamefont {Brooks}}, \bibinfo {author} {\bibfnamefont {H.~M.}\
  \bibnamefont {R$\o{}$nnow}}, \bibinfo {author} {\bibfnamefont
  {M.}~\bibnamefont {Grioni}}, \bibinfo {author} {\bibfnamefont
  {C.}~\bibnamefont {R\"uegg}}, \bibinfo {author} {\bibfnamefont
  {T.}~\bibnamefont {Schmitt}}, \ and\ \bibinfo {author} {\bibfnamefont
  {J.}~\bibnamefont {Chang}},\ }\href {\doibase 10.1103/PhysRevB.91.155104}
  {\bibfield  {journal} {\bibinfo  {journal} {Phys. Rev. B Condens. Matter
  Mater. Phys.}\ }\textbf {\bibinfo {volume} {91}},\ \bibinfo {pages} {1}
  (\bibinfo {year} {2015})}\BibitemShut {NoStop}%
\bibitem [{\citenamefont {Wang}\ \emph {et~al.}(2011)\citenamefont {Wang},
  \citenamefont {Wu},\ and\ \citenamefont {Jiang}}]{Wang2011}%
  \BibitemOpen
  \bibfield  {author} {\bibinfo {author} {\bibfnamefont {H.}~\bibnamefont
  {Wang}}, \bibinfo {author} {\bibfnamefont {F.}~\bibnamefont {Wu}}, \ and\
  \bibinfo {author} {\bibfnamefont {H.}~\bibnamefont {Jiang}},\ }\href@noop {}
  {\bibfield  {journal} {\bibinfo  {journal} {J. Phys. Chem.}\ }\textbf
  {\bibinfo {volume} {3}},\ \bibinfo {pages} {16180} (\bibinfo {year}
  {2011})}\BibitemShut {NoStop}%
\bibitem [{\citenamefont {Shishkin}\ \emph {et~al.}(2007)\citenamefont
  {Shishkin}, \citenamefont {Marsman},\ and\ \citenamefont
  {Kresse}}]{PhysRevLett.99.246403}%
  \BibitemOpen
  \bibfield  {author} {\bibinfo {author} {\bibfnamefont {M.}~\bibnamefont
  {Shishkin}}, \bibinfo {author} {\bibfnamefont {M.}~\bibnamefont {Marsman}}, \
  and\ \bibinfo {author} {\bibfnamefont {G.}~\bibnamefont {Kresse}},\ }\href
  {\doibase 10.1103/PhysRevLett.99.246403} {\bibfield  {journal} {\bibinfo
  {journal} {Phys. Rev. Lett.}\ }\textbf {\bibinfo {volume} {99}},\ \bibinfo
  {pages} {246403} (\bibinfo {year} {2007})}\BibitemShut {NoStop}%
\bibitem [{\citenamefont {Curtarolo}\ \emph {et~al.}(2013)\citenamefont
  {Curtarolo}, \citenamefont {Hart}, \citenamefont {Nardelli}, \citenamefont
  {Mingo}, \citenamefont {Sanvito},\ and\ \citenamefont {Levy}}]{Curtarolo}%
  \BibitemOpen
  \bibfield  {author} {\bibinfo {author} {\bibfnamefont {S.}~\bibnamefont
  {Curtarolo}}, \bibinfo {author} {\bibfnamefont {G.~L.~W.}\ \bibnamefont
  {Hart}}, \bibinfo {author} {\bibfnamefont {M.~B.}\ \bibnamefont {Nardelli}},
  \bibinfo {author} {\bibfnamefont {N.}~\bibnamefont {Mingo}}, \bibinfo
  {author} {\bibfnamefont {S.}~\bibnamefont {Sanvito}}, \ and\ \bibinfo
  {author} {\bibfnamefont {O.}~\bibnamefont {Levy}},\ }\href {\doibase
  10.1038/nmat3568} {\bibfield  {journal} {\bibinfo  {journal} {Nat Mater}\
  }\textbf {\bibinfo {volume} {12}},\ \bibinfo {pages} {191} (\bibinfo {year}
  {2013})}\BibitemShut {NoStop}%
\bibitem [{\citenamefont {Paier}\ \emph
  {et~al.}(2008{\natexlab{b}})\citenamefont {Paier}, \citenamefont {Marsman},\
  and\ \citenamefont {Kresse}}]{paier}%
  \BibitemOpen
  \bibfield  {author} {\bibinfo {author} {\bibfnamefont {J.}~\bibnamefont
  {Paier}}, \bibinfo {author} {\bibfnamefont {M.}~\bibnamefont {Marsman}}, \
  and\ \bibinfo {author} {\bibfnamefont {G.}~\bibnamefont {Kresse}},\ }\href
  {\doibase 10.1103/PhysRevB.78.121201} {\bibfield  {journal} {\bibinfo
  {journal} {Phys. Rev. B}\ }\textbf {\bibinfo {volume} {78}},\ \bibinfo
  {pages} {121201} (\bibinfo {year} {2008}{\natexlab{b}})}\BibitemShut
  {NoStop}%
\bibitem [{\citenamefont {He}\ and\ \citenamefont {Franchini}(2017)}]{jghe}%
  \BibitemOpen
  \bibfield  {author} {\bibinfo {author} {\bibfnamefont {J.}~\bibnamefont
  {He}}\ and\ \bibinfo {author} {\bibfnamefont {C.}~\bibnamefont {Franchini}},\
  }\href {http://stacks.iop.org/0953-8984/29/i=45/a=454004} {\bibfield
  {journal} {\bibinfo  {journal} {Journal of Physics: Condensed Matter}\
  }\textbf {\bibinfo {volume} {29}},\ \bibinfo {pages} {454004} (\bibinfo
  {year} {2017})}\BibitemShut {NoStop}%
\bibitem [{\citenamefont {Mostofi}\ \emph {et~al.}(2014)\citenamefont
  {Mostofi}, \citenamefont {Yates}, \citenamefont {Pizzi}, \citenamefont {Lee},
  \citenamefont {Souza}, \citenamefont {Vanderbilt},\ and\ \citenamefont
  {Marzari}}]{wannier90}%
  \BibitemOpen
  \bibfield  {author} {\bibinfo {author} {\bibfnamefont {A.~A.}\ \bibnamefont
  {Mostofi}}, \bibinfo {author} {\bibfnamefont {J.~R.}\ \bibnamefont {Yates}},
  \bibinfo {author} {\bibfnamefont {G.}~\bibnamefont {Pizzi}}, \bibinfo
  {author} {\bibfnamefont {Y.-S.}\ \bibnamefont {Lee}}, \bibinfo {author}
  {\bibfnamefont {I.}~\bibnamefont {Souza}}, \bibinfo {author} {\bibfnamefont
  {D.}~\bibnamefont {Vanderbilt}}, \ and\ \bibinfo {author} {\bibfnamefont
  {N.}~\bibnamefont {Marzari}},\ }\href {\doibase
  https://doi.org/10.1016/j.cpc.2014.05.003} {\bibfield  {journal} {\bibinfo
  {journal} {Comput. Phys. Commun}\ }\textbf {\bibinfo {volume} {185}},\
  \bibinfo {pages} {2309 } (\bibinfo {year} {2014})}\BibitemShut {NoStop}%
\bibitem [{\citenamefont {{Lee}}\ and\ \citenamefont {Seo}(2010)}]{Lee2010}%
  \BibitemOpen
  \bibfield  {author} {\bibinfo {author} {\bibfnamefont {D.}~\bibnamefont
  {{Lee}}, \bibfnamefont {Y}}\ and\ \bibinfo {author} {\bibfnamefont {Y.~K.}\
  \bibnamefont {Seo}},\ }\href {\doibase 10.3938/jkps.56.366} {\bibfield
  {journal} {\bibinfo  {journal} {J. Korean Phys. Soc}\ }\textbf {\bibinfo
  {volume} {56}},\ \bibinfo {pages} {366} (\bibinfo {year} {2010})}\BibitemShut
  {NoStop}%
\bibitem [{\citenamefont {A.M.~Mamedov}(1984)}]{KTO}%
  \BibitemOpen
  \bibfield  {author} {\bibinfo {author} {\bibfnamefont {L.~G.}\ \bibnamefont
  {A.M.~Mamedov}},\ }\href@noop {} {\bibfield  {journal} {\bibinfo  {journal}
  {Fiz. Tverd. Tela (Leningrad)}\ }\textbf {\bibinfo {volume} {26}},\ \bibinfo
  {pages} {583} (\bibinfo {year} {1984})}\BibitemShut {NoStop}%
\bibitem [{\citenamefont {Jung}\ \emph {et~al.}(1997)\citenamefont {Jung},
  \citenamefont {Kim}, \citenamefont {Eom}, \citenamefont {Noh}, \citenamefont
  {Choi}, \citenamefont {Yu}, \citenamefont {Kwon},\ and\ \citenamefont
  {Chung}}]{PhysRevB.55.15489}%
  \BibitemOpen
  \bibfield  {author} {\bibinfo {author} {\bibfnamefont {J.~H.}\ \bibnamefont
  {Jung}}, \bibinfo {author} {\bibfnamefont {K.~H.}\ \bibnamefont {Kim}},
  \bibinfo {author} {\bibfnamefont {D.~J.}\ \bibnamefont {Eom}}, \bibinfo
  {author} {\bibfnamefont {T.~W.}\ \bibnamefont {Noh}}, \bibinfo {author}
  {\bibfnamefont {E.~J.}\ \bibnamefont {Choi}}, \bibinfo {author}
  {\bibfnamefont {J.}~\bibnamefont {Yu}}, \bibinfo {author} {\bibfnamefont
  {Y.~S.}\ \bibnamefont {Kwon}}, \ and\ \bibinfo {author} {\bibfnamefont
  {Y.}~\bibnamefont {Chung}},\ }\href {\doibase 10.1103/PhysRevB.55.15489}
  {\bibfield  {journal} {\bibinfo  {journal} {Phys. Rev. B}\ }\textbf {\bibinfo
  {volume} {55}},\ \bibinfo {pages} {15489} (\bibinfo {year}
  {1997})}\BibitemShut {NoStop}%
\bibitem [{\citenamefont {Jung}\ \emph {et~al.}(2003)\citenamefont {Jung},
  \citenamefont {Fang}, \citenamefont {He}, \citenamefont {Kaneko},
  \citenamefont {Okimoto},\ and\ \citenamefont
  {Tokura}}]{PhysRevLett.91.056403}%
  \BibitemOpen
  \bibfield  {author} {\bibinfo {author} {\bibfnamefont {J.~H.}\ \bibnamefont
  {Jung}}, \bibinfo {author} {\bibfnamefont {Z.}~\bibnamefont {Fang}}, \bibinfo
  {author} {\bibfnamefont {J.~P.}\ \bibnamefont {He}}, \bibinfo {author}
  {\bibfnamefont {Y.}~\bibnamefont {Kaneko}}, \bibinfo {author} {\bibfnamefont
  {Y.}~\bibnamefont {Okimoto}}, \ and\ \bibinfo {author} {\bibfnamefont
  {Y.}~\bibnamefont {Tokura}},\ }\href {\doibase 10.1103/PhysRevLett.91.056403}
  {\bibfield  {journal} {\bibinfo  {journal} {Phys. Rev. Lett.}\ }\textbf
  {\bibinfo {volume} {91}},\ \bibinfo {pages} {056403} (\bibinfo {year}
  {2003})}\BibitemShut {NoStop}%
\bibitem [{\citenamefont {Pavarini}\ and\ \citenamefont
  {Koch}(2010)}]{PhysRevLett.104.086402}%
  \BibitemOpen
  \bibfield  {author} {\bibinfo {author} {\bibfnamefont {E.}~\bibnamefont
  {Pavarini}}\ and\ \bibinfo {author} {\bibfnamefont {E.}~\bibnamefont
  {Koch}},\ }\href {\doibase 10.1103/PhysRevLett.104.086402} {\bibfield
  {journal} {\bibinfo  {journal} {Phys. Rev. Lett.}\ }\textbf {\bibinfo
  {volume} {104}},\ \bibinfo {pages} {086402} (\bibinfo {year}
  {2010})}\BibitemShut {NoStop}%
\bibitem [{\citenamefont {He}\ \emph {et~al.}(2012)\citenamefont {He},
  \citenamefont {Chen}, \citenamefont {Chen},\ and\ \citenamefont
  {Franchini}}]{PhysRevB.85.195135}%
  \BibitemOpen
  \bibfield  {author} {\bibinfo {author} {\bibfnamefont {J.}~\bibnamefont
  {He}}, \bibinfo {author} {\bibfnamefont {M.-X.}\ \bibnamefont {Chen}},
  \bibinfo {author} {\bibfnamefont {X.-Q.}\ \bibnamefont {Chen}}, \ and\
  \bibinfo {author} {\bibfnamefont {C.}~\bibnamefont {Franchini}},\ }\href
  {\doibase 10.1103/PhysRevB.85.195135} {\bibfield  {journal} {\bibinfo
  {journal} {Phys. Rev. B}\ }\textbf {\bibinfo {volume} {85}},\ \bibinfo
  {pages} {195135} (\bibinfo {year} {2012})}\BibitemShut {NoStop}%
\bibitem [{\citenamefont {Franchini}\ \emph {et~al.}(2011)\citenamefont
  {Franchini}, \citenamefont {Archer}, \citenamefont {He}, \citenamefont
  {Chen}, \citenamefont {Filippetti},\ and\ \citenamefont
  {Sanvito}}]{PhysRevB.83.220402}%
  \BibitemOpen
  \bibfield  {author} {\bibinfo {author} {\bibfnamefont {C.}~\bibnamefont
  {Franchini}}, \bibinfo {author} {\bibfnamefont {T.}~\bibnamefont {Archer}},
  \bibinfo {author} {\bibfnamefont {J.}~\bibnamefont {He}}, \bibinfo {author}
  {\bibfnamefont {X.-Q.}\ \bibnamefont {Chen}}, \bibinfo {author}
  {\bibfnamefont {A.}~\bibnamefont {Filippetti}}, \ and\ \bibinfo {author}
  {\bibfnamefont {S.}~\bibnamefont {Sanvito}},\ }\href {\doibase
  10.1103/PhysRevB.83.220402} {\bibfield  {journal} {\bibinfo  {journal} {Phys.
  Rev. B}\ }\textbf {\bibinfo {volume} {83}},\ \bibinfo {pages} {220402}
  (\bibinfo {year} {2011})}\BibitemShut {NoStop}%
\bibitem [{\citenamefont {Mravlje}\ \emph {et~al.}(2012)\citenamefont
  {Mravlje}, \citenamefont {Aichhorn},\ and\ \citenamefont
  {Georges}}]{PhysRevLett.108.197202}%
  \BibitemOpen
  \bibfield  {author} {\bibinfo {author} {\bibfnamefont {J.}~\bibnamefont
  {Mravlje}}, \bibinfo {author} {\bibfnamefont {M.}~\bibnamefont {Aichhorn}}, \
  and\ \bibinfo {author} {\bibfnamefont {A.}~\bibnamefont {Georges}},\ }\href
  {\doibase 10.1103/PhysRevLett.108.197202} {\bibfield  {journal} {\bibinfo
  {journal} {Phys. Rev. Lett.}\ }\textbf {\bibinfo {volume} {108}},\ \bibinfo
  {pages} {197202} (\bibinfo {year} {2012})}\BibitemShut {NoStop}%
\bibitem [{\citenamefont {Vale}\ \emph {et~al.}()\citenamefont {Vale},
  \citenamefont {Calder}, \citenamefont {Donnerer}, \citenamefont {Pincini},
  \citenamefont {Shi}, \citenamefont {Tsujimoto}, \citenamefont {Yamaura},
  \citenamefont {Sala}, \citenamefont {van~den Brink}, \citenamefont
  {Christianson},\ and\ \citenamefont {McMorrow}}]{NOO}%
  \BibitemOpen
  \bibfield  {author} {\bibinfo {author} {\bibfnamefont {J.~G.}\ \bibnamefont
  {Vale}}, \bibinfo {author} {\bibfnamefont {S.}~\bibnamefont {Calder}},
  \bibinfo {author} {\bibfnamefont {C.}~\bibnamefont {Donnerer}}, \bibinfo
  {author} {\bibfnamefont {D.}~\bibnamefont {Pincini}}, \bibinfo {author}
  {\bibfnamefont {Y.~G.}\ \bibnamefont {Shi}}, \bibinfo {author} {\bibfnamefont
  {Y.}~\bibnamefont {Tsujimoto}}, \bibinfo {author} {\bibfnamefont
  {K.}~\bibnamefont {Yamaura}}, \bibinfo {author} {\bibfnamefont {M.~M.}\
  \bibnamefont {Sala}}, \bibinfo {author} {\bibfnamefont {J.}~\bibnamefont
  {van~den Brink}}, \bibinfo {author} {\bibfnamefont {A.~D.}\ \bibnamefont
  {Christianson}}, \ and\ \bibinfo {author} {\bibfnamefont {D.~F.}\
  \bibnamefont {McMorrow}},\ }\href {arXiv:1707.05551 [cond-mat.str-el]}
  {\bibinfo  {journal} {arXiv:1707.05551 [cond-mat.str-el]}\ }\BibitemShut
  {NoStop}%
\bibitem [{\citenamefont {Sponza}\ \emph {et~al.}(2013)\citenamefont {Sponza},
  \citenamefont {V\'eniard}, \citenamefont {Sottile}, \citenamefont
  {Giorgetti},\ and\ \citenamefont {Reining}}]{PhysRevB.87.235102}%
  \BibitemOpen
\bibfield  {journal} {  }\bibfield  {author} {\bibinfo {author} {\bibfnamefont
  {L.}~\bibnamefont {Sponza}}, \bibinfo {author} {\bibfnamefont
  {V.}~\bibnamefont {V\'eniard}}, \bibinfo {author} {\bibfnamefont
  {F.}~\bibnamefont {Sottile}}, \bibinfo {author} {\bibfnamefont
  {C.}~\bibnamefont {Giorgetti}}, \ and\ \bibinfo {author} {\bibfnamefont
  {L.}~\bibnamefont {Reining}},\ }\href {\doibase 10.1103/PhysRevB.87.235102}
  {\bibfield  {journal} {\bibinfo  {journal} {Phys. Rev. B}\ }\textbf {\bibinfo
  {volume} {87}},\ \bibinfo {pages} {235102} (\bibinfo {year}
  {2013})}\BibitemShut {NoStop}%
\end{thebibliography}%

\end{document}